\documentclass[12pt]{article}
\usepackage{epsfig}

\begin{document}


\newcommand{\psl}{p\hskip-0.21cm\slash}
\newcommand{\be}{\begin{equation}}
\newcommand{\ee}{\end{equation}}
\newcommand{\ba}{\begin{eqnarray}}
\newcommand{\ea}{\end{eqnarray}}
\newcommand{\non}{\nonumber}

\def\A{ {\cal A }}
\def\B{ {\cal B }}
\def\C{ {\cal C }}
\def\D{ {\cal D }}
\def\N{ {\cal N }}
\def\O{ {\cal O }}
\def\L{ {\cal L }}
\def\Bcal{\tilde{\cal B}}
\def\Del{\tilde{\Delta}}
\def\sw{s_W}
\def\cw{c_W}
\def\swd{s^2_W}
\def\cwd{c^2_W}
\def\ed{e^2}
\def\mwd{M_W^2}
\def\mw{M_W}
\def\mz{M_Z}
\def\mzd{M_Z^2}
\def\mf{M_f}
\def\mh{M_h}
\def\vtau{\mbox{\boldmath $\tau$}}
\def\vW{\mbox{\boldmath $W$}}
\def\Sn#1{\mathrm{Sign} #1 }
\def\tchi{\tilde \chi}
\def\stop{\tilde t}
\def\cbeta{\cos\beta}
\def\sbeta{\sin \beta}
\def\ssf{s_{\tilde \theta_f}}
\def\csf{c_{\tilde \theta_f}}
\def\ssu{s_{\tilde \theta_u}}
\def\csu{c_{\tilde \theta_u}}
\def\ssd{s_{\tilde \theta_d}}
\def\csd{c_{\tilde \theta_d}}



\begin{titlepage}
\pagenumbering{arabic}

\vspace*{0.5cm}
{
\flushright
LC-TH-2005-005\\
}
\vspace*{1.5cm}
\begin{center}
\Large
{\bf
{\boldmath 
Complete electroweak one loop contributions to the \\
pair production cross section of MSSM charged \\
and neutral Higgs bosons in $e^{+}e^{-}$ collisions}\\
}
\vspace*{1.cm}
\normalsize{
{\bf M. Beccaria} \\
{\footnotesize INFN and Dipartimento di Fisica, 
Universit\`a di Lecce (Italy)} \\
\vspace*{0.5cm}
{\bf A. Ferrari}~(Ed.) \\
{\footnotesize Department of Radiation Sciences, 
Uppsala University (Sweden)} \\
\vspace*{0.5cm}
{\bf F.M. Renard} \\
{\footnotesize Laboratoire de Physique Th\'eorique et Astroparticules,\\ 
UMR 5207, Universit\'e Montpellier II (France)} \\
\vspace*{0.5cm}
{\bf C. Verzegnassi} \\
{\footnotesize INFN and Dipartimento di Fisica Teorica, 
Universit\`a di Trieste (Italy)} \\
}
\end{center}
\vspace{\fill}
\begin{abstract}
\noindent
In this paper, we review the production cross section for charged 
and neutral Higgs bosons pairs in $e^{+}e^{-}$ collisions beyond the 
tree level, in the framework of the Minimal Supersymmetric Standard 
Model (MSSM). A complete list of formulas for all electroweak contributions 
at the one loop level is given. A numerical code has been developed 
in order to compute them accurately and, in turn, to compare the 
MSSM Higgs bosons pair production cross sections at tree level and 
at the one loop level.
~\\

\end{abstract}
\vspace{\fill}
\begin{center}
\today
\end{center}

\end{titlepage}

\pagebreak

\setcounter{page}{1}


\section{Introduction}

One major task of a future $e^{+}e^{-}$ linear collider will be the 
exploration of the Higgs sector, in the Standard Model and beyond it. 
Despite its success, the Standard Model suffers from the appearance 
of quadratically divergent contributions to the Higgs boson mass. 
However, this problem is solved by Supersymmetry. The theoretical framework 
of our study is the minimal supersymmetric extension of the Standard Model 
(MSSM), where the Higgs spectrum consists of three unphysical Goldstone modes 
($G^{+}$, $G^{-}$ and $G^0$) as well as five physical states. Two of these 
are charged ($H^{+}$ and $H^{-}$) and, among the three neutral Higgs bosons, 
two are CP-even states, $h^{0}$ and $H^{0}$, and one is CP-odd, $A^{0}$. 
Several other aspects of the MSSM Higgs boson phenomenology are reviewed 
in~\cite{heinemeyer}.\\ 

The processes $e^+e^- \rightarrow H^+H^-,\,H^0A^0,\,h^0A^0$ that will be 
observable at future $e^+e^-$ linear colliders, such as ILC and/or CLIC, 
are among the best places where one can accurately check the Higgs 
structure, see references~\cite{hprod2} to~\cite{eeHpHm2} for details. 
At tree level, $e^+$ and $e^-$ annihilate through a photon and a $Z$ boson 
in the case of $H^+H^-$ production, and through only a $Z$ boson in the 
case of $H^0A^0$ and $h^0A^0$ production. At this level, the amplitudes 
depend on the masses of the Higgs bosons and on the mixing angle $\alpha$. 
At the one loop level, most of the MSSM parameter space is involved through 
self-energies, triangle and box diagrams. In a previous paper~\cite{sudakov} 
it was shown that, at high energy, at the leading and sub-leading (Sudakov) 
logarithmic orders, a great simplification occurs. The gauge and the SUSY 
structures of these processes reflect directly in the coefficients of the 
quadratic and linear logarithmic terms. In this high energy range, they 
depend only on a few parameters (the Standard Model inputs, the angles 
$\alpha$ and $\beta$, as well as the SUSY scale $M_{SUSY}$). The next step 
is to study more deeply the SUSY structure by looking at sub-sub-leading 
effects. First, one should determine the energy range in which the above 
Sudakov limit is an acceptable approximation and can be accurately tested. 
Then, one can study the effects of the successive sub-sub-leading terms 
(constants, $m^2/s$, etc) and classify the various parameters which control 
each of them. We should then estimate the accuracy at which these parameters 
can be measured. For these purposes, we have developed a code allowing to 
compute numerically the complete electroweak one loop contributions to the 
pair production cross section of MSSM charged and neutral Higgs bosons in 
$e^+e^-$ collisions. The purpose of the present paper is to write in an 
explicit fashion all details of the electroweak one loop contributions 
that are computed by this numerical code.\\
 
In Section~2, we review the tree level MSSM Higgs sector and we calculate 
the production cross section for $H^{+}H^{-}$, $H^{0}A^{0}$ and $h^{0}A^{0}$ 
pairs in $e^{+}e^{-}$ collisions at tree level. In the rest of the paper, we 
focus on the various one loop terms. The contributions of the initial vertices 
and of $e^{\pm}$ self-energy are given in Section~3, the intermediate gauge 
boson self-energies are discussed in Section~4, the contributions of final 
vertices and of Higgs self-energies are calculated in Section~5, and the 
effect of box diagrams are presented in Section~6. Finally, a summary and 
some outlooks (in particular a more detailed description of our numerical 
code) are given in Section~7. 

\section{Tree level calculations}

\subsection{Tree level structure of the MSSM Higgs sector}

In the MSSM, two complex scalar Higgs doublets are responsible for the 
breaking of the electroweak symmetry: 

\begin{equation}
H_1 = 
\left(
\begin{array}{c}
(v_1+\phi_1^0-i\chi_1^0)/\sqrt{2} \\
-\phi_1^{-}
\end{array}
\right),~H_2 = 
\left(
\begin{array}{c}
\phi_2^{+} \\
(v_2+\phi_2^0+i\chi_2^0)/\sqrt{2}
\end{array}
\right).
\end{equation}

They have opposite hypercharge ($Y_1 = -1$ and $Y_2 = +1$) 
and their vacuum expectation values are respectively $v_1$ 
and $v_2$. After diagonalization, one obtains the following 
states:

\begin{equation}
\left(
\begin{array}{c}
H^0 \\
h^0 \\
\end{array}
\right) = 
\left(
\begin{array}{cc}
\cos\alpha & \sin\alpha \\
-\sin\alpha & \cos\alpha \\ 
\end{array}
\right) 
\left(
\begin{array}{c}
\phi_1^0 \\
\phi_2^0 \\
\end{array}
\right),
\end{equation}
\begin{equation}
\left(
\begin{array}{c}
G^0 \\
A^0 \\
\end{array}
\right) = 
\left(
\begin{array}{cc}
\cos\beta & \sin\beta \\
-\sin\beta & \cos\beta \\ 
\end{array}
\right) 
\left(
\begin{array}{c}
\chi_1^0 \\
\chi_2^0 \\
\end{array}
\right),
\end{equation}
\begin{equation}
\left(
\begin{array}{c}
G^{\pm} \\
H^{\pm} \\
\end{array}
\right) = 
\left(
\begin{array}{cc}
\cos\beta & \sin\beta \\
-\sin\beta & \cos\beta \\ 
\end{array}
\right) 
\left(
\begin{array}{c}
\phi_1^{\pm} \\
\phi_2^{\pm} \\
\end{array}
\right).
\end{equation}

Here, $G^{+}$, $G^{-}$ and $G^0$ describe three unphysical Goldstone modes. 
The five physical states are two charged bosons ($H^{+}$ and $H^{-}$), two 
neutral scalar bosons with CP = $+1$ ($h^{0}$ and $H^{0}$) and one 
pseudoscalar neutral boson with CP = $-1$ ($A^{0}$).\\

The quadratic part of the Higgs potential, which contains the soft breaking 
masses and the gauge couplings, depends on two independent parameters, which 
are usually chosen as the mass $M_A$ of the $A^0$ boson and the ratio between 
the vacuum expectation values $\tan\beta = v_2/v_1$. The masses of the other 
physical states are expressed as follows:

\begin{equation}
M_{H}^{2} = M_{A}^{2} + M_{W}^{2},
\end{equation}
\begin{equation}
M_{H^{0},h^{0}}^{2} = 
\frac{1}{2} \left(
M_{A}^{2} + M_{Z}^{2}
\pm\sqrt{(M_{A}^{2} + M_{Z}^{2})^2-4M_{A}^{2}M_{Z}^{2}\cos^{2}2\beta}
\right).
\end{equation}

As for the mixing angle between $H^{0}$ and $h^{0}$, it is given by:

\begin{equation}
\tan2\alpha = \tan2\beta \times 
\frac{M_{A}^{2} + M_{Z}^{2}}{M_{A}^{2} 
- M_{Z}^{2}},~-\frac{\pi}{2}\le \alpha \le 0.
\end{equation}

Note that these results are only valid at tree level and they become slightly 
different when one includes radiative corrections.

\subsection{Production cross section at tree level}

\begin{figure}[ht]
\begin{center}
\epsfig{file=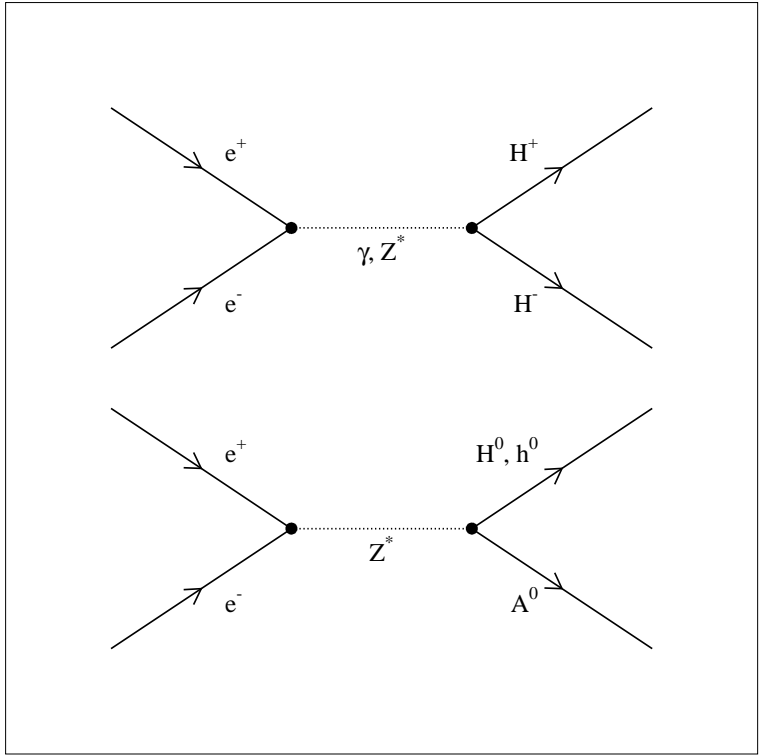,height=10cm}
\caption[]
{Feynman diagrams for the pair production of MSSM charged and neutral Higgs 
bosons in $e^{+}e^{-}$ collisions.}
\label{feynmann}
\end{center}
\end{figure}

In $e^{+}e^{-}$ collisions, charged Higgs bosons are pair produced through 
virtual photon and $Z$ boson exchange (and in top decays if $M_{H}$ is 
small enough). As for the neutral Higgs bosons, they can be produced through 
several mechanisms: $WW$ and $ZZ$ fusion processes, Higgsstrahlung or pair 
production. In this paper, we only focus on this latter process (note that 
CP conservation forbids virtual photon exchange). The Feynman diagrams of 
interest are shown in Figure~\ref{feynmann}. More details about the various 
production mechanisms and decay modes of MSSM Higgs bosons can be found 
in~\cite{zphysc74}. Here, we only focus on the total pair production cross 
sections and we ignore the different contributions of the decay channels.

The tree level production cross section can be easily derived using the 
Feynman rules. If $s_W \equiv \sin\theta_W$, $c_W \equiv \cos\theta_W$ 
and $\eta \equiv {q^2/(q^2-M^2_Z)}$, then the Born amplitudes 
$a^{Born}_{L,R}$ are:
\begin{itemize}
\item for charged Higgs bosons:
\begin{equation}
\label{born-charged}
a^{Born}_{L,R}(H^+H^-)=1-{(1-2s^2_W)\over4s^2_Wc^2_W}\eta g_{L,R}
\end{equation}
\item for neutral Higgs bosons:
\begin{equation}
\label{born-neutral}
a^{Born}_{L,R}(H^0A^0/h^0A^0)=
-{i\over4s^2_Wc^2_W}\eta g_{L,R}\left[Z_{ab}\right]
\end{equation}
\end{itemize}
where $g_{L} = 2s^2_W-1$, $g_{R} = 2s^2_W$ and 
$\left[Z_{ab}\right]= \left[-\sin(\beta-\alpha);\,\cos(\beta-\alpha)\right]$ 
for $H^0A^0$ and $h^0A^0$ final states, respectively.\\

In this paper, we use the following renormalization for the amplitude:
\begin{equation}
A={2e^2\over q^2}~\bar{v}(e^+)(\psl)(a_L P_L + a_R P_R)u(e^-),~P_{L,R} = 
\displaystyle{\frac{1 \mp \gamma_5}{2}}.
\end{equation}

The differential tree level cross sections are then given by:
\begin{equation}
\frac{d\sigma^{Born}_{L,R}}{d\cos\theta} = 
\frac{\pi\alpha_{em}^2\beta^3_H}{8q^2} 
\times (1-\cos^2\theta) \times |a^{Born}_{L,R}|^2.
\end{equation}

Here, $\beta_H$ is the velocity of the outgoing Higgs bosons. If $M_1$ 
and $M_2$ are the masses of the two outgoing Higgs bosons, then 
$\beta_H(M_1,M_2)$ is defined by:
\ba
\label{betah}
\beta_H & = & {2|p| \over \sqrt{s}} =
{1 \over s} \times \sqrt{\left( s+M^2_1-M^2_2 \right)^2 - 4sM^2_1}
\nonumber \\
        & = & \sqrt{
\left(1+\frac{M_2+M_1}{\sqrt{s}}\right)
\left(1-\frac{M_2+M_1}{\sqrt{s}}\right)
\left(1+\frac{M_2-M_1}{\sqrt{s}}\right)
\left(1-\frac{M_2-M_1}{\sqrt{s}}\right)}.~~~~~
\ea

After integration over $\cos\theta$, one gets:
\begin{itemize}
\item for charged Higgs bosons:
\begin{equation}
\sigma^{Born}_{H^{+}H^{-}} = \frac{e^{4}}{48 \pi s}
\left( 1 - \frac{4M_{H}^{2}}{s} \right)^{3/2}
\times
\left( 1 + \frac{2c'_{V}c_{V}}{1-M_{Z}^{2}/s} +
\frac{{c'_{V}}^{2}(c_{V}^{2}+c_{A}^{2})}{(1-M_{Z}^{2}/s)^{2}} \right)
\end{equation}
with $\displaystyle{
c_{V} = \frac{-1+4s^{2}_{W}}{4s_{W}c_{W}},~
c_{A} = \frac{-1}{4s_{W}c_{W}},~
c'_{V} = \frac{-1+2s^{2}_{W}}{2s_{W}c_{W}}
}$.
\item for neutral Higgs bosons:
\begin{equation}
\sigma^{Born}_{H^{0}A^{0}/h^{0}A^{0}} = \frac{e^{4}}{48 \pi s}
\times
\left(\frac{8s^4_W-4s^2_W+1}{32s^4_Wc^4_W}\right)
\times
\left[Z_{ab}\right]^2
\times
\frac{\beta^3_H(M_{H^0/h^0},M_A)}{(1-M_Z^2/s)^2}.
\end{equation}
\end{itemize}

In the decoupling limit ($M_A \gg M_Z$ and $M_A \simeq M_{H^0}$), 
$\cos(\beta-\alpha) \rightarrow 0$ and $e^{+}e^{-} \rightarrow h^0A^0$ is 
strongly suppressed, i.e. only the $H^0A^0$ pairs can be produced in 
$e^+e^-$ collisions, with a tree level cross section given by:
\begin{equation}
\sigma^{Born}_{H^{0}A^{0}} \rightarrow \frac{e^{4}}{48 \pi s}
\left( 1 - \frac{4M_{A}^{2}}{s} \right)^{3/2}
\times
\left(\frac{8s^4_W-4s^2_W+1}{32s^4_Wc^4_W}\right)
\times
\frac{1}{(1-M_Z^2/s)^2}.
\end{equation}

\begin{figure}[h!]
\begin{center}
\epsfig{file=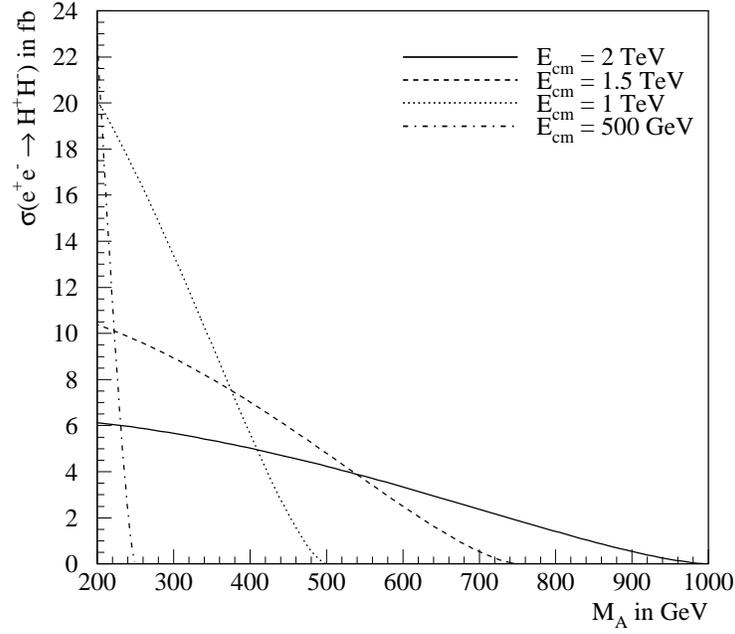,height=9.6cm}
\epsfig{file=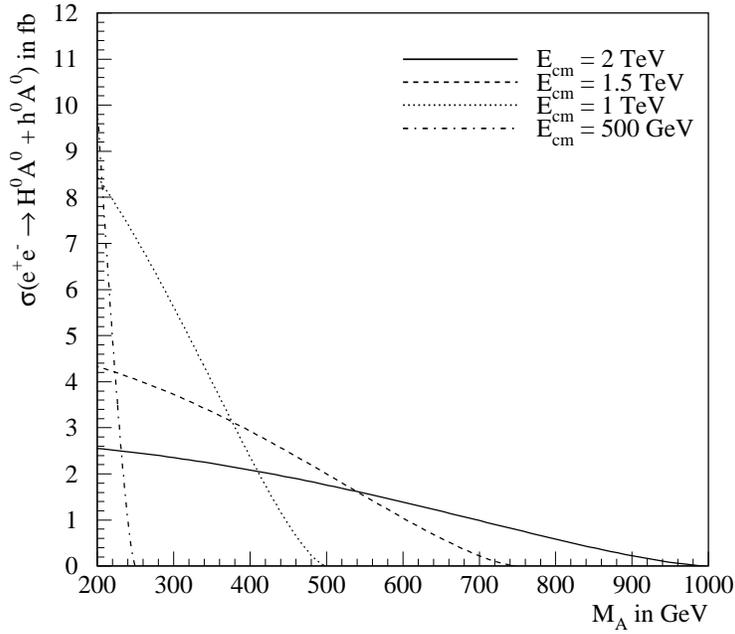,height=9.6cm}
\caption[]
{Tree level pair production cross section for charged (top) and neutral 
(bottom) Higgs bosons in $e^{+}e^{-}$ collisions, as a function of the 
MSSM parameter $M_A$ and for various centre-of-mass energies $\sqrt{s}$. 
For simplicity, we performed our calculations in the decoupling limit, 
where $H$, $H^0$ and $A^0$ are almost degenerate in mass.}
\label{cross}
\end{center}
\end{figure}

Figure~\ref{cross} shows the pair production cross section for the 
MSSM charged and neutral Higgs bosons in $e^+e^-$ collisions, at tree level, 
as a function of $M_A$ and for various values of the centre-of-mass energy.

\subsection{Complete amplitude for calculations at the one loop level}

The analytical expressions of all electroweak one loop contributions to the 
MSSM Higgs bosons pair production cross sections are given in the following: 
$e^{\pm}$ self-energy and initial vertices in Section~3, $\gamma$ and $Z$ 
self-energies with counter terms in Section~4, final vertices and Higgs 
self-energy in Section~5, and box diagrams in Section~6. The complete 
renormalized amplitude used to calculate the cross section is the sum 
of Born and one loop terms: 
\begin{eqnarray}
A(e^+e^- \to \mbox{Higgs pair}) & = & 
A^{Born}(e^+e^- \to \mbox{Higgs pair}) \nonumber\\
                                & + & 
A^{in}(e^+e^- \to \mbox{Higgs pair}) \nonumber\\
                                & + & 
A^{RG}(e^+e^- \to \mbox{Higgs pair}) + 
A^{ct}(e^+e^- \to \mbox{Higgs pair})  \nonumber\\
                                & + & 
A^{fin}(e^+e^- \to \mbox{Higgs pair}) \nonumber\\
                                & + & 
A^{box}(e^+e^- \to \mbox{Higgs pair}). 
\end{eqnarray}

The notations used in our calculations, in particular when writing vertices 
in terms of real coupling constants, are described in Appendix~A. In the 
following, we use a formalism which involves Passarino-Veltman functions, 
see Appendix~B for details.

\section{\boldmath Contribution of initial vertices and $e^{\pm}$ self-energy}

The amplitudes $a^{in}_{L,R}$ corresponding to initial triangles and 
$e^{\pm}$ self-energy are:

\begin{itemize}
\item for charged Higgs bosons:
\begin{equation}
\label{in-charged}
a^{in}_{L,R}(H^+H^-)={\Gamma^{in,\,\gamma}\over e}
+{(1-2s^2_W)\eta\over2s_Wc_W} \times {\Gamma^{in,\,Z}\over e}
\end{equation}
\item for neutral Higgs bosons:
\begin{equation}
\label{in-neutral}
a^{in}_{L,R}(H^0A^0/h^0A^0)={i\,\eta\over2s_Wc_W} \times 
{\Gamma^{in,\,Z}\over e} \times 
\left[Z_{ab}\right]
\end{equation}
\end{itemize}
where we write $\Gamma^{in,\,V} = \Gamma^{in,\,V}_LP_L + \Gamma^{in,\,V}_RP_R$ 
for $V = \gamma~\mbox{or}~Z$.\\

$\Gamma^{in,\,\gamma}$ and $\Gamma^{in,\,Z}$ are the same in 
the charged and neutral sectors, since they only depend on the 
initial state. They are obtained by summing various contributions:
\begin{eqnarray}
\Gamma^{in,\,V} & = & \Gamma^{V}_{e^+e^-}(W \nu W)  - 
\Gamma^{V}_{e^+e^-}(pinch) \nonumber\\ 
                & + & 
\Gamma^{V}_{e^+e^-}(\nu W \nu) 
+ \Gamma^{V}_{e^+e^-}(e Z e) 
+ \Gamma^{V}_{e^+e^-}(e \gamma e) \nonumber\\
                & + & 
\Gamma^{V}_{e^+e^-}(\tchi^{}_j \tilde{\nu}_L \tchi^{}_i) 
+ \Gamma^{V}_{e^+e^-}(\tchi^0_j \tilde{e}\tchi^0_i) \nonumber\\
                & + & 
\Gamma^{V}_{e^+e^-}(\tilde{\nu}_L \tchi^{}_i \tilde{\nu}_L) 
+ \Gamma^{V}_{e^+e^-}(\tilde{e}\tchi^0_i\tilde{e}) \nonumber\\
                & + & 
\Gamma^{V}_{e^+e^-}(e.s.e).
\end{eqnarray}
The particles inside the initial triangle have internal masses $m_1$, 
$m_2$ and $m_3$. They are ordered clockwise, $m_1$ being the mass of 
the particle just after the junction involving the momentum $q$. 

1) The contribution of the $W \nu W$ triangle is:
\begin{equation}
\Gamma^{V}_{e^+e^-}(W \nu W) = 
-{e\alpha_{em}\over8\pi s^2_W}f^V\tilde{C}_{WW}P_L,~f^V = 
\left\{
\begin{array}{l}
1~\mbox{for}~V=\gamma \\
c_W/s_W~\mbox{for}~V=Z
\end{array}
\right.
\end{equation}
where $\tilde C_{WW} = -12 C_{24}(W \nu W)+2-2q^2
\left[C_0(W \nu W)+C_{11}(W \nu W)+C_{23}(W \nu W)\right]$.\\

2) In $e^+e^-$ annihilations, $WW$ contributions arise in the photon and
$Z$ self-energies, as well as in the triangles connecting the photon and
the $Z$ boson to the initial $e^+e^-$ pair or to the final Higgs pair. 
Therefore, it is convenient to extract a certain part (so-called pinch) 
from such a triangle with two $W$ lines (in our case the $W \nu W$ triangle) 
and then to put it inside the photon and $Z$ self-energies contributions, 
in order to have universal charge renormalization~\cite{pinch}:
\begin{equation}
-\Gamma^{V}(pinch)=
-~{e\alpha_{em}\over4\pi s^2_W}f^VB_0(WW,q^2)P_L,~f^V = 
\left\{
\begin{array}{l}
1~\mbox{for}~V=\gamma \\
c_W/s_W~\mbox{for}~V=Z
\end{array}.
\right.
\end{equation}

3) As for the $\nu W \nu$ triangle, since neutrinos do not couple to 
photons, one has:
\begin{equation} 
\Gamma^{\gamma}_{e^+e^-}(\nu W \nu) = 0
\end{equation}
while, for the $Z$ boson, one obtains:
\begin{equation}
\Gamma^{Z}_{e^+e^-}(\nu W \nu)=-~{e\alpha_{em}\over16\pi s^3_Wc_W}
\tilde{C}_{W}P_L
\end{equation}
where $\tilde{C}_{W} = 4C_{24}(\nu W \nu)-2+2q^2
\left[C_{11}(\nu W \nu)+C_{23}(\nu W \nu)\right]$.\\

4) The contribution of the $eZe$ triangle is:
\begin{equation}
\Gamma^{\gamma}_{e^+e^-}(eZe) = 
{e\alpha_{em}\over16\pi s^2_Wc^2_W}\tilde{C}_{Z}
\left[g^2_{L}P_L+g^2_{R}P_R\right]
\end{equation}
or
\begin{equation}
\Gamma^{Z}_{e^+e^-}(eZe) = 
-~{e\alpha_{em}\over32\pi s^3_Wc^3_W}\tilde{C}_{Z}
\left[g^3_{L}P_L+g^3_{R}P_R\right]
\end{equation}
where $\tilde{C}_{Z}=4C_{24}(eZe)-2+2q^2
\left[C_{11}(eZe)+C_{23}(eZe)\right]$.\\

5) The contribution of the $e \gamma e$ triangle is:
\begin{equation}
\Gamma^{\gamma}_{e^+e^-}(e\gamma e) = 
{e\alpha_{em}\over4\pi}\tilde{C}_{\gamma}\left[P_L+P_R\right]
\end{equation}
or
\begin{equation}
\Gamma^{Z}_{e^+e^-}(e\gamma e) = 
-~{e\alpha_{em}\over8\pi s_Wc_W}\tilde{C}_{\gamma}
\left[g_{L}P_L+g_{R}P_R\right]
\end{equation}
where $\tilde{C}_{\gamma}=4C_{24}(e\gamma e)-2+2q^2
\left[C_{11}(e\gamma e)+C_{23}(e\gamma e)\right]$.\\

6) The contribution of the $\tchi^{}_j \tilde{\nu}_L\tchi^{}_i$ triangles is:
\begin{eqnarray}
\Gamma^{\gamma}_{e^+e^-}(\tchi^{}_i \tilde{\nu}_L\tchi^{}_i) & = &
\sum_{i}{e\alpha_{em}\over4\pi s^2_W}|Z^+_{1i}|^2 
(2\tilde{C}_{24}^{ii}-|M_{\tchi^{}_i}|^2\tilde{C}_0^{ii})P_L 
\end{eqnarray}
or 
\begin{eqnarray}
\Gamma^{Z}_{e^+e^-}(\tchi^{}_j \tilde{\nu}_L\tchi^{}_i)&=&
\sum_{ij}{e\alpha_{em}\over8\pi s^3_Wc_W} Z^+_{1i}Z^{+*}_{1j} \times 
\nonumber\\
&&\Big\{
\left[Z^{+*}_{1i}Z^{+}_{1j}+\delta_{ij}(c^2_W-s^2_W)\right]
2\tilde{C}_{24}^{ij} \nonumber\\
&&- \left[Z^{-}_{1i}Z^{-*}_{1j}+\delta_{ij}(c^2_W-s^2_W)\right]
M_{\tchi^{}_i}M_{\tchi^{}_j}\tilde{C}_0^{ij}\Big\}P_L 
\end{eqnarray}
where $\displaystyle{\tilde C_{24}^{ij} = 
C_{24}(\tchi^{}_j \tilde{\nu}_L\tchi^{}_i) 
- \frac{1}{4} 
+ \frac{q^2}{2}\left[C_{12}(\tchi^{}_j \tilde{\nu}_L\tchi^{}_i)+
C_{23}(\tchi^{}_j \tilde{\nu}_L\tchi^{}_i)\right]}$ 
and $\tilde{C}_0^{ij} = C_0(\tchi^{}_j \tilde{\nu}_L\tchi^{}_i)$.\\

7) As for the $\tchi^0_j \tilde{e}\tchi^0_i$ triangles, 
since neutralinos do not couple to photons, one has: 
\begin{equation}
\Gamma^{\gamma}_{e^+e^-}(\tchi^0_j \tilde{e}\tchi^0_i) = 0
\end{equation}
while, for the $Z$ boson, one obtains:
\begin{equation}
\label{chi-se-chi-1}
\Gamma^{Z}_{e^+e^-}(\tchi^0_j \tilde{e}\tchi^0_i) = 
\sum_{ij}{e\alpha_{em}\over8\pi s^3_Wc^3_W} \times
\left[K^{ij}_L P_L + K^{ij}_R P_R\right]
\end{equation}
by defining $K^{ij}_L$ and $K^{ij}_R$ as follows:
\begin{eqnarray}
K^{ij}_L & = & 
{(Z^{N*}_{1j}s_W+Z^{N*}_{2j}c_W)
(Z^{N}_{1i}s_W+Z^{N}_{2i}c_W)\over2} \times \nonumber \\
&&\left[2(Z^{N}_{3j}Z^{N*}_{3i}-Z^{N}_{4j}Z^{N*}_{4i})\tilde{C}_{24}^{ij}+
M_{\tchi^0_i}M_{\tchi^0_j}(Z^{N*}_{3j}Z^{N}_{3i}-Z^{N*}_{4j}Z^{N}_{4i})
\tilde{C}_{0}^{ij}\right] \\
K^{ij}_R & = & 
{2(Z^{N}_{1j}Z^{N*}_{1i})s^2_W} \times \nonumber \\
&&\left[2(Z^{N*}_{4j}Z^{N}_{4i}-Z^{N*}_{3j}Z^{N}_{3i})\tilde{C}_{24}^{ij}+
M_{\tchi^0_i}M_{\tchi^0_j}(Z^{N}_{4j}Z^{N*}_{4i}-Z^{N}_{3j}Z^{N*}_{3i})
\tilde{C}_{0}^{ij}\right]
\end{eqnarray}
where $\displaystyle{\tilde C_{24}^{ij} = 
C_{24}(\tchi^0_j \tilde{e} \tchi^0_i) 
- \frac{1}{4} 
+ \frac{q^2}{2}\left[C_{12}(\tchi^0_j \tilde{e} \tchi^0_i)+
C_{23}(\tchi^0_j \tilde{e} \tchi^0_i)\right]}$ 
and $\tilde{C}_0^{ij} = C_0(\tchi^0_j \tilde{e} \tchi^0_i)$.\\

8) As for the $\tilde{\nu}_L \tchi^{}_i \tilde{\nu}_L$ triangles, 
since sneutrinos do not couple to photons, one has: 
\begin{equation}
\Gamma^{\gamma}_{e^+e^-}(\tilde{\nu}_L \tchi^{}_i \tilde{\nu}_L) = 0 
\end{equation}
while, for the $Z$ boson, one obtains:
\begin{equation}
\Gamma^{Z}_{e^+e^-}(\tilde{\nu}_L \tchi^{}_i \tilde{\nu}_L)=
-~\sum_i{e\alpha_{em}\over4\pi s^3_Wc_W}|Z^+_{1i}|^2 \tilde{C}_{24}^i P_L
\end{equation}
where $\tilde{C}_{24}^i = C_{24}(\tilde{\nu}_L\tchi^{}_i\tilde{\nu}_L)$.\\

9) The contribution of the $\tilde{e}\tchi^0_i\tilde{e}$ triangles is:
\begin{eqnarray}
\label{se-chi-se-1}
\Gamma^{\gamma}_{e^+e^-}(\tilde{e}\tchi^0_i\tilde{e})&=&
\sum_i{e\alpha_{em}\over4\pi s^2_Wc^2_W} \times \nonumber\\
&&\left[ |Z^N_{1i}s_W+Z^N_{2i}c_W|^2 \tilde{C}_{24}^{i} P_L + 
4s^2_W|Z^N_{1i}|^2 \tilde{C}_{24}^{i} P_R \right] 
\end{eqnarray}
or
\begin{eqnarray}
\label{se-chi-se-2}
\Gamma^{Z}_{e^+e^-}(\tilde{e}\tchi^0_i\tilde{e})&=&
-~\sum_i{e\alpha_{em}\over4\pi s^3_Wc^3_W} \times \nonumber\\
&&\left[ (s^2_W-\frac{1}{2})
{|Z^N_{1i}s_W+Z^N_{2i}c_W|^2} \tilde{C}_{24}^{i} P_L + 
4s^4_W{|Z^N_{1i}|^2}\tilde{C}_{24}^{i} P_R \right] 
\end{eqnarray}
where $\tilde{C}_{24}^i = C_{24}(\tilde{e} \tchi^0_i \tilde{e})$.\\

10) The electron self-energy ($e.s.e$) contributions are obtained as follows:
\begin{equation}
\Gamma^{\gamma}_{e^+e^-}(e.s.e)=-e\left[\delta_LP_L + \delta_RP_R\right]
\end{equation}
or
\begin{equation}
\Gamma^{Z}_{e^+e^-}(e.s.e)={e\over2s_Wc_W}\left[\delta_L g_{L} P_L + 
\delta_R g_{R} P_R\right]
\end{equation}
where the following loops are taken into account: $(W\nu)$, $(Ze)$, 
$(\gamma e)$, $(\tchi \tilde\nu)$, $(\tchi^0 \tilde e)$.\\

For the $(W\nu)$ loop, one has:
\begin{equation}
\delta_L(W\nu)=-{\alpha_{em}\over4\pi s^2_W}
\left(B_1(W\nu,0)+{1\over2}\right),
\end{equation}
\begin{equation}
\delta_R(W\nu)=0.
\end{equation}

For the $(Ze)$ loop, one has:
\begin{equation}
\delta_{L}(Ze)=-{\alpha_{em} g^2_{L}\over8\pi s^2_Wc^2_W}
\left(B_1(Ze,0)+{1\over2}\right),
\end{equation}
\begin{equation}
\delta_{R}(Ze)=-{\alpha_{em} g^2_{R}\over8\pi s^2_Wc^2_W}
\left(B_1(Ze,0)+{1\over2}\right).
\end{equation}

For the $(\gamma e)$ loop, one has:
\begin{equation}
\delta_{L}(\gamma e) = \delta_{R}(\gamma e) = 
-~{\alpha_{em} \over2\pi }\left(B_1(\gamma e,0)+{1\over2}\right).
\end{equation}

For each $(\tchi^{}_i \tilde{\nu})$ loop, one has:
\begin{equation}
\delta_L(\tchi^{}_i \tilde{\nu})=-{\alpha_{em}\over4\pi s^2_W}
|{Z^+_{1i}}|^2B_1(\tchi^{}_i \tilde{\nu}_L,0),
\end{equation}
\begin{equation}
\delta_R(\tchi^{}_i \tilde{\nu})=0.
\end{equation}

For each $(\tchi^0_i \tilde{e})$ loop, one has:
\begin{equation}
\label{loop1}
\delta_L(\tchi^0_i \tilde{e})=-{\alpha_{em}\over8\pi s^2_Wc^2_W}
|{Z^N_{1i}s_W+Z^N_{2i}c_W}|^2B_1(\tchi^0_i \tilde{e}_{L},0),
\end{equation}
\begin{equation}
\label{loop2}
\delta_R(\tchi^0_i \tilde{e})=-~{\alpha_{em}\over2\pi c^2_W}
|{Z^{N*}_{1i}}|^2B_1(\tchi^0_i \tilde{e}_{R},0).
\end{equation}

\section{\boldmath Contribution of $\gamma$ and $Z$ self-energies}

\subsection{Definition of gauge self-energy functions}

The on-shell renormalization procedure~\cite{hollik,dabelstein} that 
allows full determination of the MSSM Higgs sector at one loop, as 
well as of the corresponding counter terms, makes use of several 
gauge self-energy functions, which are detailed in the following of 
this section~\cite{dabelstein,hep0207273,hep9807427}. Let us first 
define several useful expressions:
\begin{eqnarray}
PV1(XY,q^2) & = & \frac{M_X^2+M_Y^2}{2} - 
2B_{22}(XY,q^2)-\frac{q^2}{6}-q^2\left[B_{1}(XY,q^2)+B_{21}(XY,q^2)\right],
~~~~~~\\
PV2(XY,q^2) & = & 10B_{22}(XY,q^2) + (4q^2+M_X^2+M_Y^2)B_0(XY,q^2) \nonumber 
~~~~~~\\
& + & A(M_X^2) + A(M_Y^2) -2\left(M_X^2+M_Y^2-\frac{q^2}{3}\right),
~~~~~~\\
PV3(XY,q^2) & = & 2B_{22}(XY,q^2)-\frac{A(M^2_{X})+A(M^2_{Y})}{2}+
\frac{(q^2-M_{X}^2-M_{Y}^2)}{2}B_0(XY,q^2).
~~~~~~ 
\end{eqnarray}

\underline{a) Photon self-energies:}\\

The photon self-energy is defined as:
\begin{equation}
A_{\gamma\gamma}(q^2)=\Sigma_{\gamma\gamma}(q^2)+A_{\gamma\gamma}(pinch).
\end{equation}

The pinch term is given by:
\begin{equation}
A_{\gamma\gamma}(pinch)=-{\alpha_{em}\over\pi}q^2B_0(WW,q^2).
\end{equation}

The self-energy term without pinch $\Sigma_{\gamma\gamma}(q^2)$ is 
the sum of various contributions:
\begin{eqnarray}
\Sigma_{\gamma\gamma}(q^2) & = & \Sigma_{\gamma\gamma}(\mbox{g+H}) + 
\Sigma_{\gamma\gamma}(ff) \nonumber \\ 
& + & \Sigma_{\gamma\gamma}(\tchi \tchi) + 
\Sigma_{\gamma\gamma}(\tilde{f}\tilde{f}). 
\end{eqnarray}

The contribution of the gauge and Higgs sectors is:
\begin{eqnarray}
\Sigma_{\gamma \gamma}(\mbox{g+H}) & 
= -\displaystyle{\frac{\alpha_{em}}{2\pi}} &
\Big\{2B_{22}(HH,q^2)-A(M^2_{H})+6B_{22}(WW,q^2) \nonumber \\
&&-3A(M^2_W)+2q^2B_0(WW,q^2)+ \frac{q^2}{3}\Big\}.
\end{eqnarray}

The contribution of the fermion pairs is:
\begin{eqnarray}
\Sigma_{\gamma \gamma}(f\bar f) & = & 
\sum_f \frac{\alpha_{em}N_c^f Q_f^2}{\pi} 
\Big\{ PV{1}(ff,q^2) + M_f^2B_0(ff,q^2) \Big\}.
\end{eqnarray}

The contribution of the chargino pairs is:
\begin{eqnarray}
\Sigma_{\gamma \gamma}(\tchi \tchi) & = & 
\sum_i  \frac{\alpha_{em}}{\pi} \Big\{ PV{1}(\tchi^{}_i \tchi^{}_i,q^2) + 
M^2_{\tchi^{}_i}B_0(\tchi^{}_i \tchi^{}_i,q^2) \Big\}.
\end{eqnarray}

The contribution of the sfermion pairs is:
\begin{eqnarray}
\Sigma_{\gamma \gamma}(\tilde{f}\tilde{f}) & = & 
-\sum_{\tilde{f}} \frac{\alpha_{em} N_c^f Q_f^2}{2\pi}  \sum_{i=1,2}
\Big\{2B_{22}(\tilde{f}_i\tilde{f}_i,q^2)-A(M^2_{\tilde{f}_i})\Big\}.
\end{eqnarray}

Here, $\tilde{f}_1,\,\tilde{f}_2$ account for $\tilde{f}_L,\,\tilde{f}_R$ in 
the case of unmixed sfermions, or for the physical states obtained after 
mixing (i.e. $\tilde{t}_1$, $\tilde{t}_2$, $\tilde{b}_1$, $\tilde{b}_2$ in the 
case of third generation squarks). The coupling between a photon and a
sfermion pair is the same with and without mixing.\\

\underline{b) $Z$ self-energies:}\\

The $Z$ boson self-energy is defined as:
\begin{equation}
A_{ZZ}(q^2)=\Sigma_{ZZ}(q^2)+A_{ZZ}(pinch).
\end{equation}

The pinch term is given by:
\begin{equation}
A_{ZZ}(pinch)=-{\alpha_{em}c^2_W\over\pi s^2_W}(q^2-M^2_Z)B_0(WW,q^2).
\end{equation}

The self-energy term without pinch $\Sigma_{ZZ}(q^2)$ is 
the sum of various contributions:
\begin{eqnarray}
\Sigma_{ZZ}(q^2) & = & \Sigma_{ZZ}(\mbox{g+H}) + \Sigma_{ZZ}(ff) \nonumber \\ 
& + & \Sigma_{ZZ}(\tchi \tchi) + \Sigma_{ZZ}(\tchi^0\tchi^0) + 
\Sigma_{ZZ}(\tilde{f}\tilde{f}). 
\end{eqnarray}

The contribution of the gauge and Higgs sectors is:
\begin{eqnarray}
\Sigma_{ZZ}(\mbox{g+H}) & 
= \displaystyle{\frac{\alpha_{em}}{4 \pi \swd \cwd}} &
\Big\{ 
\frac{1}{4}\left[A(M^2_{h^0})+A(M^2_{H^0})+A(M^2_{A})+A(M^2_{Z})\right] 
\nonumber \\
&&
+\sin^2(\beta-\alpha)\left[\mzd B_0(Zh^0,q^2)-B_{22}(Zh^0,q^2)-
B_{22}(A^0H^0,q^2)\right] \nonumber \\
&&
+\cos^2(\beta-\alpha)\left[\mzd B_0(ZH^0,q^2)-B_{22}(ZH^0,q^2)-
B_{22}(A^0h^0,q^2)\right] \nonumber \\ 
&&
-\frac{1}{2}\cos^2(2\theta_W)
\left[2B_{22}(HH,q^2) - A(M^2_{H}) \right] 
\nonumber \\
&&
-\left[8\cw^4+\cos^2(2\theta_W)\right]B_{22}(WW,q^2) \nonumber \\
&&
-\left[4\cw^4 q^2+2\mwd \cos(2\theta_W)\right]B_0(WW,q^2) \nonumber \\
&&
+\frac{1}{2}\left[12\cw^4-4\cwd+1\right]A(M^2_W)-\frac{2}{3}\cw^4 q^2 
\Big\}.
\end{eqnarray}

The contribution of the fermion pairs is:
\begin{eqnarray}
\Sigma_{ZZ}(f\bar f) & = & \sum_f \frac{\alpha_{em}N_c^f}{4\pi s^2_W c^2_W} 
\Big\{
(g_{Vf}^2+g_{Af}^2)PV1(M_f^2,q^2)+(g_{Vf}^2-g_{Af}^2)M_f^2B_0(f\bar f,q^2)
\Big\}~~~
\end{eqnarray}
where $g_{Vf} = T^3_{f}(1-4|Q_f|s^2_W)$ and $g_{Af} = T^3_{f}$.\\

The contribution of the chargino pairs is:
\begin{eqnarray}
\Sigma_{ZZ}(\tchi \tchi) & =\displaystyle{{2\over16\pi^2}}~
\displaystyle{\sum_{ij}}&
\Big\{PV1(\tchi^{}_i \tchi^{}_j,q^2)
\left[\O^{ZL*}_{ij}\O^{ZL}_{ij}+\O^{ZR*}_{ij}\O^{ZR}_{ij}\right] \nonumber\\
&&
+ M_{\tchi^{}_i}M_{\tchi^{}_j}
\left[\O^{ZL*}_{ij}\O^{ZR}_{ij}+\O^{ZL}_{ij}\O^{ZR*}_{ij}\right]
B_0(\tchi^{}_i \tchi^{}_j,q^2)\Big\}.
\end{eqnarray}

The contribution of the neutralino pairs is:
\begin{eqnarray}
\Sigma_{ZZ}(\tchi^0\tchi^0) &=\displaystyle{{1\over16\pi^2}}~
\displaystyle{\sum_{ij}}&
\Big\{PV1(\tchi^0_i\tchi^0_j,q^2)
\left[\O^{0L*}_{ij}\O^{0L}_{ij}+\O^{0R*}_{ij}\O^{0R}_{ij}\right] 
\nonumber\\
&&
+ M_{\tchi^0_i}M_{\tchi^0_j}
\left[\O^{0L*}_{ij}\O^{0R}_{ij}+\O^{0L}_{ij}\O^{0R*}_{ij}\right]
B_0(\tchi^0_i\tchi^0_j,q^2)\Big\}.
\end{eqnarray}

The contribution of sfermion pairs is:
\begin{itemize}
\item for unmixed sfermions:
\begin{equation}
\Sigma_{ZZ}^{light} (\tilde{f}\tilde{f})= 
-\sum_{\tilde{f}_{L,R}}{\alpha_{em}N_c^f\over2\pi}
\left(\frac{g^0_{Z \tilde{f} \tilde{f}}}{e}\right)^2
\Big\{2B_{22}(\tilde{f}\tilde{f},q^2)-A(M^2_{\tilde{f}})\Big\},
\end{equation}
\item with sfermion mixing (third generation squarks):
\begin{eqnarray}
\Sigma_{ZZ}^{heavy}(\tilde{f}\tilde{f}) & = 
-\displaystyle{{3\alpha_{em}\over2\pi s^2_Wc^2_W}}~
\displaystyle{\sum_{\tilde{f}=\tilde{t},\tilde{b}}} &
\Big\{ 
{c^2_{\tilde{f}}s^2_{\tilde{f}}\over2}\left[B_{22}(\tilde{f}_1\tilde{f}_2,q^2)
+B_{22}(\tilde{f}_2\tilde{f}_1,q^2)\right] \nonumber \\
&&
+2(T^3_{f_{L}}c^2_{\tilde{f}}-s^2_WQ_{f})^2
B_{22}(\tilde{f}_1\tilde{f}_1,q^2) \nonumber \\
&&
-\left[c^2_{\tilde{f}}(T^3_{f_{L}}-s^2_WQ_{f})^2
+s^2_{\tilde{f}}Q_{f}^2s^4_W \right]A(M^2_{\tilde{f}_1}) \nonumber \\ 
&&
+2(T^3_{f_{L}}s^2_{\tilde{f}}-s^2_WQ_{f})^2
B_{22}(\tilde{f}_2\tilde{f}_2,q^2) \nonumber \\
&&
-\left[s^2_{\tilde{f}}(T^3_{f_{L}}-s^2_WQ_{f})^2
+c^2_{\tilde{f}}Q_{f}^2s^4_W \right]A(M^2_{\tilde{f}_2})\Big\}.~~~~~~
\end{eqnarray}
\end{itemize}

\underline{c) Mixed $\gamma Z$ self-energies:}\\

The mixed $\gamma Z$ self-energy is defined as:
\begin{equation}
A_{\gamma Z}(q^2)=\Sigma_{\gamma Z}(q^2)+A_{\gamma Z}(pinch).
\end{equation}

The pinch term is given by:
\begin{equation}
A_{\gamma Z}(pinch)=
-{\alpha_{em}c_W\over\pi s_W}(q^2-\frac{M^2_Z}{2})B_0(WW,q^2).
\end{equation}

The self-energy term without pinch $\Sigma_{\gamma Z}(q^2)$ is 
the sum of various contributions:
\begin{eqnarray}
\Sigma_{\gamma Z}(q^2) & = & 
\Sigma_{\gamma Z}(\mbox{g+H}) + \Sigma_{\gamma Z}(ff) \nonumber \\ 
& + & 
\Sigma_{\gamma Z}(\tchi \tchi) + \Sigma_{\gamma Z}(\tilde{f}\tilde{f}). 
\end{eqnarray}

The contribution of the gauge and Higgs sectors can be expressed in several
ways. Here, we choose the definition given in~\cite{hep9807427}:
\begin{eqnarray}
\Sigma_{\gamma Z}(\mbox{g+H}) &
=\displaystyle{\frac{\alpha_{em}}{4\pi}} &
\Big\{ 
-2\frac{c_W^2-s_W^2}{s_W c_W} 
\left[ B_{22}(HH,q^2)+B_{22}(WW,q^2) \right] \nonumber \\
&&
+\frac{c_W^2-s_W^2}{s_W c_W}\left[A(M^2_{H})+A(M^2_W)\right] + 
\frac{c_W}{s_W}\left[6A(M^2_W)-4M_W^2\right] \nonumber \\
&&
+2\frac{c_W}{s_W} B_{22}(WW,q^2)-2s_Wc_WM_Z^2B_0(WW,q^2) \nonumber \\
&&
-\frac{c_W}{s_W} PV2(WW,q^2) \Big\}.
\end{eqnarray}

The contribution of the fermion pairs is:
\begin{eqnarray}
\Sigma_{\gamma Z}(f\bar f) & = & 
\sum_f \frac{\alpha_{em}N_c^f Q_f g_{Vf}}{2\pi s_W c_W} 
\Big\{ PV{1}(ff,q^2) + M_f^2B_0(ff,q^2) \Big\}.
\end{eqnarray}

The contribution of the chargino pairs is:
\begin{eqnarray}
\Sigma_{\gamma Z}(\tchi \tchi) & 
\displaystyle{{2\over16\pi^2}}~\displaystyle{\sum_{ij}} &
\Big\{PV1(\tchi^{}_i\tchi^{}_j,q^2)
\left[\O^{\gamma L*}_{ij}\O^{ZL}_{ij}+\O^{\gamma R*}_{ij}\O^{ZR}_{ij}\right]
\nonumber\\
&&
+ M_{\tchi^{}_i}M_{\tchi^{}_j}
\left[\O^{\gamma L*}_{ij}\O^{ZR}_{ij}+\O^{ZL}_{ij}\O^{\gamma R*}_{ij}\right]
B_0(\tchi^{}_i\tchi^{}_j,q^2)\Big\}.
\end{eqnarray}

The contribution of the sfermion pairs is:
\begin{itemize}
\item for unmixed sfermions:
\begin{equation}
\Sigma_{\gamma Z}^{light}(\tilde{f}\tilde{f}) = 
-{\alpha_{em}\over2\pi}\sum_{\tilde{f}_{L,R}}N_c^fQ_f 
\left(-\frac{g^0_{Z \tilde{f} \tilde{f}}}{e}\right) \times
\Big\{2B_{22}(\tilde{f}\tilde{f},q^2)-A(M^2_{\tilde{f}})\Big\},
\end{equation}
\item with sfermion mixing (third generation squarks):
\begin{eqnarray}
\Sigma_{\gamma Z}^{heavy}(\tilde{f}\tilde{f}) & 
= -\displaystyle{{3\alpha_{em}\over2\pi s_W c_W}}
\displaystyle{\sum_{\tilde{f}=\tilde{t},\tilde{b}}} Q_f&
\Big\{ (T^3_{f_{L}}c^2_{\tilde{f}}-s^2_WQ_{f})
\left[2B_{22}(\tilde{f}_1\tilde{f}_1,q^2) -A(M^2_{\tilde{f}_1}) \right] 
\nonumber \\
&&
+ (T^3_{f_{L}}s^2_{\tilde{f}}-s^2_WQ_{f})
\left[2B_{22}(\tilde{f}_2\tilde{f}_2,q^2) -A(M^2_{\tilde{f}_2}) \right] 
\Big\}.~~~~~~
\end{eqnarray}
\end{itemize}

\underline{d) $W$ self-energies:}\\

The self-energy term without pinch $\Sigma_{WW}(q^2)$ is the sum of various 
contributions:
\begin{eqnarray}
\Sigma_{WW}(q^2) & = & \Sigma_{WW}(\mbox{g+H}) + \Sigma_{WW}(ff') \nonumber \\ 
& + & \Sigma_{WW}(\tchi \tchi^0) + \Sigma_{WW}(\tilde{f}\tilde{f}'). 
\end{eqnarray}

The contribution of the gauge and Higgs sectors can be expressed in several
ways. Here, we choose the definition given in~\cite{hep9807427}:
\begin{eqnarray}
\Sigma_{WW}(\mbox{g+H}) &
= \displaystyle{\frac{\alpha_{em}}{4\pi s_W^2}} &
\Big\{ -\sin^2(\beta-\alpha) 
\left[ 
B_{22}(H H^0,q^2)+B_{22}(Wh^0,q^2) 
\right] \nonumber \\
&& 
-\cos^2(\beta-\alpha)  
\left[ 
B_{22}(H h^0,q^2)+B_{22}(WH^0,q^2) 
\right] \nonumber\\
&& 
-B_{22}(WZ,q^2) - B_{22}(H A^0,q^2) \nonumber\\
&& 
+2s_W^2 B_{22}(\gamma W,q^2) +2c_W^2 B_{22}(WZ,q^2) \nonumber\\
&& 
+\frac{1}{4} \left[ A(M^2_{H^0})+A(M^2_{h^0})+A(M^2_Z)+A(M^2_{A}) \right] 
\nonumber \\
&& 
+ \frac{1}{2} \left[A(M^2_W)+ A(M^2_{H}) \right] \nonumber\\
&& 
+ M_W^2
\left[
\sin^2(\beta-\alpha)B_0(h^0W,q^2)+\cos^2(\beta-\alpha)B_0(H^0W,q^2)
\right] \nonumber \\
&& 
+ M_W^2
\left[
s_W^2B_0(W\gamma,q^2)+\frac{s_W^4}{c_W^2} B_0(WZ,q^2)
\right] \nonumber \\
&& 
+\left[3A(M^2_W)-2M_W^2\right] + 
c_W^2\left[3A(M^2_Z)-2M_Z^2\right] \nonumber \\
&& 
- c_W^2 PV2(ZW,q^2) - s_W^2PV2(\gamma W,q^2) 
\Big\}.
\end{eqnarray}

The contribution of the fermion pairs is:
\begin{equation}
\Sigma_{WW}(ff') = \sum_{(ff')} \frac{\alpha_{em} N_c^f}{4 \pi\swd} 
PV_3(ff',q^2).
\end{equation}

The contribution of the gaugino pairs is:
\begin{eqnarray}
\Sigma_{WW}(\tchi \tchi^0) & 
= \displaystyle {\frac{\alpha_{em}}{2\pi \swd}~\sum_{ij}} &
\Big\{
\left(\O_{ij}^{WL}\O_{ij}^{WL*}+\O_{ij}^{WR}\O_{ij}^{WR*}\right)
PV_3(\tchi^{}_i \tchi_j^0,q^2) \nonumber \\
&&
+\left(\O_{ij}^{WL}\O_{ij}^{WR*}+\O_{ij}^{WL*}\O_{ij}^{WR}\right)
M_{\tchi^{}_i}M_{\tchi_j^0}B_0(\tchi^{}_i \tchi_j^0,q^2)
\Big \}.
\end{eqnarray}

The contribution of the sfermion pairs is:
\begin{itemize}
\item for unmixed sfermions:
\begin{equation}
\Sigma_{WW}^{light}(\tilde{f}\tilde{f}') = - \frac{\alpha_{em}}{2 \pi\swd}
\sum_{(ff')} 
N_c^f \left[
B_{22}(\tilde{f}\tilde{f}',q^2)
-\frac{A(M^2_{\tilde{f}})+A(M^2_{\tilde{f}'})}{4}
\right],
\end{equation}
\item with sfermion mixing (third generation squarks):
\begin{eqnarray}
\Sigma_{WW}^{heavy}(\tilde{f}\tilde{f}') & = 
-\displaystyle{\frac{3\alpha_{em}}{2 \pi\swd}} & \Big \{
c^2_{\tilde{t}} c^2_{\tilde{b}} B_{22}(\tilde t_1 \tilde b_1,q^2) +
c^2_{\tilde{t}} s^2_{\tilde{b}} B_{22}(\tilde t_1 \tilde b_2,q^2)  
\nonumber \\
&& 
+s^2_{\tilde{t}} c^2_{\tilde{b}} B_{22}(\tilde t_2 \tilde b_1,q^2) +
s^2_{\tilde{t}} s^2_{\tilde{b}} B_{22}(\tilde t_2 \tilde b_2,q^2)  
\nonumber \\
&& -\frac{1}{4}\left[
c^2_{\tilde{t}} A(M^2_{\tilde t_1})+s^2_{\tilde{t}} A(M^2_{\tilde t_2})
+c^2_{\tilde{b}} A(M^2_{\tilde b_1})+s^2_{\tilde{b}} A(M^2_{\tilde b_2})
\right]
\Big \}.~~~~~~
\end{eqnarray}
\end{itemize}

\subsection{Charged Higgs sector}

For $e^+e^- \rightarrow H^+H^-$, the on-shell renormalization 
procedure leads to the following $RG$ terms:
\begin{equation}
\label{se-charged}
a^{RG}_{L,R}(H^+H^-)=
\left[
{\eta^2  (1-2s^2_W)g_{L,R} \over 4s^2_Wc^2_W}
\right]
\frac{A_{ZZ}(q^2)}{q^2} 
-
\left[
{\eta (1-2s^2_W-g_{L,R}) \over 2s_Wc_W} \right]
\frac{A_{\gamma Z}(q^2)}{q^2}
-
\frac{A_{\gamma\gamma}(q^2)}{q^2}
\end{equation}
with the corresponding counter terms:
\begin{eqnarray}
a^{ct}_{L,R}(H^+H^-) & = & 
\label{ct-charged}
\left[\Pi_{\gamma\gamma}(0)+
{2s_W\Sigma_{\gamma Z}(0)\over c_WM^2_Z}
\right] \times \left[
1+\eta g_{L,R}{2s^2_W-1\over4s^2_Wc^2_W}
\right] \nonumber \\
& + & {\eta g_{L,R}(2s^2_W-1)\over4s^2_Wc^2_W} \times
\left[{\Sigma_{ZZ}(M^2_Z)\over q^2-M^2_Z}\right] \nonumber\\
& + & {\eta \over4s^2_Wc^2_W} \times
\left[
{\Sigma_{ZZ}(M^2_Z)\over M^2_Z}-
{\Sigma_{WW}(M^2_W)\over M^2_W})
\right] \times
\left[{g_{L}\over s^2_W}P_L+g_{R}P_R\right].
\end{eqnarray}

Here, $\Pi_{\gamma\gamma}(q^2) \equiv 
\displaystyle{\frac{\Sigma_{\gamma\gamma}(q^2)}{q^2}}$ 
(no pinch term) and $\Pi_{\gamma\gamma}(0)$ 
is thus simply obtained as follows:
\begin{equation}
\Pi_{\gamma\gamma}(0) = 
\left( \frac{d\Sigma_{\gamma\gamma}}{dq^2}\right)_{q^2=0}.
\end{equation}

\subsection{Neutral Higgs sector}

For $e^+e^- \rightarrow H^0A^0/h^0A^0$, the on-shell renormalization 
procedure leads to the following $RG$ terms:
\begin{equation}
\label{se-neutral}
a^{RG}_{L,R}(H^0A^0/h^0A^0)=
{i \over q^2} \left[Z_{ab}\right] \times \left(
{\eta^2g_{L,R}\over 4s^2_Wc^2_W} A_{ZZ}(q^2) - 
{\eta \over 2s_Wc_W} A_{\gamma Z}(q^2) \right).
\end{equation}

As for the counter terms, we only consider those corresponding to electroweak
couplings and gauge boson masses here (the counter terms corresponding to 
$H^0A^0$ or $h^0A^0$ final states will be calculated in Section~\ref{nhct}):
\begin{eqnarray}
\label{ct-neutral}
a^{ct}_{L,R}(H^0A^0/h^0A^0) & 
= & i \eta \left[Z_{ab}\right] 
\left[
{1-2s^2_W+2s^4_W\over4s^4_Wc^2_W}P_L+{1\over2c^2_W}P_R
\right]
\left[
{\Sigma_{WW}(M^2_W)\over M^2_W}-
{\Sigma_{ZZ}(M^2_Z)\over M^2_Z}
\right] \nonumber \\ 
& - & 
i \eta \left[Z_{ab}\right]
{g_{L,R}\over4s^2_Wc^2_W}
\left[\Pi_{\gamma\gamma}(0)+{\Sigma_{ZZ}(M^2_Z)\over q^2-M^2_Z}+
{2s_W\Sigma_{\gamma Z}(0)\over c_WM^2_Z}\right].
\end{eqnarray}

\section{Contribution of final vertices and Higgs self-energies}

\subsection{Diagram structures for final triangles}

Several useful expressions are needed when estimating the contributions of 
the final vertices. The particles inside the final triangle have internal 
masses $m_1$, $m_2$ and $m_3$. They are ordered clockwise, $m_1$ being the 
mass of the particle just after the junction involving the momentum $q$.\\

Let $P_{f1}$ and $P_{f2}$ (respectively $M_1$ and $M_2$) be the momenta 
(respectively the masses) of the two outgoing Higgs bosons (i.e. $H^+H^-$ 
or $H^0A^0$ or $h^0A^0$), then one has:
\begin{eqnarray}
P^2_{f1} & = & M^2_{1}, \\
P^2_{f2} & = & M^2_{2}, \\
P_{f1}P_{f2} & = & \frac{q^2 -(M^2_{1}+M^2_{2})}{2}.
\end{eqnarray}

\underline{a) Tri1-type triangles:}
\begin{eqnarray}
\C_1 & = & \frac{1}{6}+6(C_{001}-C_{002})+P^2_{f1}C_{111}-P^2_{f2}C_{222}
\nonumber \\
& + & (2P_{f1}P_{f2}-P^2_{f1})C_{112}+(P^2_{f2}-2P_{f1}P_{f2})C_{122}
\nonumber\\
& + & 
2\left[
P_{f1}P_{f2}C_{21}-P^2_{f2}C_{22}+(P^2_{f2}-P_{f1}P_{f2})C_{23}-C_{24}
\right] 
\nonumber\\
& - & (2P_{f1}P_{f2}+P^2_{f1})(C_{11}-C_{12}).
\end{eqnarray}

\underline{b) Tri2-type triangles:}
\begin{eqnarray}
\C_2 & = & (8P_{f1}P_{f2}+6P^2_{f1}+2P^2_{f2})C_0 +
(8P_{f1}P_{f2}+7P^2_{f1}+P^2_{f2})C_{11}  
\nonumber\\
& + & (P^2_{f1}-P^2_{f2})C_{12} + 
(2P_{f1}P_{f2}+2P^2_{f1})C_{21} + 
(2P_{f1}P_{f2}+2P^2_{f2})C_{22} \nonumber\\ 
& + & (4P_{f1}P_{f2}+2P^2_{f1}+2P^2_{f2})C_{23} + 12C_{24} - 2.
\end{eqnarray}

\underline{c) Tri3-type triangles:}
\begin{eqnarray}
\C_3 & = & \frac{1}{6}+6(C_{001}-C_{002}) + 
P^2_{f1}C_{111} - P^2_{f2}C_{222} + 
(2P_{f1}P_{f2}-P^2_{f1})C_{112} \nonumber\\
& + & (P^2_{f2}-2P_{f1}P_{f2})C_{122} + 
P^2_{f1}C_{21}-(2P_{f1}P_{f2}+P^2_{f2})C_{22} \nonumber\\
& - & 2P^2_{f1}C_{23} - q^2C_{12}-2C_{24}+{1\over2},
\end{eqnarray}
\begin{eqnarray}
\C'_3 & = & C_{11}-C_{12},
\end{eqnarray}
\begin{eqnarray}
\C''_3 & = & C_0+C_{11}-C_{12}.
\end{eqnarray}

\underline{d) Tri4-type triangles:}
\begin{eqnarray}
\C_4 & = & C_{12}-C_{11}-2C_0.
\end{eqnarray}

\underline{e) Tri5-type triangles:}
\begin{eqnarray}
\C_5 & = &C_{11}-C_{12}-C_0.
\end{eqnarray}

\underline{f) Tri6-type triangles:}
\begin{eqnarray}
\C_6 & = & C_{11}-C_{12}.
\end{eqnarray}

\subsection{Charged Higgs sector}

The amplitudes $a^{fin}_{L,R}$ corresponding to final vertices with 
$H^+H^-$ states are:
\begin{equation}
\label{fin-charged}
a^{fin}_{L,R}(H^+H^-)={\Gamma^{fin,\,\gamma}(H^+H^-)\over 2e}
-{\eta g_{L,R}\over2s_Wc_W} \times {\Gamma^{fin,\,Z}(H^+H^-)\over 2e}.
\end{equation}

Here, $\Gamma^{fin,\,\gamma}(H^+H^-)$ and $\Gamma^{fin,\,Z}(H^+H^-)$ are 
obtained by summing the contributions of various triangles and of charged 
Higgs self-energy terms, as detailed in the following.\\

For the photon, one has:
\begin{eqnarray}
\Gamma^{fin,\,\gamma}(H^+H^-) & = & \Gamma^{\gamma}_{H^+H^-}(\mbox{1ch}) +  
\Gamma^{\gamma}_{H^+H^-}(2) - \Gamma^{\gamma}_{H^+H^-}(2,pinch) 
\nonumber \\
& + &  \Gamma^{\gamma}_{H^+H^-}(3f) 
+ \Gamma^{\gamma}_{H^+H^-}(\tchi \tchi^0 \tchi) +
\Gamma^{\gamma}_{H^+H^-}(\mbox{6ch}) + \Gamma^{\gamma}_{H^+H^-}(6\tilde{f})  
\nonumber \\
& + &  \Gamma^{\gamma}_{H^+H^-}(\mbox{4-leg}) + 
\Gamma^{\gamma}_{H^+H^-}(H.s.e).
\end{eqnarray}

For the $Z$ boson, one has:
\begin{eqnarray}
\Gamma^{fin,\,Z}(H^+H^-) & = & \Gamma^{Z}_{H^+H^-}(\mbox{1ch}) +  
\Gamma^{Z}_{H^+H^-}(\mbox{1n}) +  
\Gamma^{Z}_{H^+H^-}(2) - \Gamma^{Z}_{H^+H^-}(2,pinch)
\nonumber \\
& + &  \Gamma^{Z}_{H^+H^-}(3f) + 
\Gamma^{Z}_{H^+H^-}(\tchi \tchi^0 \tchi) +
\Gamma^{Z}_{H^+H^-}(\tchi^0 \tchi \tchi^0) + \Gamma^{Z}_{H^+H^-}(4) 
\nonumber \\
& + &  \Gamma^{Z}_{H^+H^-}(\mbox{6ch}) +  \Gamma^{Z}_{H^+H^-}(\mbox{6n}) 
+ \Gamma^{Z}_{H^+H^-}(6\tilde{f}) 
\nonumber \\
& + &  \Gamma^{Z}_{H^+H^-}(\mbox{4-leg}) 
+ \Gamma^{Z}_{H^+H^-}(H.s.e).
\end{eqnarray}

\subsubsection{Tri1-type triangles}

The Tri1-type triangles contribute to $\Gamma^{fin,\,\gamma}(H^+H^-)$ with:
\begin{eqnarray}
\Gamma^{\gamma}_{H^+H^-}(\mbox{1ch})& = & -\frac{e^3}{8\pi^2} 
\left[
\C_1(H \gamma H)+
\left(\frac{1-2s_W^2}{2s_Wc_W}\right)^2 \C_1(H Z H)
\right].
\end{eqnarray}

The Tri1-type triangles contribute to $\Gamma^{fin,\,Z}(H^+H^-)$ with:
\begin{eqnarray}
\Gamma^{Z}_{H^+H^-}(\mbox{1ch})& = & -\frac{e^3}{8\pi^2}  
\left(\frac{1-2s_W^2}{2s_Wc_W}\right)
\left[
\C_1(H \gamma H)+
\left(\frac{1-2s_W^2}{2s_Wc_W}\right)^2 \C_1(H Z H)
\right],~~~
\end{eqnarray}
\begin{eqnarray}
\Gamma^{Z}_{H^+H^-}(\mbox{1n})& 
= \displaystyle{\frac{e^3}{64\pi^2s_W^3c_W}}~\times &
\Big\{
\sin^2(\beta-\alpha)
\left[\C_1(H^0 W A^0)+\C_1(A^0 W H^0)\right] \nonumber \\
&& + \cos^2(\beta-\alpha)
\left[\C_1(h^0 W A^0)+\C_1(A^0 W h^0)\right]
\Big\}.
\end{eqnarray}

\subsubsection{Tri2-type triangles}

With $f^V = 
\left\{
\begin{array}{l}
1~\mbox{for}~V=\gamma \\
c_W/s_W~\mbox{for}~V=Z
\end{array}
\right.$ 
\hspace*{-2mm}, the contribution of the Tri2-type triangles is:
\begin{eqnarray}
\Gamma^{V}_{H^+H^-}(2)& 
= \displaystyle{\frac{e^3 f^V}{64 \pi^2 s_W^2}}~\times &
\Big\{
\sin^2(\beta-\alpha)\,\C_2(W H^0 W) \nonumber \\
&& + \cos^2(\beta-\alpha)\,\C_2(W h^0 W) \nonumber \\ 
&& + \C_2(W A^0 W)
\Big\}.
\end{eqnarray}

However, one must also take into account the pinch term:
\begin{eqnarray}
-\Gamma^{V}_{H^+H^-}(2,pinch)& = & -\frac{e^3 f^V}{8 \pi^2 s_W^2} 
\times B_0(WW,q^2).
\end{eqnarray}
 
\subsubsection{Tri3-type triangles}

The Tri3-type triangles contribute to both $\Gamma^{fin,\,\gamma}(H^+H^-)$ and 
$\Gamma^{fin,\,Z}(H^+H^-)$ with:
\begin{eqnarray}
\Gamma^{V}_{H^+H^-}(3f) & = & -\frac{1}{8\pi^2}\sum_{(ff')}
\frac{N_c^f e^3}{2 s_W^2 M_W^2} \times \nonumber \\
&& \Big\{ 
\left[ g_{V R f} M_f^2 \mbox{cot}^2\beta +  
g_{V L f} M_{f'}^2 \tan^2\beta \right] \C_3(ff'f) \nonumber \\
&& 
-\left[ g_{V R f'} M_{f'}^2 \tan^2\beta +  
g_{V L f'} M_f^2  \mbox{cot}^2\beta \right] \C_3(f'ff') \nonumber \\
&&
+ 2 M_f^2 M_{f'}^2 (g_{V R f}+g_{V L f})\,\C_3'(ff'f) \nonumber \\ 
&&
- 2 M_f^2 M_{f'}^2 (g_{V R f'}+g_{V L f'})\,\C_3'(f'ff') \nonumber \\
&&
+ M_f^2 \left[ g_{V L f} M_f^2 \mbox{cot}^2\beta +  
g_{V R f} M_{f'}^2 \tan^2\beta \right] \C_3''(ff'f) \nonumber \\
&&
- M_{f'}^2 \left[ g_{V L f'} M_{f'}^2 \tan^2\beta +  
g_{V R f'} M_f^2 \mbox{cot}^2\beta \right] \C_3''(f'ff')\Big\}
\end{eqnarray}
\begin{eqnarray}
\Gamma^{V}_{H^+H^-}(\tchi \tchi^0 \tchi)& 
= \displaystyle{\frac{1}{8\pi^2} \sum_{ijk}} &
\Big\{ 
\left[ \O^{V L}_{kj} c^{R}_{Hji} c^{R*}_{Hki} +
\O^{V R}_{kj} c^{L}_{Hji} c^{L*}_{Hki} \right] 
\C_3(\tchi^{}_k \tchi_i^0 \tchi^{}_j) \nonumber \\
&& 
+ M_{ \tchi^{}_j}M_{\tchi_i^0}
\left[ \O^{V L}_{kj} c^{L}_{Hji} c^{R*}_{Hki} +
\O^{V R}_{kj} c^{R}_{Hji} c^{L*}_{Hki} \right] 
\C_3'(\tchi^{}_k \tchi_i^0 \tchi^{}_j) \nonumber \\
&& 
+ M_{\tchi_i^0}M_{ \tchi^{}_k}
\left[ \O^{V L}_{kj} c^{R}_{Hji} c^{L*}_{Hki} +
\O^{V R}_{kj} c^{L}_{Hji} c^{R*}_{Hki} \right] 
\C_3'(\tchi^{}_k \tchi_i^0 \tchi^{}_j) \nonumber \\
&& 
+ M_{ \tchi^{}_j}M_{ \tchi^{}_k}
\left[ \O^{V L}_{kj} c^{L}_{Hji} c^{L*}_{Hki} +
\O^{V R}_{kj} c^{R}_{Hji} c^{R*}_{Hki} \right] 
\C_3''(\tchi^{}_k \tchi_i^0 \tchi^{}_j)
\Big\},~~~~~~
\end{eqnarray}
where $V = \gamma~\mbox{or}~Z$, and where $(f,f')$ is 
either $(q_u,q_d)$ or $(\nu_\ell,\ell)$ for each fermion 
generation.\\

In addition, Tri3-type triangles contribute to $\Gamma^{fin,\,Z}(H^+H^-)$ 
with the following term:
\begin{eqnarray}
\Gamma^{Z}_{H^+H^-}(\tchi^0 \tchi \tchi^0)& 
= -\displaystyle{\frac{1}{8\pi^2} \sum_{ijk}} & 
\Big\{ 
\left[ \O^{0L}_{jk} c^{L}_{Hij} c^{L*}_{Hik} +
\O^{0R}_{jk} c^{R}_{Hij} c^{R*}_{Hik} \right] 
\C_3(\tchi_j^0 \tchi^{}_i \tchi_k^0) \nonumber \\
&& 
+ M_{\tchi_k^0}M_{ \tchi^{}_i}
\left[ \O^{0L}_{jk} c^{L}_{Hij} c^{R*}_{Hik} +
\O^{0R}_{jk} c^{R}_{Hij} c^{L*}_{Hik} \right] 
\C_3'(\tchi_j^0 \tchi^{}_i \tchi_k^0) \nonumber \\
&& 
+ M_{ \tchi^{}_i}M_{\tchi_j^0}
\left[ \O^{0L}_{jk} c^{R}_{Hij} c^{L*}_{Hik} +
\O^{0R}_{jk} c^{L}_{Hij} c^{R*}_{Hik} \right] 
\C_3'(\tchi_j^0 \tchi^{}_i \tchi_k^0) \nonumber \\
&& 
+ M_{\tchi_k^0}M_{\tchi_j^0}
\left[ \O^{0L}_{jk} c^{R}_{Hij} c^{R*}_{Hik} +
\O^{0R}_{jk} c^{L}_{Hij} c^{L*}_{Hik} \right] 
\C_3''(\tchi_j^0 \tchi^{}_i \tchi_k^0)
\Big\}.~~~~~~
\end{eqnarray}

\subsubsection{Tri4-type triangles}

The Tri4-type triangles contribute to $\Gamma^{fin,\,Z}(H^+H^-)$ with the 
following term:
\begin{eqnarray}
\Gamma^{Z}_{H^+H^-}(4)& = & -\frac{1}{16\pi^2} 
\left(
\frac{e^2 M_W (1-2s_W^2)}{2s_W^2c_W^3} 
\right) \times \nonumber \\
&&\Big\{
g_{H^0 H H}\cos(\beta-\alpha)
\left[\C_4(H^0 H Z) + \C_4(Z H H^0) \right] 
\nonumber \\
&&
+ g_{h^0 H H}\sin(\beta-\alpha)
\left[\C_4(h^0 H Z) + \C_4(Z H h^0) \right]
\Big\}.
\end{eqnarray}

\subsubsection{Tri5-type triangles}

There is no contribution from Tri5-type triangles in the production 
of $H^+H^-$ pairs.
 
\subsubsection{Tri6-type triangles}

The Tri6-type triangles contribute to both $\Gamma^{fin,\,\gamma}(H^+H^-)$ and 
$\Gamma^{fin,\,Z}(H^+H^-)$ with:
\begin{eqnarray}
\Gamma^{V}_{H^+H^-}(\mbox{6ch})& 
= -\displaystyle{\frac{1}{8\pi^2}} &
\Big\{
g_{VGG}g^2_{A^0 G H} \C_6(G A^0 G) \nonumber \\
&&
+ g_{VGG}g^2_{H^0 G H} \C_6(G H^0 G) + 
g_{VGG}g^2_{h^0 G H} \C_6(G h^0 G) \nonumber \\
&&
+ g_{VHH}g^2_{H^0 H H} \C_6(H H^0 H) + 
g_{VHH}g^2_{h^0 H H} \C_6(H h^0 H)  
\Big\}
\end{eqnarray}
and
\begin{equation}
\Gamma^{V}_{H^+H^-}(6\tilde{f}) = 
\Gamma^{V,\,heavy}_{H^+H^-}(6\tilde{f}) + 
\Gamma^{V,\,light}_{H^+H^-}(6\tilde{f}).
\end{equation}

The third generation squark contribution, with sfermion mixing, is:
\begin{eqnarray}
\Gamma^{V,\,heavy}_{H^+H^-}(6\tilde{f}) & 
= \displaystyle{\frac{3}{8\pi^2} \sum_{ijk = 1,2}} & 
\Big\{
g_{V \tilde{b}_i \tilde{b}_k} 
g_{H \tilde{t}_j \tilde{b}_i}
g_{H \tilde{t}_j \tilde{b}_k}
\C_6(\tilde{b}_i\tilde{t}_j\tilde{b}_k) \nonumber \\
&&
- g_{V \tilde{t}_i \tilde{t}_k} 
g_{H \tilde{t}_i \tilde{b}_j}
g_{H \tilde{t}_k \tilde{b}_j}
\C_6(\tilde{t}_i\tilde{b}_j\tilde{t}_k)
\Big\}.
\end{eqnarray}

The coupling of L-sfermions to the charged Higgs boson does not vanish like 
the fermion mass, so they also contribute to $\Gamma^{V}(6\tilde{f})$. 
With $(f,f') = (u,d), (c,s)~\mbox{or}~3\times(\nu_\ell,\ell)$, one has:
\begin{eqnarray}
\Gamma^{V,\,light}_{H^+H^-}(6\tilde{f}) & = & \frac{1}{8\pi^2} 
\sum_{(\tilde{f} \tilde{f}')} 
N_c^f g^2_{H \tilde{f}_L \tilde{f}'_L}
\Big\{
g^0_{V \tilde{f}'_L \tilde{f}'_L} 
\C_6(\tilde{f}'_L \tilde{f}_L \tilde{f}'_L) - 
g^0_{V \tilde{f}_L  \tilde{f}_L}\, 
\C_6(\tilde{f}_L \tilde{f}'_L \tilde{f}_L)
\Big\}.~~~~~~
\end{eqnarray}

In addition, Tri6-type triangles contribute to $\Gamma^{fin,\,Z}(H^+H^-)$ 
with the following term:
\begin{eqnarray}
\Gamma^{Z}_{H^+H^-}(\mbox{6n})& = & \frac{1}{8\pi^2} 
\left(
\frac{e^2 M_W}{4s_W^2c_W}
\right)
\times \nonumber \\
&&
\Big\{g_{H^0 G H} \sin(\beta-\alpha) 
\left[ \C_6(H^0 G A^0)+\C_6(A^0 G H^0) \right] \nonumber \\
&&
-g_{h^0 G H} \cos(\beta-\alpha) 
\left[ \C_6(h^0 G A^0)+\C_6(A^0 G h^0) \right]
\Big\}.
\end{eqnarray}

\subsubsection{4-leg diagrams}

The 4-leg diagrams contribute to $\Gamma^{fin,\,\gamma}(H^+H^-)$ with the 
following term:
\begin{eqnarray}
\Gamma^{\gamma}_{H^+H^-}(\mbox{4-leg})& 
= -\displaystyle{\frac{1}{8\pi^2}} &
\Big\{ 
2e^3 
\left[ B_0(H \gamma,M_{H}^2) - B_1(H \gamma,M_{H}^2) \right] 
\nonumber \\
&&
+ \frac{e^3(1-2s_W^2)^2}{2s_W^2c_W^2} 
\left[ B_0(H Z,M_{H}^2) - B_1(H Z,M_{H}^2) \right] 
\nonumber \\
&&
+ \frac{e^3\sin^2(\beta-\alpha)}{4s_W^2} 
\left[ B_0(H^0 W,M_{H}^2) - B_1(H^0 W,M_{H}^2) \right] 
\nonumber \\
&&
+ \frac{e^3\cos^2(\beta-\alpha)}{4s_W^2} 
\left[ B_0(h^0 W,M_{H}^2) - B_1(h^0 W,M_{H}^2) \right] 
\nonumber \\
&&
+ \frac{e^3}{4s_W^2} 
\left[ B_0(A^0 W,M_{H}^2) - B_1(A^0 W,M_{H}^2) \right] 
\Big\}.
\end{eqnarray}

The 4-leg diagrams contribute to $\Gamma^{fin,\,Z}(H^+H^-)$ with the 
following term:
\begin{eqnarray}
\Gamma^{Z}_{H^+H^-}(\mbox{4-leg})& 
= -\displaystyle{\frac{1}{8\pi^2}} &
\Big\{ 
\frac{2e^3(1-2s_W^2)}{2s_Wc_W} 
\left[ B_0(H \gamma,M_{H}^2) - B_1(H \gamma,M_{H}^2) \right] 
\nonumber \\
&&
+ \frac{e^3(1-2s_W^2)^3}{4s_W^3c_W^3} 
\left[ B_0(H Z,M_{H}^2) - B_1(H Z,M_{H}^2) \right] 
\nonumber \\
&&
- \frac{e^3\sin^2(\beta-\alpha)}{4s_Wc_W} 
\left[ B_0(H^0 W,M_{H}^2) - B_1(H^0 W,M_{H}^2) \right] 
\nonumber \\
&&
- \frac{e^3\cos^2(\beta-\alpha)}{4s_Wc_W} 
\left[ B_0(h^0 W,M_{H}^2) - B_1(h^0 W,M_{H}^2) \right] 
\nonumber \\
&&
- \frac{e^3}{4s_Wc_W} 
\left[ B_0(A^0 W,M_{H}^2) - B_1(A^0 W,M_{H}^2) \right] 
\Big\}.
\end{eqnarray}

\subsubsection{Charged Higgs self-energies}

The charged Higgs self-energies contribute to 
$\Gamma^{fin,\,\gamma}(H^+H^-)$ and $\Gamma^{fin,\,Z}(H^+H^-)$ with:
\begin{equation}
\label{hse1}
\Gamma^{\gamma}_{H^+H^-}(H.s.e) = 2e \times 
\left( 
\frac{d\Sigma_{H^+H^-}}{dq^2}
\right)_{q^2 = M_{H}^2}
\end{equation}
and 
\begin{equation}
\label{hse2}
\Gamma^{Z}_{H^+H^-}(H.s.e) = \frac{2e(1-2s_W^2)}{2s_Wc_W} \times 
\left( \frac{d\Sigma_{H^+H^-}}{dq^2}
\right)_{q^2 = M_{H}^2}
\end{equation}
where $\Sigma_{H^+H^-}(q^2)$ is the sum of various bubble terms.\\

These terms contain some combinations of Passarino-Veltman 
functions, such as:
\begin{eqnarray}
SE^\pm_1(XY,q^2) & = & 4B_{22}(XY,q^2) + q^2 
\left[ B_0(XY,q^2)+B_{21}(XY,q^2)-2B_1(XY,q^2) \right],~~~~\\
SE^\pm_2(XY,q^2) & = & 4B_{22}(XY,q^2) + q^2 
\left[ B_1(XY,q^2)+B_{21}(XY,q^2) \right].~~~~
\end{eqnarray}

Here, we consider only the contributions which depend on $q^2$, because 
$\Gamma^{fin,\,\gamma}(H^+H^-)$ and $\Gamma^{fin,\,Z}(H^+H^-)$ depend 
on the derivate of $\Sigma_{H^+H^-}(q^2)$. Four types of bubbles are 
taken into account when calculating this ``reduced'' self-energy, 
which we refer to as $\tilde{\Sigma}_{H^+H^-}(q^2)$:
\begin{equation}
\tilde{\Sigma}_{H^+H^-}(q^2) = \tilde{\Sigma}_{H^+H^-}(VS,q^2) + 
\tilde{\Sigma}_{H^+H^-}(SS',q^2) + 
\tilde{\Sigma}_{H^+H^-}(ff',q^2) + 
\tilde{\Sigma}_{H^+H^-}(\tchi \tchi^0,q^2).~
\end{equation}

The $VS$ bubbles contribute to $\tilde{\Sigma}_{H^+H^-}(q^2)$ with:
\begin{eqnarray}
\tilde{\Sigma}_{H^+H^-}(VS,q^2) & 
= \displaystyle{\frac{1}{16\pi^2}} &
\Big\{ 
e^2 SE^\pm_1(H \gamma,q^2)
+ \frac{e^2(1-2s_W^2)^2}{4s_W^2c_W^2} SE^\pm_1(H Z,q^2)
\nonumber \\
&&
+ \frac{e^2\sin^2(\beta-\alpha)}{4s_W^2} SE^\pm_1(H^0 W,q^2)
+ \frac{e^2\cos^2(\beta-\alpha)}{4s_W^2} SE^\pm_1(h^0 W,q^2)
\nonumber \\
&&
+ \frac{e^2}{4s_W^2} SE^\pm_1(A^0 W,q^2)
\Big\}.
\end{eqnarray}

The $SS'$ bubbles contribute to $\tilde{\Sigma}_{H^+H^-}(q^2)$ with:
\begin{eqnarray}
\tilde{\Sigma}_{H^+H^-}(SS',q^2) & 
= -\displaystyle{\frac{1}{16\pi^2}} &
\Big\{ 
\hspace*{-3mm}
\sum_{light~(\tilde{f} \tilde{f}')} 
\hspace*{-3mm}
N_c^f g^2_{H \tilde{f}_L \tilde{f}'_L} 
B_0(\tilde{f}_L \tilde{f}'_L, q^2) + 
\sum_{ij = 1,2}
3g^2_{H \tilde{t}_i \tilde{b}_j} B_0(\tilde{t}_i \tilde{b}_j, q^2) 
\nonumber \\
&&
+\,g^2_{H^0 H H} B_0(H H^0, q^2) + 
g^2_{h^0 H H} B_0(H h^0, q^2) \nonumber \\
&&
+\,g^2_{H^0 G H} B_0(G H^0, q^2) + 
g^2_{h^0 G H} B_0(G h^0, q^2) \nonumber \\
&& 
+\,g^2_{A^0 G H} B_0(G A^0, q^2) 
\Big\}.
\end{eqnarray}

The fermion and gaugino bubbles contribute to $\tilde{\Sigma}_{H^+H^-}(q^2)$ 
with respectively:
\begin{eqnarray}
\tilde{\Sigma}_{H^+H^-}(ff',q^2) & 
= \displaystyle{\frac{e^2}{16\pi^2s_W^2M_W^2} \sum_{(ff')}} N_c^f & 
\hspace*{-2mm}
\Big\{ 
(M_{f'}^2 \tan^2 \beta + M_f^2 \mbox{cot}^2 \beta)\,SE^\pm_2(ff',q^2) 
\nonumber \\
&&
\hspace*{-2mm}
+ 2M_f^2 M_{f'}^2 B_0(ff',q^2)
\Big\}
\end{eqnarray}
and 
\begin{eqnarray}
\tilde{\Sigma}_{H^+H^-}(\tchi \tchi^0,q^2) & 
= \displaystyle{\frac{1}{8\pi^2} \sum_{ij}} &
\Big\{ \left[
c_{H ij}^{R*}c_{H ij}^{R} + c_{H ij}^{L*}c_{H ij}^{L}
\right]
SE^\pm_2(\tchi^{}_i \tchi_j^0,q^2)
\nonumber \\
&& 
+ M_{\tchi^{}_i} M_{\tchi_j^0} 
\left[
c_{H ij}^{R*}c_{H ij}^{L} + c_{H ij}^{L*}c_{H ij}^{R}
\right]
B_0(\tchi^{}_i \tchi_j^0,q^2) 
\Big\}.
\end{eqnarray}

After having computed the full expression for $\tilde{\Sigma}_{H^+H^-}(q^2)$, 
one simply needs to calculate its derivative at $q^2 = M_{H}^2$ and then 
insert it into equations~(\ref{hse1}) and~(\ref{hse2}).

\subsection{Neutral Higgs sector}

The amplitudes $a^{fin}_{L,R}$ corresponding to final vertices with 
$H^0A^0$ or $h^0A^0$ are:
\begin{equation}
\label{fin-neutral}
a^{fin}_{L,R}(H^0A^0/h^0A^0)={i\Gamma^{fin,\,\gamma}(H^0A^0/h^0A^0)\over 2e}
-{\eta g_{L,R}\over2s_Wc_W} \times {i\Gamma^{fin,\,Z}(H^0A^0/h^0A^0)\over 2e}.
\end{equation}

$\Gamma^{fin,\,\gamma}(H^0A^0/h^0A^0)$ and 
$\Gamma^{fin,\,Z}(H^0A^0/h^0A^0)$ depend on the 
final state. They are obtained by summing 
the contributions of various triangles and of neutral 
Higgs self-energy terms.\\

For the photon, one has:
\begin{eqnarray}
\Gamma^{fin,\,\gamma}(H^0A^0/h^0A^0) & = & 
\Gamma^{\gamma}_{H^0A^0/h^0A^0}(\mbox{1ch}) +  
\Gamma^{\gamma}_{H^0A^0/h^0A^0}(2) - 
\Gamma^{\gamma}_{H^0A^0/h^0A^0}(2,pinch) 
\nonumber \\
& + &  
\Gamma^{\gamma}_{H^0A^0/h^0A^0}(3f) + 
\Gamma^{\gamma}_{H^0A^0/h^0A^0}(\tchi \tchi \tchi) +
\Gamma^{\gamma}_{H^0A^0/h^0A^0}(\mbox{4ch}) 
\nonumber \\
& + &  
\Gamma^{\gamma}_{H^0A^0/h^0A^0}(6\tilde{f}) + 
\Gamma^{\gamma}_{H^0A^0/h^0A^0}(\mbox{6ch}) + 
\Gamma^{\gamma}_{H^0A^0/h^0A^0}(\mbox{4-leg}).~~~~
\end{eqnarray}

For the $Z$ boson, one has:
\begin{eqnarray}
\Gamma^{fin,\,Z}(H^0A^0/h^0A^0) & = & 
\Gamma^{Z}_{H^0A^0/h^0A^0}(\mbox{1ch}) +  
\Gamma^{Z}_{H^0A^0/h^0A^0}(\mbox{1n}) \nonumber \\
& + &  
\Gamma^{Z}_{H^0A^0/h^0A^0}(2) - \Gamma^{Z}_{H^0A^0/h^0A^0}(2,pinch) + 
\Gamma^{Z}_{H^0A^0/h^0A^0}(3f) +
\nonumber \\
& + &  \Gamma^{Z}_{H^0A^0/h^0A^0}(\tchi \tchi \tchi) +
\Gamma^{Z}_{H^0A^0/h^0A^0}(\tchi^0 \tchi^0 \tchi^0) + 
\Gamma^{Z}_{H^0A^0/h^0A^0}(\mbox{4n}) \nonumber \\
& + &  
\Gamma^{Z}_{H^0A^0/h^0A^0}(\mbox{4ch}) + \Gamma^{Z}_{H^0A^0/h^0A^0}(5) +
\Gamma^{Z}_{H^0A^0/h^0A^0}(6\tilde{f}) \nonumber \\
& + &  
\Gamma^{Z}_{H^0A^0/h^0A^0}(\mbox{6ch}) +  
\Gamma^{Z}_{H^0A^0/h^0A^0}(\mbox{6n}) + 
\Gamma^{Z}_{H^0A^0/h^0A^0}(\mbox{4-leg}) + \nonumber \\
& + &  
\Gamma^{Z}_{H^0A^0/h^0A^0}(H.s.e) + 
\Gamma^{Z}_{H^0A^0/h^0A^0}(H.c.t).
\end{eqnarray}

\subsubsection{Tri1-type triangles}

The Tri1-type triangles contribute to $\Gamma^{fin,\,\gamma}(H^0A^0/h^0A^0)$ 
with charged terms only:
\begin{eqnarray}
\Gamma^{\gamma}_{H^0A^0/h^0A^0}(\mbox{1ch})& = & \frac{1}{8\pi^2} \times 
\frac{e^3}{2s_W^2} \left[ Z_{ab} \right] 
\times \C_1(H W H).
\end{eqnarray}

The Tri1-type triangles contribute to $\Gamma^{fin,\,Z}(H^0A^0/h^0A^0)$ with 
the following charged terms:
\begin{eqnarray}
\Gamma^{Z}_{H^0A^0/h^0A^0}(\mbox{1ch})& = & \frac{1}{8\pi^2}   
\times \frac{e^3(1-2s_W^2)}{4s_W^3c_W} \left[ Z_{ab} \right]
\times \C_1(H W H)
\end{eqnarray}
and with the following neutral terms:
\begin{eqnarray}
\Gamma^{Z}_{H^0A^0}(\mbox{1n})& = 
\displaystyle{\frac{e^3\sin(\beta-\alpha)}{64\pi^2s_W^3c_W^3}}&
\hspace*{-2mm}
\Big\{
\sin^2(\beta-\alpha)\,\C_1(A^0 Z H^0) 
+ \cos^2(\beta-\alpha)\,\C_1(G^0 Z H^0) 
\nonumber \\
&&
\hspace*{-2mm}
+ \cos^2(\beta-\alpha)\,\C_1(A^0 Z h^0) 
- \cos^2(\beta-\alpha)\,\C_1(G^0 Z h^0) 
\Big\},~~~~~~~
\end{eqnarray}
\begin{eqnarray}
\Gamma^{Z}_{h^0A^0}(\mbox{1n})& = 
-\displaystyle{\frac{e^3\cos(\beta-\alpha)}{64\pi^2s_W^3c_W^3}}&
\hspace*{-2mm}
\Big\{
\sin^2(\beta-\alpha)\,\C_1(A^0 Z H^0) 
- \sin^2(\beta-\alpha)\,\C_1(G^0 Z H^0) 
\nonumber \\
&&
\hspace*{-2mm}
+ \cos^2(\beta-\alpha)\,\C_1(A^0 Z h^0) 
+ \sin^2(\beta-\alpha)\,\C_1(G^0 Z h^0) 
\Big\}.~~~~~~~
\end{eqnarray}

\subsubsection{Tri2-type triangles}

With $f^V = 
\left\{
\begin{array}{l}
1~\mbox{for}~V=\gamma \\
c_W/s_W~\mbox{for}~V=Z
\end{array}
\right.$ 
\hspace*{-2mm}, the contribution of the Tri2-type triangles is:
\begin{eqnarray}
\Gamma^{V}_{H^0A^0/h^0A^0}(2)& = & \frac{e^3 f^V}{32 \pi^2 s_W^2} 
\left[ Z_{ab} \right] \times \C_2(W H W).
\end{eqnarray}

However, one must also take into account the pinch term:
\begin{eqnarray}
-\Gamma^{V}_{H^0A^0/h^0A^0}(2,pinch)& = & -\frac{e^3 f^V}{8 \pi^2 s_W^2} 
\left[ Z_{ab} \right] \times B_0(WW,q^2).
\end{eqnarray}
 
\subsubsection{Tri3-type triangles}

Let $V$ be either the photon or the $Z$ boson. The Tri3-type triangles 
contribute to both $\Gamma^{fin,\,\gamma}(H^0A^0/h^0A^0)$ and 
$\Gamma^{fin,\,Z}(H^0A^0/h^0A^0)$ with two different terms.\\

The first term, with fermion triangles at the final vertex, 
is given by:
\begin{equation}
\Gamma^{V}_{H^0A^0/h^0A^0}(3f) = -{e^3\over16\pi^2s^2_WM^2_W}
\sum_f N_c^f M^2_f y_f (g_{V L f}-g_{V R f}) 
\left[ \C_3(fff)-M^2_f\,\C_3''(fff) \right].~
\end{equation}

In the previous equation, the term $y_f$ depends both on the 
fermion in the triangle and on the final state. More explicitely, 
for $(q_u,q_d)$ or $(\nu_\ell,\ell)$ doublets, one writes $y_f$ 
as follows:
\begin{eqnarray}
y_f & = & \left( 
{\sin\alpha\,\mbox{cot}\beta\over\sin\beta},\,
{\cos\alpha \tan\beta\over\cos\beta}
\right)~\mbox{for}~H^0A^0, \\
&~~& \nonumber\\
y_f & = & \left( 
{\cos\alpha\,\mbox{cot}\beta\over\sin\beta},\,
-{\sin\alpha \tan\beta\over\cos\beta}
\right)~\mbox{for}~h^0A^0.
\end{eqnarray}

The second term corresponds to chargino triangles at the final vertex.\\

For $H^0A^0$ final states:
\begin{eqnarray}
\Gamma^{V}_{H^0A^0}(\tchi \tchi \tchi) & 
= \displaystyle{{1\over8\pi^2}~\sum_{ijk}} & 
\Big\{
\left[\O^{VL}_{ik}c^{R}_{A^0kj}c^{L}_{H^0ji}+
\O^{VR}_{ik}c^{L}_{A^0kj}c^{R}_{H^0ji}\right]
\C_3(\tchi^{}_i\tchi^{}_j\tchi^{}_k)\nonumber\\
&&
+M_{\tchi^{}_j}M_{\tchi^{}_k}\left[\O^{VL}_{ik}c^{L}_{A^0kj}c^{L}_{H^0ji}+
\O^{VR}_{ik}c^{R}_{A^0kj}c^{R}_{H^0ji}\right]
\C_3'(\tchi^{}_i\tchi^{}_j\tchi^{}_k)\nonumber\\
&&
+M_{\tchi^{}_i}M_{\tchi^{}_j}\left[\O^{VL}_{ik}c^{R}_{A^0kj}c^{R}_{H^0ji}+
\O^{VR}_{ik}c^{L}_{A^0kj}c^{L}_{H^0ji}\right]
\C_3'(\tchi^{}_i\tchi^{}_j\tchi^{}_k)\nonumber\\
&&
+M_{\tchi^{}_i}M_{\tchi^{}_k}\left[\O^{VL}_{ik}c^{L}_{A^0kj}c^{R}_{H^0ji}
+\O^{VR}_{ik}c^{R}_{A^0kj}c^{L}_{H^0ji}\right]
\C_3''(\tchi^{}_i\tchi^{}_j\tchi^{}_k)
\Big\}
\nonumber\\
& 
- \displaystyle{{1\over8\pi^2}~\sum_{ijk}} & 
\Big\{
\left[\O^{VR}_{ki}c^{R}_{A^0jk}c^{L}_{H^0ij}+
\O^{VL}_{ki}c^{L}_{A^0jk}c^{R}_{H^0ij}\right]
\C_3(\tchi^{}_i\tchi^{}_j\tchi^{}_k)\nonumber\\
&&
+M_{\tchi^{}_j}M_{\tchi^{}_k}\left[\O^{VR}_{ki}c^{L}_{A^0jk}c^{L}_{H^0ij}+
\O^{VL}_{ki}c^{R}_{A^0jk}c^{R}_{H^0ij}\right]
\C_3'(\tchi^{}_i\tchi^{}_j\tchi^{}_k)\nonumber\\
&&
+M_{\tchi^{}_i}M_{\tchi^{}_j}\left[\O^{VR}_{ki}c^{R}_{A^0jk}c^{R}_{H^0ij}+
\O^{VL}_{ki}c^{L}_{A^0jk}c^{L}_{H^0ij}\right]
\C_3'(\tchi^{}_i\tchi^{}_j\tchi^{}_k)\nonumber\\
&&
+M_{\tchi^{}_i}M_{\tchi^{}_k}\left[\O^{VR}_{ki}c^{L}_{A^0jk}c^{R}_{H^0ij}
+\O^{VL}_{ki}c^{R}_{A^0jk}c^{L}_{H^0ij}\right]
\C_3''(\tchi^{}_i\tchi^{}_j\tchi^{}_k)
\Big\},~~~~~~~
\end{eqnarray}

For $h^0A^0$ final states:
\begin{eqnarray}
\Gamma^{V}_{h^0A^0}(\tchi \tchi \tchi) & 
= \displaystyle{{1\over8\pi^2}~\sum_{ijk}} & 
\Big\{
\left[\O^{VL}_{ik}c^{R}_{A^0kj}c^{L}_{h^0ji}+
\O^{VR}_{ik}c^{L}_{A^0kj}c^{R}_{h^0ji}\right]
\C_3(\tchi^{}_i\tchi^{}_j\tchi^{}_k)\nonumber\\
&&
+M_{\tchi^{}_j}M_{\tchi^{}_k}\left[\O^{VL}_{ik}c^{L}_{A^0kj}c^{L}_{h^0ji}+
\O^{VR}_{ik}c^{R}_{A^0kj}c^{R}_{h^0ji}\right]
\C_3'(\tchi^{}_i\tchi^{}_j\tchi^{}_k)\nonumber\\
&&
+M_{\tchi^{}_i}M_{\tchi^{}_j}\left[\O^{VL}_{ik}c^{R}_{A^0kj}c^{R}_{h^0ji}+
\O^{VR}_{ik}c^{L}_{A^0kj}c^{L}_{h^0ji}\right]
\C_3'(\tchi^{}_i\tchi^{}_j\tchi^{}_k)\nonumber\\
&&
+M_{\tchi^{}_i}M_{\tchi^{}_k}\left[\O^{VL}_{ik}c^{L}_{A^0kj}c^{R}_{h^0ji}
+\O^{VR}_{ik}c^{R}_{A^0kj}c^{L}_{h^0ji}\right]
\C_3''(\tchi^{}_i\tchi^{}_j\tchi^{}_k)
\Big\}
\nonumber\\
& 
- \displaystyle{{1\over8\pi^2}~\sum_{ijk}} &
\Big\{
\left[\O^{VR}_{ki}c^{R}_{A^0jk}c^{L}_{h^0ij}+
\O^{VL}_{ki}c^{L}_{A^0jk}c^{R}_{h^0ij}\right]
\C_3(\tchi^{}_i\tchi^{}_j\tchi^{}_k)\nonumber\\
&&
+M_{\tchi^{}_j}M_{\tchi^{}_k}\left[\O^{VR}_{ki}c^{L}_{A^0jk}c^{L}_{h^0ij}+
\O^{VL}_{ki}c^{R}_{A^0jk}c^{R}_{h^0ij}\right]
\C_3'(\tchi^{}_i\tchi^{}_j\tchi^{}_k)\nonumber\\
&&
+M_{\tchi^{}_i}M_{\tchi^{}_j}\left[\O^{VR}_{ki}c^{R}_{A^0jk}c^{R}_{h^0ij}+
\O^{VL}_{ki}c^{L}_{A^0jk}c^{L}_{h^0ij}\right]
\C_3'(\tchi^{}_i\tchi^{}_j\tchi^{}_k)\nonumber\\
&&
+M_{\tchi^{}_i}M_{\tchi^{}_k}\left[\O^{VR}_{ki}c^{L}_{A^0jk}c^{R}_{h^0ij}
+\O^{VL}_{ki}c^{R}_{A^0jk}c^{L}_{h^0ij}\right]
\C_3''(\tchi^{}_i\tchi^{}_j\tchi^{}_k)
\Big\}.~~~~~~~
\end{eqnarray}

In addition, neutralino triangles at the final vertex give no contribution 
to $\Gamma^{fin,\,\gamma}(H^0A^0/h^0A^0)$ but they enter into the expression 
of $\Gamma^{fin,\,Z}(H^0A^0/h^0A^0)$.\\

For $H^0A^0$ final states:
\begin{eqnarray}
\Gamma^{V}_{H^0A^0}(\tchi^0 \tchi^0 \tchi^0) & 
= \displaystyle{{1\over16\pi^2}\sum_{ijk}} & 
\hspace*{-3mm}\Big\{
\left[\O^{0L}_{ik}n^{R}_{A^0kj}n^{L}_{H^0ji}+
\O^{0R}_{ik}n^{L}_{A^0kj}n^{R}_{H^0ji}\right]
\C_3(\tchi^0_i\tchi^0_j\tchi^0_k)\nonumber\\
&&
\hspace*{-3mm}+M_{\tchi^0_j}M_{\tchi^0_k}
\left[\O^{0L}_{ik}n^{L}_{A^0kj}n^{L}_{H^0ji}+
\O^{0R}_{ik}n^{R}_{A^0kj}n^{R}_{H^0ji}\right]
\C_3'(\tchi^0_i\tchi^0_j\tchi^0_k)\nonumber\\
&&
\hspace*{-3mm}+M_{\tchi^0_i}M_{\tchi^0_j}
\left[\O^{0L}_{ik}n^{R}_{A^0kj}n^{R}_{H^0ji}+
\O^{0R}_{ik}n^{L}_{A^0kj}n^{L}_{H^0ji}\right]
\C_3'(\tchi^0_i\tchi^0_j\tchi^0_k)\nonumber\\
&&
\hspace*{-3mm}+M_{\tchi^0_i}M_{\tchi^0_k}
\left[\O^{0L}_{ik}n^{L}_{A^0kj}n^{R}_{H^0ji}
+\O^{0R}_{ik}n^{R}_{A^0kj}n^{L}_{H^0ji}\right]
\C_3''(\chi^0_i\chi^0_j\chi^0_k)
\Big\}
\nonumber\\
& 
- \displaystyle{{1\over16\pi^2}\sum_{ijk}} & 
\hspace*{-3mm}\Big\{
\left[\O^{0R}_{ki}n^{R}_{A^0jk}n^{L}_{H^0ij}+
\O^{0L}_{ki}n^{L}_{A^0jk}n^{R}_{H^0ij}\right]
\C_3(\tchi^0_i\tchi^0_j\tchi^0_k)\nonumber\\
&&
\hspace*{-3mm}+M_{\tchi^0_j}M_{\tchi^0_k}
\left[\O^{0R}_{ki}n^{L}_{A^0jk}n^{L}_{H^0ij}+
\O^{0L}_{ki}n^{R}_{A^0jk}n^{R}_{H^0ij}\right]
\C_3'(\tchi^0_i\tchi^0_j\tchi^0_k)\nonumber\\
&&
\hspace*{-3mm}+M_{\tchi^0_i}M_{\tchi^0_j}
\left[\O^{0R}_{ki}n^{R}_{A^0jk}n^{R}_{H^0ij}+
\O^{0L}_{ki}n^{L}_{A^0jk}n^{L}_{H^0ij}\right]
\C_3'(\tchi^0_i\tchi^0_j\tchi^0_k)\nonumber\\
&&
+\hspace*{-3mm}M_{\tchi^0_i}M_{\tchi^0_k}
\left[\O^{0R}_{ki}n^{L}_{A^0jk}n^{R}_{H^0ij}
+\O^{0L}_{ki}n^{R}_{A^0jk}n^{L}_{H^0ij}\right]
\C_3''(\tchi^0_i\tchi^0_j\tchi^0_k)
\Big\}.~~~~~~~~
\end{eqnarray}

For the $h^0A^0$ final states:
\begin{eqnarray}
\Gamma^{V}_{h^0A^0}(\tchi^0 \tchi^0 \tchi^0) & 
= \displaystyle{{1\over16\pi^2}\sum_{ijk}} & 
\hspace*{-3mm}\Big\{
\left[\O^{0L}_{ik}n^{R}_{A^0kj}n^{L}_{h^0ji}+
\O^{0R}_{ik}n^{L}_{A^0kj}n^{R}_{h^0ji}\right]
\C_3(\tchi^0_i\tchi^0_j\tchi^0_k)\nonumber\\
&&
\hspace*{-3mm}+M_{\tchi^0_j}M_{\tchi^0_k}
\left[\O^{0L}_{ik}n^{L}_{A^0kj}n^{L}_{h^0ji}+
\O^{0R}_{ik}n^{R}_{A^0kj}n^{R}_{h^0ji}\right]
\C_3'(\tchi^0_i\tchi^0_j\tchi^0_k)\nonumber\\
&&
\hspace*{-3mm}+M_{\tchi^0_i}M_{\tchi^0_j}
\left[\O^{0L}_{ik}n^{R}_{A^0kj}n^{R}_{h^0ji}+
\O^{0R}_{ik}n^{L}_{A^0kj}n^{L}_{h^0ji}\right]
\C_3'(\tchi^0_i\tchi^0_j\tchi^0_k)\nonumber\\
&&
\hspace*{-3mm}+M_{\tchi^0_i}M_{\tchi^0_k}
\left[\O^{0L}_{ik}n^{L}_{A^0kj}n^{R}_{h^0ji}
+\O^{0R}_{ik}n^{R}_{A^0kj}n^{L}_{h^0ji}\right]
\C_3''(\chi^0_i\chi^0_j\chi^0_k)
\Big\}
\nonumber\\
& 
- \displaystyle{{1\over16\pi^2}\sum_{ijk}} & 
\hspace*{-3mm}\Big\{
\left[\O^{0R}_{ki}n^{R}_{A^0jk}n^{L}_{h^0ij}+
\O^{0L}_{ki}n^{L}_{A^0jk}n^{R}_{h^0ij}\right]
\C_3(\tchi^0_i\tchi^0_j\tchi^0_k)\nonumber\\
&&
\hspace*{-3mm}+M_{\tchi^0_j}M_{\tchi^0_k}
\left[\O^{0R}_{ki}n^{L}_{A^0jk}n^{L}_{h^0ij}+
\O^{0L}_{ki}n^{R}_{A^0jk}n^{R}_{h^0ij}\right]
\C_3'(\tchi^0_i\tchi^0_j\tchi^0_k)\nonumber\\
&&
\hspace*{-3mm}+M_{\tchi^0_i}M_{\tchi^0_j}
\left[\O^{0R}_{ki}n^{R}_{A^0jk}n^{R}_{h^0ij}+
\O^{0L}_{ki}n^{L}_{A^0jk}n^{L}_{h^0ij}\right]
\C_3'(\tchi^0_i\tchi^0_j\tchi^0_k)\nonumber\\
&&
\hspace*{-3mm}+M_{\tchi^0_i}M_{\tchi^0_k}
\left[\O^{0R}_{ki}n^{L}_{A^0jk}n^{R}_{h^0ij}
+\O^{0L}_{ki}n^{R}_{A^0jk}n^{L}_{h^0ij}\right]
\C_3''(\tchi^0_i\tchi^0_j\tchi^0_k)
\Big\}.~~~~~~~~
\end{eqnarray}

\subsubsection{Tri4-type triangles}

Let $V$ denote either the photon or the $Z$ boson. The Tri4-type 
triangles contribute to both $\Gamma^{fin,\,\gamma}(H^0A^0/h^0A^0)$ 
and $\Gamma^{fin,\,Z}(H^0A^0/h^0A^0)$ with charged terms, which are 
given by:
\begin{equation}
\Gamma^{V}_{H^0A^0}(\mbox{4ch}) = \frac{1}{8\pi^2} \,g_{VWG} \times
\Big\{
g_{WHA^0}g_{H^0GH}
\,\C_4(G H W) 
+ g_{A^0GH}g_{WHH^0}
\,\C_4(W H G)
\Big\}
\end{equation}
and
\begin{equation}
\Gamma^{V}_{h^0A^0}(\mbox{4ch}) = \frac{1}{8\pi^2} \,g_{VWG} \times
\Big\{
g_{WHA^0}g_{h^0GH}
\,\C_4(G H W) 
+ g_{A^0GH}g_{WHh^0}
\,\C_4(W H G)
\Big\}.
\end{equation}

The coupling constants $g_{WHH^0}$, $g_{WHh^0}$ and $g_{A^0GH}$ 
depend on the charge of the particles at the vertex. Note that, 
in the triangles considered here, $W$ and $G$ carry the same 
charge.\\

Tri4-type triangles also contribute to 
$\Gamma^{fin,\,Z}(H^0A^0/h^0A^0)$ with 
neutral terms:
\begin{eqnarray}
\Gamma^{Z}_{H^0A^0}(\mbox{4n}) & 
= -\displaystyle{\frac{1}{16\pi^2}} &
\hspace*{-3mm}
\Big\{
g_{ZZH^0}
\left[g_{H^0H^0H^0}g_{ZH^0A^0}\,\C_4(H^0H^0Z)
+g_{h^0H^0H^0}g_{Zh^0A^0}\,\C_4(H^0h^0Z)\right]
\nonumber\\
&&
\hspace*{-3mm}
+g_{ZZh^0}
\left[g_{h^0H^0H^0}g_{ZH^0A^0}\,\C_4(h^0H^0Z)
+g_{H^0h^0h^0}g_{Zh^0A^0}\,\C_4(h^0h^0Z)\right]
\nonumber\\
&&
\hspace*{-3mm}
+g_{ZZH^0}
\left[g_{H^0A^0A^0}g_{ZH^0A^0}\,\C_4(H^0A^0Z)
+g_{H^0A^0G^0}g_{ZH^0G^0}\,\C_4(H^0G^0Z)\right]
\nonumber\\
&&
\hspace*{-3mm}
+g_{ZZh^0}
\left[g_{h^0A^0A^0}g_{ZH^0A^0}\,\C_4(h^0A^0Z)
+g_{h^0A^0G^0}g_{ZH^0G^0}\,\C_4(h^0G^0Z)\right]
\Big\},~~~~~~~~
\end{eqnarray}
\begin{eqnarray}
\Gamma^{Z}_{h^0A^0}(\mbox{4n}) & 
= -\displaystyle{\frac{1}{16\pi^2}} &
\hspace*{-3mm}
\Big\{
g_{ZZH^0}
\left[g_{h^0H^0H^0}g_{ZH^0A^0}\,\C_4(H^0H^0Z)
+g_{H^0h^0h^0}g_{Zh^0A^0}\,\C_4(H^0h^0Z)\right]
\nonumber\\
&&
\hspace*{-3mm}
+g_{ZZh^0}
\left[g_{H^0h^0h^0}g_{ZH^0A^0}\,\C_4(h^0H^0Z)
+g_{h^0h^0h^0}g_{Zh^0A^0}\,\C_4(h^0h^0Z)\right]
\nonumber\\
&&
\hspace*{-3mm}
+g_{ZZH^0}
\left[g_{H^0A^0A^0}g_{Zh^0A^0}\,\C_4(H^0A^0Z)
+g_{H^0A^0G^0}g_{Zh^0G^0}\,\C_4(H^0G^0Z)\right]
\nonumber\\
&&
\hspace*{-3mm}
+g_{ZZh^0}
\left[g_{h^0A^0A^0}g_{Zh^0A^0}\,\C_4(h^0A^0Z)
+g_{h^0A^0G^0}g_{Zh^0G^0}\,\C_4(h^0G^0Z)\right]
\Big\}.~~~~~~~~
\end{eqnarray}

\subsubsection{Tri5-type triangles}

The Tri5-type triangles contribute only to $\Gamma^{fin,\,Z}(H^0A^0/h^0A^0)$.
\begin{equation}
\Gamma^{Z}_{H^0A^0}(5) = {1\over16\pi^2}
\left[
g^2_{ZZH^0}g_{ZH^0A^0}\,\C_5(ZZH^0)+
g_{ZZh^0}g_{ZZH^0}g_{Zh^0A^0}\,\C_5(ZZh^0)
\right],
\end{equation}
\begin{equation}
\Gamma^{Z}_{h^0A^0}(5) = {1\over16\pi^2}
\left[
g_{ZZH^0}g_{ZZh^0}g_{ZH^0A^0}\,\C_5(ZZH^0)+
g^2_{ZZh^0}g_{Zh^0A^0}\,\C_5(ZZh^0)
\right].
\end{equation}

\subsubsection{Tri6-type triangles}

Two Tri6-type triangles contribute to both 
$\Gamma^{fin,\,\gamma}(H^0A^0/h^0A^0)$ and 
$\Gamma^{fin,\,Z}(H^0A^0/h^0A^0)$.\\

Let $V$ denote either the photon or the $Z$ boson, we focus on 
$\Gamma^{V}_{H^0A^0/h^0A^0}(6\tilde{f})$ first. The coupling of 
$A^0$ to sfermions is proportional to the corresponding fermion 
mass and it is thus negligible, except in the case of third 
generation squarks, with sfermion mixing.\\
 
For $H^0A^0$ final states, one has:
\begin{eqnarray}
\Gamma^{V}_{H^0A^0}(6\tilde f) & 
= \displaystyle{\frac{3}{8\pi^2}~\sum_{ijk=1,2}} &
\hspace*{-3mm}
\Big\{
g_{V\tilde b_i\tilde b_k}
g_{H^0\tilde b_i\tilde b_j}
g_{A^0\tilde b_j\tilde b_k}
\C_6(\tilde b_i\tilde b_j\tilde b_k) 
+ g_{V\tilde b_i\tilde b_k}
g_{A^0\tilde b_i\tilde b_j}
g_{H^0\tilde b_j\tilde b_k}
\C_6(\tilde b_i\tilde b_j\tilde b_k)
\Big\} \nonumber \\
& + \displaystyle{\frac{3}{8\pi^2}~\sum_{ijk=1,2}} &
\hspace*{-3mm}
\Big\{
g_{V\tilde t_i\tilde t_k}
g_{H^0\tilde t_i\tilde t_j}
g_{A^0\tilde t_j\tilde t_k}
\C_6(\tilde t_i\tilde t_j\tilde t_k) 
+ g_{V\tilde t_i\tilde t_k}
g_{A^0\tilde t_i\tilde t_j}
g_{H^0\tilde t_j\tilde t_k}
\C_6(\tilde t_i\tilde t_j\tilde t_k)
\Big\}.~~~~~~~
\end{eqnarray}

For $h^0A^0$ final states, one has:
\begin{eqnarray}
\Gamma^{V}_{h^0A^0}(6\tilde f) & 
= \displaystyle{\frac{3}{8\pi^2}~\sum_{ijk=1,2}} &
\hspace*{-3mm}
\Big\{
g_{V\tilde b_i\tilde b_k}
g_{h^0\tilde b_i\tilde b_j}
g_{A^0\tilde b_j\tilde b_k}
\C_6(\tilde b_i\tilde b_j\tilde b_k) 
+ g_{V\tilde b_i\tilde b_k}
g_{A^0\tilde b_i\tilde b_j}
g_{h^0\tilde b_j\tilde b_k}
\C_6(\tilde b_i\tilde b_j\tilde b_k)
\Big\} \nonumber \\
& + \displaystyle{\frac{3}{8\pi^2}~\sum_{ijk=1,2}} &
\hspace*{-3mm}
\Big\{
g_{V\tilde t_i\tilde t_k}
g_{h^0\tilde t_i\tilde t_j}
g_{A^0\tilde t_j\tilde t_k}
\C_6(\tilde t_i\tilde t_j\tilde t_k) 
+ g_{V\tilde t_i\tilde t_k}
g_{A^0\tilde t_i\tilde t_j}
g_{h^0\tilde t_j\tilde t_k}
\C_6(\tilde t_i\tilde t_j\tilde t_k)
\Big\}.~~~~~~~
\end{eqnarray}

The 6ch triangles also contribute to both 
$\Gamma^{fin,\,\gamma}(H^0A^0/h^0A^0)$ and 
$\Gamma^{fin,\,Z}(H^0A^0/h^0A^0)$:
\begin{eqnarray}
\Gamma^{V}_{H^0A^0}(\mbox{6ch}) & = & 
{eM_W\over8\pi^2s_W} 
g_{H^0GH} \left[ g_{VHH}\,\C_6(H G H)-g_{VGG}\,\C_6(G H G) \right],
\end{eqnarray}
\begin{eqnarray}
\Gamma^{V}_{h^0A^0}(\mbox{6ch}) & = & 
{eM_W\over8\pi^2s_W}
g_{h^0GH} \left[ g_{VHH}\,\C_6(H G H)-g_{VGG}\,\C_6(G H G) \right].
\end{eqnarray}

Since neutral Higgs or Goldstone bosons do not couple to a photon, the 6n 
triangles contribute to $\Gamma^{fin,\,Z}(H^0A^0/h^0A^0)$ only.\\

For $H^0A^0$ final states, one has:\\
\begin{eqnarray}
\Gamma^{Z}_{H^0A^0}(\mbox{6n}) & 
= \displaystyle{\frac{1}{8\pi^2}} & 
\hspace*{-3mm}
\Big\{
g_{Zh^0G^0}
\left[
g_{h^0H^0H^0}g_{H^0A^0G^0}\C_6(h^0H^0G^0)+
g_{H^0h^0h^0}g_{h^0A^0G^0}\C_6(h^0h^0G^0) \right]
\nonumber\\
&&
\hspace*{-3mm}
+ g_{ZH^0G^0}
\left[ g_{H^0H^0H^0}g_{H^0A^0G^0}\C_6(H^0H^0G^0)+
g_{h^0H^0H^0}g_{h^0A^0G^0}\C_6(H^0h^0G^0) \right]
\nonumber\\
&&
\hspace*{-3mm}
+ g_{Zh^0A^0}
\left[ g_{h^0H^0H^0}g_{H^0A^0A^0}\C_6(h^0H^0A^0)+
g_{H^0h^0h^0}g_{h^0A^0A^0}\C_6(h^0h^0A^0) \right] 
\nonumber\\
&&
\hspace*{-3mm}
+ g_{ZH^0A^0}
\left[ g_{H^0H^0H^0}g_{H^0A^0A^0}\C_6(H^0H^0A^0)+
g_{h^0H^0H^0}g_{h^0A^0A^0}\C_6(H^0h^0A^0) \right] 
\Big\} \nonumber\\
& - \displaystyle{\frac{1}{8\pi^2}} & 
\hspace*{-3mm}
\Big\{
g_{Zh^0G^0}
\left[g_{H^0G^0G^0}g_{h^0A^0G^0}\C_6(G^0G^0h^0)
+g_{H^0A^0G^0}g_{h^0A^0A^0}\C_6(G^0A^0h^0) \right]
\nonumber\\
&&
\hspace*{-3mm}
+ g_{ZH^0G^0}
\left[g_{H^0G^0G^0}g_{H^0A^0G^0}\C_6(G^0G^0H^0)
+g_{H^0A^0G^0}g_{H^0A^0A^0}\C_6(G^0A^0H^0) \right] 
\nonumber\\
&&
\hspace*{-3mm}
+ g_{Zh^0A^0}
\left[g_{H^0A^0A^0}g_{h^0A^0A^0}\C_6(A^0A^0h^0)
+g_{H^0A^0G^0}g_{h^0A^0G^0}\C_6(A^0G^0h^0) \right]
\nonumber\\
&&
\hspace*{-3mm}
+ g_{ZH^0A^0}
\left[g^2_{H^0A^0A^0}\C_6(A^0A^0H^0)
+g^2_{H^0A^0G^0}\C_6(A^0G^0H^0) \right] 
\Big\}.
\end{eqnarray}

For $h^0A^0$ final states, one has:\\
\begin{eqnarray}
\Gamma^{Z}_{h^0A^0}(\mbox{6n}) & 
=  \displaystyle{\frac{1}{8\pi^2}} &
\hspace*{-3mm}
\Big\{
g_{ZH^0G^0}
\left[ g_{h^0H^0H^0}g_{H^0A^0G^0}\C_6(H^0H^0G^0)+
g_{H^0h^0h^0}g_{h^0A^0G^0}\C_6(H^0h^0G^0) \right] 
\nonumber\\
&&
\hspace*{-3mm}
+ g_{Zh^0G^0}
\left[
g_{H^0h^0h^0}g_{H^0A^0G^0}\C_6(h^0H^0G^0)+
g_{h^0h^0h^0}g_{h^0A^0G^0}\C_6(h^0h^0G^0) \right]
\nonumber\\
&&
\hspace*{-3mm}
+ g_{ZH^0A^0}
\left[ g_{h^0H^0H^0}g_{H^0A^0A^0}\C_6(H^0H^0A^0)+
g_{H^0h^0h^0}g_{h^0A^0A^0}\C_6(H^0h^0A^0) \right]  
\nonumber\\
&&
\hspace*{-3mm}
+ g_{Zh^0A^0}
\left[ g_{H^0h^0h^0}g_{H^0A^0A^0}\C_6(h^0H^0A^0)+
g_{h^0h^0h^0}g_{h^0A^0A^0}\C_6(h^0h^0A^0) \right] 
\Big\} \nonumber\\
& - \displaystyle{\frac{1}{8\pi^2}} & 
\hspace*{-3mm}
\Big\{
g_{ZH^0G^0}
\left[g_{h^0G^0G^0}g_{H^0A^0G^0}\C_6(G^0G^0H^0)
+g_{h^0A^0G^0}g_{H^0A^0A^0}\C_6(G^0A^0H^0) \right] 
\nonumber\\
&&
\hspace*{-3mm}
+ g_{Zh^0G^0}
\left[g_{h^0G^0G^0}g_{h^0A^0G^0}\C_6(G^0G^0h^0)
+g_{h^0A^0G^0}g_{h^0A^0A^0}\C_6(G^0A^0h^0) \right]
\nonumber\\
&&
\hspace*{-3mm}
+ g_{ZH^0A^0}
\left[g_{h^0A^0A^0}g_{H^0A^0A^0}\C_6(A^0A^0H^0)
+g_{H^0A^0G^0}g_{h^0A^0G^0}\C_6(A^0G^0H^0) \right] 
\nonumber\\
&&
\hspace*{-3mm}
+ g_{Zh^0A^0}
\left[g^2_{h^0A^0A^0}\C_6(A^0A^0h^0)
+g^2_{h^0A^0G^0}\C_6(A^0G^0h^0) \right] 
\Big\}.
\end{eqnarray}

\subsubsection{4-leg diagrams}

For $H^0A^0$ final states, the 4-leg diagrams give the following 
contributions:
\begin{eqnarray}
\Gamma^{\gamma}_{H^0A^0}(\mbox{4-leg}) & 
= -\displaystyle{{\sin(\beta-\alpha)\over16\pi^2} 
\left({e^3\over2s^2_W}\right)} \times &
\hspace*{-2mm}
\Big\{
\left[B_0(H W, M^2_{H^0})-B_1(H W, M^2_{H^0})\right] \nonumber \\
&&  
\hspace*{-2mm}
+ \left[B_0(H W, M^2_{A})-B_1(H W, M^2_{A})\right] 
\Big\}
\end{eqnarray}
\begin{eqnarray}
\Gamma^{Z}_{H^0A^0}(\mbox{4-leg}) & = 
-\displaystyle{{s_W\over c_W}\Gamma^{\gamma}(\mbox{4-leg})} & 
+~{\sin(\beta-\alpha)\over16\pi^2} 
\left({e^3\over4s^3_Wc^3_W}\right) \times
\nonumber \\
&&
\Big\{
\left[B_0(A^0 Z, M^2_{H^0})-B_1(A^0 Z, M^2_{H^0})\right]  
\nonumber \\
&&
+ \left[B_0(H^0 Z, M^2_{A})-B_1(H^0 Z, M^2_{A})\right] 
\Big\}.
\end{eqnarray}

For $h^0A^0$ final states, the 4-leg diagrams give the following 
contributions:
\begin{eqnarray}
\Gamma^{\gamma}_{h^0A^0}(\mbox{4-leg}) & 
= \displaystyle{{\cos(\beta-\alpha)\over16\pi^2} 
\left({e^3\over2s^2_W}\right)} \times &
\hspace*{-2mm}
\Big\{
\left[B_0(H W, M^2_{h^0})-B_1(H W, M^2_{h^0})\right] 
\nonumber \\
&&
\hspace*{-2mm}
+ \left[B_0(H W, M^2_{A})-B_1(H W, M^2_{A})\right] 
\Big\}
\end{eqnarray}
\begin{eqnarray}
\Gamma^{Z}_{h^0A^0}(\mbox{4-leg}) & = 
-\displaystyle{{s_W\over c_W}\Gamma^{\gamma}(\mbox{4-leg})} &
-~{\cos(\beta-\alpha)\over16\pi^2} 
\left({e^3\over4s^3_Wc^3_W}\right) \times 
\nonumber \\
&&
\Big\{
\left[B_0(A^0 Z, M^2_{h^0})-B_1(A^0 Z, M^2_{h^0})\right] 
\nonumber \\
&&
+ \left[B_0(H^0 Z, M^2_{A})-B_1(H^0 Z, M^2_{A})\right] 
\Big\}.
\end{eqnarray}

\subsubsection{Neutral Higgs self-energies}

Before estimating the self-energy terms, let us 
first define several useful expressions:
\begin{eqnarray}
\label{defv1}
v_1 = \sin 2\alpha \sin 2\beta - \frac{s^2_W}{c^2_W} \cos2\alpha \cos2\beta. \\
\label{defv2}
v_2 = \cos 2\alpha \sin 2\beta + \frac{s^2_W}{c^2_W} \sin2\alpha \cos2\beta.
\end{eqnarray}
\begin{eqnarray}
SE^0_1(XY,q^2) & = & 2q^2B_1(XY,q^2) - A(M^2_Y) - (q^2+M^2_X)B_0(XY,q^2),\\
SE^0_2(ff,q^2) & = & 2M^2_fB_0(ff,q^2)+A(M^2_f)+q^2B_1(ff,q^2),\\
SE^0_3(XY,q^2,a,b,c,d) & = & 8\Big\{ 
(ad+bc) \left[q^2B_1(XY,q^2)+ A(M^2_X)+M^2_YB_0(XY,q^2)\right]
\nonumber \\
&&~~ + (ac+bd)M_XM_YB_0(XY,q^2) \Big\}, \\
SE^0_4(XY,q^2) & = & B_1(XY,q^2)-B_0(XY,q^2),\\
SE^0_5(XY,q^2) & = & 2B_1(XY,q^2)+B_0(XY,q^2),\\
SE^0_6(XY,q^2,a,b,c,d) & = & -8\Big\{ 
(ad+bc)M_XB_1(XY,q^2)\nonumber \\
&&~~~~ + (ac+bd)M_Y\left[B_0(XY,q^2)+B_1(XY,q^2)\right] \Big\},
\end{eqnarray}

In the following, all neutral Higgs self-energies $\Sigma(q^2)$ and 
the Higgs tadpoles $T_{H^0/h^0}$ are written as the sum of various 
contributions coming from the gauge and Higgs sectors, fermion pairs, 
gaugino pairs and sfermion pairs (where we consider separately 
the unmixed case and the third generation squarks with mixing):
\begin{eqnarray}
\Sigma(q^2) & = & \Sigma(\mbox{g+H}) + \Sigma(ff) + 
\Sigma(\tchi \tchi) + \Sigma(\tchi^0\tchi^0) + 
\Sigma(\tilde{f}\tilde{f}), \\
T_{H^0/h^0} & = & T_{H^0/h^0}(\mbox{g+H}) + T_{H^0/h^0}(ff) + 
T_{H^0/h^0}(\tchi \tchi) + T_{H^0/h^0}(\tchi^0\tchi^0) + 
T_{H^0/h^0}(\tilde{f}\tilde{f}).~~~~~~
\end{eqnarray}

\underline{a) $H^0$ self-energies:}\\


The contribution of the gauge and Higgs sectors is:
\begin{eqnarray}
\Sigma_{H^0H^0}(\mbox{g+H}) & = \displaystyle{\frac{1}{16\pi^2}} & 
\Big\{
g^2_{Z H^0 A^0}SE^0_1(A^0 Z,q^2)+ 
g^2_{Z H^0 G^0}SE^0_1(ZZ,q^2) \nonumber \\
&&
+ 2g^2_{W H H^0}SE^0_1(H W,q^2) + 
2g^2_{W G H^0}SE^0_1(WW,q^2) \nonumber \\
&&
+ 2g^2_{W W H^0} \left[ 2B_0(WW,q^2) -1 \right] +
g^2_{Z Z H^0} \left[ 2B_0(ZZ,q^2) -1 \right] \nonumber \\
&&
+ g^2_{H^0 H H}B_0(HH,q^2) + g^2_{H^0 G G}B_0(WW,q^2) 
+ 2g^2_{H^0 G H}B_0(WH,q^2)  \nonumber \\
&&
+ \frac{1}{2} \left[ 
g^2_{H^0 h^0 h^0}B_0(h^0 h^0,q^2) + 
g^2_{H^0 H^0 H^0}B_0(H^0 H^0,q^2) 
\right] 
\nonumber \\
&&
+ \frac{1}{2} \left[ 
g^2_{H^0 A^0 A^0}B_0(A^0 A^0,q^2) + 
g^2_{H^0 G^0 G^0}B_0(ZZ,q^2) 
\right] \nonumber \\
&&
+ g^2_{h^0 H^0 H^0}B_0(H^0 h^0,q^2) 
+ g^2_{H^0 A^0 G^0}B_0(A^0 Z,q^2) \nonumber \\
&&
- \frac{e^2M^2_W\cos^2(\beta - \alpha)}{2s^2_W}  
\left[
B_0(WW,q^2) + \frac{1}{2c^4_W}B_0(ZZ,q^2)
\right] \nonumber \\
&&
+ \frac{e^2}{s^2_W} \left( 
\left[ 2A(M_W^2) - M_W^2 \right] +
\frac{1}{2c^2_W} \left[ 2A(M_Z^2) - M_Z^2 \right] 
\right) \nonumber \\
&&
+ \frac{e^2}{8s^2_Wc^2_W} 
\left[
(3\sin^2 2\alpha -1)A(M^2_{h^0}) + 
3\cos^2 2\alpha A(M^2_{H^0})\right] \nonumber \\
&&
- \frac{e^2}{8s^2_Wc^2_W}\cos 2\beta \cos 2\alpha 
\left[ A(M^2_{A}) - A(M^2_{Z}) \right] \nonumber \\
&&
+ \frac{e^2}{4s^2_W} \left[ (1-v_1)A(M^2_{W}) + (1+v_1)A(M^2_{H}) \right] 
\Big\}.
\end{eqnarray}

The contribution of the fermion pairs is:
\begin{eqnarray}
\Sigma_{H^0H^0}(ff) & = & - \displaystyle{\frac{1}{4\pi^2}} 
\sum_f N_c^f \times (c_{H^0f}^L)^2 \times SE^0_2(ff,q^2).
\end{eqnarray}

The contributions of the gaugino pairs are:
\begin{eqnarray}
\Sigma_{H^0H^0}(\tchi \tchi) & = & - \displaystyle{\frac{1}{64\pi^2}} 
\sum_{ij} SE^0_3(\tchi^{}_i\tchi^{}_j,q^2,
c^{L}_{H^0ji},c^{L}_{H^0ij},c^{L}_{H^0ij},c^{L}_{H^0ji}), \\
\Sigma_{H^0H^0}(\tchi^0\tchi^0) & = & - \displaystyle{\frac{1}{128\pi^2}} 
\sum_{ij} SE^0_3(\tchi^0_i\tchi^0_j,q^2,
n^{L}_{H^0ji},n^{L}_{H^0ij},n^{L}_{H^0ij},n^{L}_{H^0ji}).
\end{eqnarray}

The contribution of sfermion pairs consists of two terms:
\begin{eqnarray}
\Sigma_{H^0H^0}^{light}(\tilde{f}\tilde{f}) & 
= \displaystyle{\frac{1}{16\pi^2} \sum_{\tilde{f}}N_c^f} & 
\Big\{
\left[
g^2_{H^0 \tilde{f}_L \tilde{f}_L}+g^2_{H^0 \tilde{f}_R \tilde{f}_R}
\right] 
B_0(\tilde{f}\tilde{f},q^2) \nonumber \\
&&
- \left[
g_{H^0 H^0 \tilde{f}_L \tilde{f}_L} + g_{H^0 H^0 \tilde{f}_R \tilde{f}_R}
\right] 
A(M^2_{\tilde{f}})
\Big\},
\end{eqnarray}
\begin{eqnarray}
\Sigma_{H^0H^0}^{heavy}(\tilde{f}\tilde{f}) & 
= \displaystyle{\frac{3}{16\pi^2}} &
\sum_{ij=1,2} \Big\{
g^2_{H^0 \tilde{t}_i \tilde{t}_j} B_0(\tilde{t}_i\tilde{t}_j,q^2) + 
g^2_{H^0 \tilde{b}_i \tilde{b}_j} B_0(\tilde{b}_i\tilde{b}_j,q^2)  
\Big\} \nonumber \\
& - \displaystyle{\frac{3}{16\pi^2}} & 
\sum_{\tilde{f}=\tilde{t},\tilde{b}} 
\Big\{
c^2_{\tilde{f}}
\left[ g_{H^0 H^0 \tilde{f}_L \tilde{f}_L} A(M^2_{\tilde{f}_1}) +
g_{H^0 H^0 \tilde{f}_R \tilde{f}_R} A(M^2_{\tilde{f}_2}) \right]
\Big\} \nonumber \\
& - \displaystyle{\frac{3}{16\pi^2}} & 
\sum_{\tilde{f}=\tilde{t},\tilde{b}} 
\Big\{
s^2_{\tilde{f}}
\left[ g_{H^0 H^0 \tilde{f}_R \tilde{f}_R} A(M^2_{\tilde{f}_1}) +
g_{H^0 H^0 \tilde{f}_L \tilde{f}_L} A(M^2_{\tilde{f}_2}) \right]
\Big\}.
\end{eqnarray}

\underline{b) $h^0$ self-energies:}\\


The contribution of the gauge and Higgs sectors is:
\begin{eqnarray}
\Sigma_{h^0h^0}(\mbox{g+H}) & = \displaystyle{\frac{1}{16\pi^2}} & 
\Big\{
g^2_{Z h^0 A^0}SE^0_1(A^0 Z,q^2)+ 
g^2_{Z h^0 G^0}SE^0_1(ZZ,q^2) \nonumber \\
&&
+ 2g^2_{W H h^0}SE^0_1(HW,q^2) + 
2g^2_{W G h^0}SE^0_1(WW,q^2) \nonumber \\
&&
+ 2g^2_{W W h^0} \left[ 2B_0(WW,q^2) -1 \right] +
g^2_{Z Z h^0} \left[ 2B_0(ZZ,q^2) -1 \right] \nonumber \\
&&
+ g^2_{h^0 H H}B_0(HH,q^2) + g^2_{h^0 G G}B_0(WW,q^2) + 
2g^2_{h^0 G H}B_0(WH,q^2) \nonumber \\
&&
+\frac{1}{2} \left[ 
g^2_{h^0 H^0 H^0}B_0(H^0 H^0,q^2) + 
g^2_{h^0 h^0 h^0}B_0(h^0 h^0,q^2) 
\right] \nonumber \\
&&
+\frac{1}{2}\left[ 
g^2_{h^0 A^0 A^0}B_0(A^0 A^0,q^2) + 
g^2_{h^0 G^0 G^0}B_0(ZZ,q^2)
\right] \nonumber \\
&&
+ g^2_{H^0 h^0 h^0}B_0(H^0 h^0,q^2) + 
g^2_{h^0 A^0 G^0}B_0(A^0 Z,q^2) \nonumber \\
&&
- \frac{e^2M^2_W\sin^2(\beta - \alpha)}{2s^2_W}  
\left[
B_0(WW,q^2) + \frac{1}{2c^4_W}B_0(ZZ,q^2)
\right] \nonumber \\
&&
+ \frac{e^2}{s^2_W} \left( 
\left[ 2A(M_W^2) - M_W^2 \right] +
\frac{1}{2c^2_W} \left[ 2A(M_Z^2) - M_Z^2 \right] 
\right) \nonumber \\
&&
+ \frac{e^2}{8s^2_Wc^2_W} 
\left[
(3\sin^2 2\alpha -1)A(M^2_{H^0}) + 
3\cos^2 2\alpha A(M^2_{h^0})\right] \nonumber \\
&&
+ \frac{e^2}{8s^2_Wc^2_W}\cos 2\beta \cos 2\alpha 
\left[ A(M^2_{A}) - A(M^2_{Z}) \right] \nonumber \\
&&
+ \frac{e^2}{4s^2_W} \left[ (1+v_1)A(M^2_{W}) + (1-v_1)A(M^2_{H}) \right] 
\Big\}.
\end{eqnarray}

The contribution of the fermion pairs is:
\begin{eqnarray}
\Sigma_{h^0h^0}(ff) & = & - \displaystyle{\frac{1}{4\pi^2}} 
\sum_f N_c^f \times (c_{h^0f}^L)^2 \times SE^0_2(ff,q^2).
\end{eqnarray}

The contributions of the gaugino pairs are:
\begin{eqnarray}
\Sigma_{h^0h^0}(\tchi \tchi) & = & - \displaystyle{\frac{1}{64\pi^2}} 
\sum_{ij} SE^0_3(\tchi^{}_i\tchi^{}_j,q^2,
c^{L}_{h^0ji},c^{L}_{h^0ij},c^{L}_{h^0ij},c^{L}_{h^0ji}), \\
\Sigma_{h^0h^0}(\tchi^0\tchi^0) & = & - \displaystyle{\frac{1}{128\pi^2}} 
\sum_{ij} SE^0_3(\tchi^0_i\tchi^0_j,q^2,
n^{L}_{h^0ji},n^{L}_{h^0ij},n^{L}_{h^0ij},n^{L}_{h^0ji}).
\end{eqnarray}

The contribution of sfermion pairs consists of two terms:
\begin{eqnarray}
\Sigma_{h^0h^0}^{light}(\tilde{f}\tilde{f})_{\mbox{light}} & 
= \displaystyle{\frac{1}{16\pi^2} \sum_{\tilde{f}}N_c^f} & 
\Big\{
\left[
g^2_{h^0 \tilde{f}_L \tilde{f}_L}+g^2_{h^0 \tilde{f}_R \tilde{f}_R}
\right] 
B_0(\tilde{f}\tilde{f},q^2) \nonumber \\
&&
- \left[
g_{h^0 h^0 \tilde{f}_L \tilde{f}_L} + g_{h^0 h^0 \tilde{f}_R \tilde{f}_R}
\right] 
A(M^2_{\tilde{f}})
\Big\},
\end{eqnarray}
\begin{eqnarray}
\Sigma_{h^0h^0}^{heavy}(\tilde{f}\tilde{f}) & 
= \displaystyle{\frac{3}{16\pi^2}} &
\sum_{ij=1,2} \Big\{
g^2_{h^0 \tilde{t}_i \tilde{t}_j} B_0(\tilde{t}_i\tilde{t}_j,q^2) + 
g^2_{h^0 \tilde{b}_i \tilde{b}_j} B_0(\tilde{b}_i\tilde{b}_j,q^2)  
\Big\} \nonumber \\
& - \displaystyle{\frac{3}{16\pi^2}} & 
\sum_{\tilde{f}=\tilde{t},\tilde{b}} 
\Big\{
c^2_{\tilde{f}}
\left[ g_{h^0 h^0 \tilde{f}_L \tilde{f}_L} A(M^2_{\tilde{f}_1}) +
g_{h^0 h^0 \tilde{f}_R \tilde{f}_R} A(M^2_{\tilde{f}_2}) \right]
\Big\} \nonumber \\
& - \displaystyle{\frac{3}{16\pi^2}} & 
\sum_{\tilde{f}=\tilde{t},\tilde{b}} 
\Big\{
s^2_{\tilde{f}}
\left[ g_{h^0 h^0 \tilde{f}_R \tilde{f}_R} A(M^2_{\tilde{f}_1}) +
g_{h^0 h^0 \tilde{f}_L \tilde{f}_L} A(M^2_{\tilde{f}_2}) \right]
\Big\}.
\end{eqnarray}

\underline{c) Mixed $H^0h^0$ self-energies:}\\


The contribution of the gauge and Higgs sectors is:
\begin{eqnarray}
\Sigma_{H^0h^0}(\mbox{g+H}) & = \displaystyle{\frac{1}{16\pi^2}} & 
\Big\{
g_{Z H^0 A^0}g_{Z h^0 A^0}SE^0_1(A^0 Z,q^2)+ 
g_{Z H^0 G^0}g_{Z h^0 G^0}SE^0_1(ZZ,q^2) \nonumber \\
&&
+ 2g_{W H H^0}g_{W H h^0}SE^0_1(H W,q^2) + 
2g_{W G H^0}g_{W G h^0}SE^0_1(WW,q^2) \nonumber \\
&&
+ 2g_{W W H^0}g_{W W h^0} \left[ 2B_0(WW,q^2) -1 \right] \nonumber \\
&& 
+ g_{Z Z H^0}g_{Z Z h^0} \left[ 2B_0(ZZ,q^2) -1 \right] \nonumber \\
&&
+ g_{H^0 H H}g_{h^0 HH}B_0(HH,q^2) + 
g_{H^0 G G}g_{h^0 G G}B_0(WW,q^2) \nonumber \\
&&
+2g_{H^0 G H}g_{h^0 G H}B_0(WH,q^2) \nonumber \\
&&
+ \frac{1}{2} \left[ 
g_{H^0 H^0 H^0}g_{h^0 H^0 H^0}B_0(H^0 H^0,q^2) + 
g_{H^0 h^0 h^0}g_{h^0 h^0 h^0}B_0(h^0 h^0,q^2) 
\right] \nonumber \\
&&
+ \frac{1}{2} \left[ 
g_{H^0 A^0 A^0}g_{h^0 A^0 A^0}B_0(A^0 A^0,q^2) + 
g_{H^0 G^0 G^0}g_{h^0 G^0 G^0}B_0(ZZ,q^2) 
\right] \nonumber \\
&&
+ g_{h^0 H^0 H^0}g_{H^0 h^0 h^0}B_0(H^0 h^0,q^2) 
+ g_{H^0 A^0 G^0}g_{h^0 A^0 G^0}B_0(A^0 Z,q^2) \nonumber \\
&&
- \frac{e^2M^2_W\sin(\beta - \alpha)\cos(\beta - \alpha)}{2s^2_W}  
\left[
B_0(WW,q^2) + \frac{1}{2c^4_W}B_0(ZZ,q^2)
\right] \nonumber \\
&&
+ \frac{3 e^2 \sin2\alpha \cos2\alpha}{8s^2_Wc^2_W} 
\left[A(M^2_{h^0}) - A(M^2_{H^0})\right] \nonumber \\
&&
+ \frac{e^2}{8s^2_Wc^2_W}\cos 2\beta \sin 2\alpha 
\left[ A(M^2_{A}) - A(M^2_{Z}) \right] \nonumber \\
&&
+ \frac{e^2 v_2}{4s^2_W} 
\left[ A(M^2_{H}) - A(M^2_{W}) \right] \Big\}.
\end{eqnarray}

The contribution of the fermion pairs is:
\begin{eqnarray}
\Sigma_{H^0h^0}(ff) & = & - \displaystyle{\frac{1}{4\pi^2}} 
\sum_f N_c^f \times c_{H^0f}^L c_{h^0f}^L \times SE^0_2(ff,q^2).
\end{eqnarray}

The contributions of the gaugino pairs are:
\begin{eqnarray}
\Sigma_{H^0h^0}(\tchi \tchi) & = & - \displaystyle{\frac{1}{64\pi^2}} 
\sum_{ij} SE^0_3(\tchi^{}_i\tchi^{}_j,q^2,
c^{L}_{h^0ji},c^{L}_{h^0ij},c^{L}_{H^0ij},c^{L}_{H^0ji}), \\
\Sigma_{H^0h^0}(\tchi^0\tchi^0) & = & - \displaystyle{\frac{1}{128\pi^2}} 
\sum_{ij} SE^0_3(\tchi^0_i\tchi^0_j,q^2,
n^{L}_{h^0ji},n^{L}_{h^0ij},n^{L}_{H^0ij},n^{L}_{H^0ji}).
\end{eqnarray}

The contribution of the sfermion pairs consists of two terms: 
\begin{eqnarray}
\Sigma_{H^0h^0}^{light}(\tilde{f}\tilde{f}) & 
= \displaystyle{\frac{1}{16\pi^2} \sum_{\tilde{f}} N_c^f} &
\Big\{
\left[
g_{H^0 \tilde{f}_L \tilde{f}_L}g_{h^0 \tilde{f}_L \tilde{f}_L} + 
g_{H^0 \tilde{f}_R \tilde{f}_R}g_{h^0 \tilde{f}_R \tilde{f}_R}
\right] 
B_0(\tilde{f}\tilde{f},q^2) \nonumber \\
&&
- \left[
g_{H^0 h^0 \tilde{f}_L \tilde{f}_L} + g_{H^0 h^0 \tilde{f}_R \tilde{f}_R}
\right] 
A(M^2_{\tilde{f}})
\Big\}.
\end{eqnarray}
\begin{eqnarray}
\Sigma_{H^0h^0}^{heavy}(\tilde{f}\tilde{f}) & 
= \displaystyle{\frac{3}{16\pi^2}} &
\sum_{ij=1,2} \Big\{
g_{H^0 \tilde{t}_i \tilde{t}_j}g_{h^0 \tilde{t}_i \tilde{t}_j} 
B_0(\tilde{t}_i\tilde{t}_j,q^2) + 
g_{H^0 \tilde{b}_i \tilde{b}_j}g_{h^0 \tilde{b}_i \tilde{b}_j} 
B_0(\tilde{b}_i\tilde{b}_j,q^2)  
\Big\} \nonumber \\
& - \displaystyle{\frac{3}{16\pi^2}} & 
\sum_{\tilde{f}=\tilde{t},\tilde{b}} 
\Big\{
c^2_{\tilde{f}}
\left[ g_{H^0 h^0 \tilde{f}_L \tilde{f}_L} A(M^2_{\tilde{f}_1}) +
g_{H^0 h^0 \tilde{f}_R \tilde{f}_R} A(M^2_{\tilde{f}_2}) \right]
\Big\} \nonumber \\
& - \displaystyle{\frac{3}{16\pi^2}} & 
\sum_{\tilde{f}=\tilde{t},\tilde{b}} 
\Big\{
s^2_{\tilde{f}}
\left[ g_{H^0 h^0 \tilde{f}_R \tilde{f}_R} A(M^2_{\tilde{f}_1}) +
g_{H^0 h^0 \tilde{f}_L \tilde{f}_L} A(M^2_{\tilde{f}_2}) \right]
\Big\}.
\end{eqnarray}

\underline{d) $A^0$ self-energies:}\\


The contribution of the gauge and Higgs sectors is:
\begin{eqnarray}
\Sigma_{A^0A^0}(\mbox{g+H}) & = \displaystyle{\frac{1}{16\pi^2}} & 
\Big\{
g^2_{Z h^0 A^0}SE^0_1(h^0 Z,q^2)+ 
g^2_{Z H^0 A^0}SE^0_1(H^0 Z,q^2) \nonumber \\
&&
+ 2g^2_{W H A^0}SE^0_1(H W,q^2) + 
2g^2_{A^0 G H}B_0(W H,q^2) \nonumber \\
&&
+ g^2_{H^0 A^0 A^0}B_0(A^0 H^0,q^2) + 
g^2_{h^0 A^0 A^0}B_0(A^0 h^0,q^2) \nonumber \\
&&
+ g^2_{H^0 A^0 G^0}B_0(Z H^0,q^2) + 
g^2_{h^0 A^0 G^0}B_0(Z h^0,q^2) \nonumber \\
&&
+ \frac{e^2}{s^2_W} \left( 
\left[ 2A(M_W^2) - M_W^2 \right] +
\frac{1}{2c^2_W} \left[ 2A(M_Z^2) - M_Z^2 \right] 
\right) \nonumber \\
&&
+ \frac{e^2\cos^2 2\beta}{4s^2_Wc^2_W}A(M^2_{H}) + 
\frac{e^2}{4s^2_W} 
\left[ 1 + \sin^2 2\beta -\frac{s^2_W}{c^2_W}\cos^2 2\beta \right]
A(M^2_{W}) \nonumber \\
&&
+ \frac{e^2}{8s^2_Wc^2_W} 
\left[
(3\sin^2 2\beta -1)A(M^2_{Z}) + 
3\cos^2 2\beta A(M^2_{A})\right] \nonumber \\
&&
+ \frac{e^2}{8s^2_Wc^2_W}\cos 2\beta \cos 2\alpha 
\left[ A(M^2_{h^0}) - A(M^2_{H^0}) \right] 
\Big\}.
\end{eqnarray}

The contribution of the fermion pairs is:
\begin{eqnarray}
\Sigma_{A^0A^0}(ff) & = \displaystyle{\frac{1}{64\pi^2}} & 
\sum_f N_c^f SE^0_3(ff,q^2,c_{A^0f}^L,-c_{A^0f}^L,c_{A^0f}^L,-c_{A^0f}^L).
\end{eqnarray}

The contributions of the gaugino pairs are:
\begin{eqnarray}
\Sigma_{A^0A^0}(\tchi \tchi) & = & \displaystyle{\frac{1}{64\pi^2}} 
\sum_{ij} SE^0_3(\tchi^{}_i\tchi^{}_j,q^2,
c^{L}_{A^0ji},-c^{L}_{A^0ij},c^{L}_{A^0ij},-c^{L}_{A^0ji}), \\ 
\Sigma_{A^0A^0}(\tchi^0\tchi^0) & = & \displaystyle{\frac{1}{128\pi^2}} 
\sum_{ij} SE^0_3(\tchi^0_i\tchi^0_j,q^2,
n^{L}_{A^0ji},-n^{L}_{A^0ij},n^{L}_{A^0ij},-n^{L}_{A^0ji}).
\end{eqnarray}

The contribution of the sfermion pairs consists of two terms: 
\begin{eqnarray}
\Sigma_{A^0A^0}^{light}(\tilde{f}\tilde{f}) & = & -\frac{1}{16\pi^2} 
\sum_{\tilde{f}}
N_c^f 
\Big\{
\left[
g_{A^0 A^0 \tilde{f}_L \tilde{f}_L} + g_{A^0 A^0 \tilde{f}_R \tilde{f}_R}
\right] 
A(M^2_{\tilde{f}})
\Big\},
\end{eqnarray}
\begin{eqnarray}
\Sigma_{A^0A^0}^{heavy}(\tilde{f}\tilde{f}) & 
= \displaystyle{\frac{3}{16\pi^2}} &
\sum_{ij=1,2} \Big\{
g^2_{A^0 \tilde{t}_i \tilde{t}_j} B_0(\tilde{t}_i\tilde{t}_j,q^2) + 
g^2_{A^0 \tilde{b}_i \tilde{b}_j} B_0(\tilde{b}_i\tilde{b}_j,q^2)  
\Big\} \nonumber \\
& - \displaystyle{\frac{3}{16\pi^2}} & 
\sum_{\tilde{f}=\tilde{t},\tilde{b}} 
\Big\{
c^2_{\tilde{f}}
\left[ g_{A^0 A^0 \tilde{f}_L \tilde{f}_L} A(M^2_{\tilde{f}_1}) +
g_{A^0 A^0 \tilde{f}_R \tilde{f}_R} A(M^2_{\tilde{f}_2}) \right]
\Big\} \nonumber \\
& - \displaystyle{\frac{3}{16\pi^2}} & 
\sum_{\tilde{f}=\tilde{t},\tilde{b}} 
\Big\{
s^2_{\tilde{f}}
\left[ g_{A^0 A^0 \tilde{f}_R \tilde{f}_R} A(M^2_{\tilde{f}_1}) +
g_{A^0 A^0 \tilde{f}_L \tilde{f}_L} A(M^2_{\tilde{f}_2}) \right]
\Big\}.
\end{eqnarray}

\underline{e) Mixed $A^0 Z$ self-energies:}\\


The contribution of the gauge and Higgs sectors is:
\begin{eqnarray}
\Sigma_{A^0 Z}(\mbox{g+H}) & = \displaystyle{\frac{e^2}{32\pi^2s^2_Wc^2_W}} & 
\Big\{
M_Z \cos(\beta-\alpha) \sin(\beta-\alpha) SE^0_4(H^0 Z,q^2) \nonumber \\
&&
-M_Z \cos(\beta-\alpha) \sin(\beta-\alpha) SE^0_4(h^0 Z,q^2) \nonumber \\
&&
+\frac{1}{2} M_Z \cos 2\beta 
\cos(\beta+\alpha)\sin(\beta-\alpha)SE^0_5(A^0H^0,q^2) \nonumber \\
&&
-\frac{1}{2} M_Z \cos 2\beta 
\sin(\beta+\alpha)\cos(\beta-\alpha)SE^0_5(A^0h^0,q^2) \nonumber \\
&&
-\frac{1}{2} M_Z \sin 2\beta 
\cos(\beta+\alpha)\cos(\beta-\alpha)SE^0_5(Z H^0,q^2) \nonumber \\
&&
+\frac{1}{2} M_Z \sin 2\beta 
\sin(\beta+\alpha)\sin(\beta-\alpha)SE^0_5(Z h^0,q^2) 
\Big\}.~~~~~~
\end{eqnarray}

The contribution of the fermion pairs is:
\begin{eqnarray}
\Sigma_{A^0 Z}(ff) & = \displaystyle{\frac{1}{64\pi^2}} & 
\sum_f N_c^f SE^0_6(ff,q^2,v_f+a_f,v_f-a_f,c_{A^0f}^L,-c_{A^0f}^L).
\end{eqnarray}
Here, $v_f$ and $a_f$ are defined as follows:
\begin{eqnarray}
v_f = \frac{1}{2s_Wc_W} \left[ +\frac{1}{2} -\frac{4}{3}s^2_W \right], &
a_f = +\displaystyle{\frac{1}{4s_Wc_W}} & 
\mbox{if}~f = q_u~\mbox{or}~\nu_\ell, \\
v_f = \frac{1}{2s_Wc_W} \left[ -\frac{1}{2} +\frac{2}{3}s^2_W \right], &
a_f = -\displaystyle{\frac{1}{4s_Wc_W}} & 
\mbox{if}~f = q_d~\mbox{or}~\ell.
\end{eqnarray} 

The contributions of the gaugino pairs are:
\begin{eqnarray}
\Sigma_{A^0 Z}(\tchi \tchi) & 
= \displaystyle{\frac{1}{8\pi^2} \sum_{ij}} &
\Big\{
M_{\tchi^{}_i}
\left( c_{A^0ji}^L \O_{ij}^{ZR} + c_{A^0ji}^R \O_{ij}^{ZL} \right)
\left[ B_0(\tchi^{}_i\tchi^{}_j,q^2) + B_1(\tchi^{}_i\tchi^{}_j,q^2) \right]
\nonumber \\
&&
+ M_{\tchi^{}_j}
\left[ c_{A^0ji}^L \O_{ij}^{ZL} + c_{A^0ji}^R \O_{ij}^{ZR} \right]
B_1(\tchi^{}_i\tchi^{}_j,q^2) \Big\},~~~ \\
\Sigma_{A^0 Z}(\tchi^0\tchi^0) & 
= \displaystyle{\frac{1}{16\pi^2} \sum_{ij}} & 
\Big\{
M_{\tchi^0_i}
\left( n_{A^0ji}^L \O_{ij}^{ZR} + n_{A^0ji}^R \O_{ij}^{ZL} \right)
\left[ B_0(\tchi^0_i\tchi^0_j,q^2) + B_1(\tchi^0_i\tchi^0_j,q^2) \right]
 \nonumber \\
&&
+ M_{\tchi^0_j}
\left[ n_{A^0ji}^L \O_{ij}^{ZL} + n_{A^0ji}^R \O_{ij}^{ZR} \right]
B_1(\tchi^0_i\tchi^0_j,q^2) \Big\}.~~~
\end{eqnarray}

Note that there is no contribution from sfermion pairs to 
$\Sigma_{A^0 Z}(q^2)$.\\

\underline{f) $H^0$ tadpole:}\\


The contribution of the gauge and Higgs sectors is:
\begin{eqnarray}
T_{H^0}(\mbox{g+H}) & = \displaystyle{\frac{1}{16\pi^2}} &
\Big\{
g_{H^0HH}A(M^2_{H}) + g_{H^0GG}A(M^2_{W}) \nonumber \\
&&
+ \frac{1}{2}g_{H^0H^0H^0}A(M^2_{H^0}) +
\frac{1}{2}g_{H^0h^0h^0}A(M^2_{h^0}) \nonumber \\
&&
+ \frac{1}{2}g_{H^0A^0A^0}A(M^2_{A^0}) 
+ \frac{1}{2}g_{H^0G^0G^0}A(M^2_{Z}) \nonumber \\
&&
- g_{WWH^0} \left[ 4A(M^2_{W})-2M^2_{W} \right] 
- \frac{1}{2}g_{ZZH^0} \left[ 4A(M^2_{Z})-2M^2_Z \right] \nonumber \\
&&
+ \frac{eM_W\cos(\beta-\alpha)}{s_W}
\left[ A(M^2_{W})+\frac{1}{2c^2_W}A(M^2_{Z}) \right]
\Big\}.
\end{eqnarray}

The contribution of the fermion pairs is:
\begin{eqnarray}
T_{H^0}(ff) & = & -\displaystyle{\frac{1}{4\pi^2}} 
\sum_f N_c^f c_{H^0f}^L M_f A(M^2_f).
\end{eqnarray}

The contributions of the gaugino pairs are:
\begin{eqnarray}
T_{H^0}(\tchi \tchi) & = & -\frac{1}{4\pi^2} 
\sum_i c_{H^0 ii}^L M_{\tchi^{}_i} A(M^2_{\tchi^{}_i}), \\
T_{H^0}(\tchi^0\tchi^0) & = & -\frac{1}{8\pi^2} 
\sum_i n_{H^0 ii}^L M_{\tchi_i^0} A(M^2_{\tchi_i^0}). 
\end{eqnarray}

The contribution of the sfermion pairs is the sum of two terms:
\begin{eqnarray}
T_{H^0}^{light}(\tilde{f}\tilde{f}) & = &
\frac{1}{16\pi^2} \sum_f N_c^f 
\Big\{
\left[ 
g_{H^0 \tilde{f}_L \tilde{f}_L}+g_{H^0 \tilde{f}_R \tilde{f}_R} 
\right]
A(M^2_{\tilde{f}})
\Big\}, \\
T_{H^0}^{heavy}(\tilde{f}\tilde{f}) & = &
\frac{3}{16\pi^2} \sum_{i=1,2}  
\Big\{
g_{H^0 \tilde{t}_i \tilde{t}_i}A(M^2_{\tilde{t}_i}) + 
g_{H^0 \tilde{b}_i \tilde{b}_i}A(M^2_{\tilde{b}_i}) 
\Big\}.
\end{eqnarray}

\underline{g) $h^0$ tadpole:}\\


The contribution of the gauge and Higgs sectors is:
\begin{eqnarray}
T_{h^0}(\mbox{g+H}) & = \displaystyle{\frac{1}{16\pi^2}} &
\Big\{
g_{h^0HH}A(M^2_{H}) + g_{h^0GG}A(M^2_{W}) \nonumber \\
&&
+ \frac{1}{2}g_{h^0H^0H^0}A(M^2_{H^0}) 
+ \frac{1}{2}g_{h^0h^0h^0}A(M^2_{h^0}) \nonumber \\
&&
+ \frac{1}{2}g_{h^0A^0A^0}A(M^2_{A^0}) 
+ \frac{1}{2}g_{h^0G^0G^0}A(M^2_{Z}) \nonumber \\
&& 
- g_{WWh^0} \left[ 4A(M^2_{W})-2M^2_{W} \right] 
- \frac{1}{2}g_{ZZh^0} \left[ 4A(M^2_{Z})-2M^2_Z \right] \nonumber \\
&&
+ \frac{eM_W\sin(\beta-\alpha)}{s_W}
\left[ A(M^2_{W})+\frac{1}{2c^2_W}A(M^2_{Z}) \right]
\Big\}.
\end{eqnarray}

The contribution of the fermion pairs is:
\begin{eqnarray}
T_{h^0}(ff) & = & -\displaystyle{\frac{1}{4\pi^2}} 
\sum_f N_c^f c_{h^0f}^L M_f A(M^2_f).
\end{eqnarray}

The contributions of the gaugino pairs are:
\begin{eqnarray}
T_{h^0}(\tchi \tchi) & = & -\frac{1}{4\pi^2} 
\sum_i c_{h^0 ii}^L M_{\tchi^{}_i} A(M^2_{\tchi^{}_i}), \\
T_{h^0}(\tchi^0\tchi^0) & = & -\frac{1}{8\pi^2} 
\sum_i n_{h^0 ii}^L M_{\tchi_i^0} A(M^2_{\tchi_i^0}). 
\end{eqnarray}

The contribution of the sfermion pairs is the sum of two terms:
\begin{eqnarray}
T_{h^0}^{light}(\tilde{f}\tilde{f}) & = &
\frac{1}{16\pi^2} \sum_f N_c^f 
\Big\{
\left[ 
g_{h^0 \tilde{f}_L \tilde{f}_L}+g_{h^0 \tilde{f}_R \tilde{f}_R} 
\right]
A(M^2_{\tilde{f}})
\Big\}, \\
T_{h^0}^{heavy}(\tilde{f}\tilde{f}) & = &
\frac{3}{16\pi^2} \sum_{i=1,2}  
\Big\{
g_{h^0 \tilde{t}_i \tilde{t}_i}A(M^2_{\tilde{t}_i}) + 
g_{h^0 \tilde{b}_i \tilde{b}_i}A(M^2_{\tilde{b}_i}) 
\Big\}.
\end{eqnarray}

\underline{h) Expressions of the renormalized self-energies:}\\

When calculating the effective contribution of the neutral Higgs 
self-energies to the pair production cross section at the one loop level, 
we must consider the renormalized self-energy terms $\hat\Sigma$, 
obtained by adding various counter terms to the unnormalized 
self-energies $\Sigma$.\\ 

Let us first define the Higgs field renormalization constants:
\begin{eqnarray}
\delta Z_{H_1} & = & -\left( \frac{d\Sigma_{A^0A^0}}{dq^2} \right)_{q^2=M^2_A} 
                   - \frac{\mbox{cot}\beta}{M_Z} \Sigma_{A^0Z}(M^2_A), \\
\delta Z_{H_2} & = & -\left( \frac{d\Sigma_{A^0A^0}}{dq^2} \right)_{q^2=M^2_A} 
                   + \frac{\tan\beta}{M_Z} \Sigma_{A^0Z}(M^2_A).
\end{eqnarray}
 
They are used in the calculation of the various mass counter terms, together 
with the Higgs tadpoles and the parameter $\delta M^2_{A^0A^0}$ defined as:
\begin{equation}
\delta M^2_{A^0A^0} = \Sigma_{A^0A^0}(M^2_A) - 
M^2_A \left( \frac{d\Sigma_{A^0A^0}}{dq^2} \right)_{q^2=M^2_A} .
\end{equation}

Indeed, one has the following expressions for the mass counter terms:
\begin{eqnarray}
\delta M^2_{H^0H^0} & = & 
\sin^2(\beta-\alpha)\,\delta M^2_{A^0A^0} + 
\cos^2(\beta+\alpha) \Sigma_{ZZ}(M^2_Z) \nonumber \\
& - &
\frac{e \cos(\beta-\alpha)}{2s_WM_W} \Big\{
\left[ 1+\sin^2(\beta-\alpha) \right] T_{H^0} 
- \cos(\beta-\alpha)\sin(\beta-\alpha)T_{h^0} 
\Big\} \nonumber \\
& + &
M^2_Z\cos(\beta+\alpha)\cos(\beta-\alpha) 
\left[ \delta Z_{H_1} - \delta Z_{H_2} \right] \nonumber \\
& + &
M^2_Z\cos^2(\beta+\alpha)
\left[ \delta Z_{H_1}\sin^2\beta + \delta Z_{H_2}\cos^2\beta \right],
\end{eqnarray}
\begin{eqnarray}
\delta M^2_{h^0h^0} & = & 
\cos^2(\beta-\alpha)\,\delta M^2_{A^0A^0} + 
\sin^2(\beta+\alpha) \Sigma_{ZZ}(M^2_Z) \nonumber \\
& + &
\frac{e\sin(\beta-\alpha)}{2s_WM_W} \Big\{
\sin(\beta-\alpha)\cos(\beta-\alpha)T_{H^0} 
- \left[ 1+\cos^2(\beta-\alpha) \right] T_{h^0} 
\Big\} \nonumber \\
& - &
M^2_Z\sin(\beta+\alpha)\sin(\beta-\alpha) 
\left[ \delta Z_{H_1} - \delta Z_{H_2} \right] \nonumber \\
& + &
M^2_Z\sin^2(\beta+\alpha)
\left[ \delta Z_{H_1}\sin^2\beta + \delta Z_{H_2}\cos^2\beta \right],
\end{eqnarray}
\begin{eqnarray}
\delta M^2_{H^0h^0} & = & 
-\sin(\beta-\alpha)\cos(\beta-\alpha) \delta M^2_{A^0A^0} - 
\cos(\beta+\alpha)\sin(\beta+\alpha) \Sigma_{ZZ}(M^2_Z) \nonumber \\
& - &
\frac{e}{2s_WM_W} \Big\{
\sin^3(\beta-\alpha)T_{H^0} + \cos^3(\beta-\alpha)T_{h^0} 
\Big\} \nonumber \\
& - &
M^2_Z\sin\alpha\cos\alpha 
\left[ \delta Z_{H_1} - \delta Z_{H_2} \right] \nonumber \\
& - &
M^2_Z\cos(\beta+\alpha)\sin(\beta+\alpha)
\left[ \delta Z_{H_1}\sin^2\beta + \delta Z_{H_2}\cos^2\beta \right].
\end{eqnarray}

There is no contribution from the neutral Higgs self-energies to 
$\Gamma^{fin,\,\gamma}(H^0A^0/h^0A^0)$. As for their contributions 
to $\Gamma^{fin,\,Z}(H^0A^0/h^0A^0)$, they can be derived using the 
renormalized self-energies, which are expressed as follows:
\begin{eqnarray}
\hat\Sigma_{H^0H^0}(q^2) & = & \Sigma_{H^0H^0}(q^2) + 
q^2\left[ \delta Z_{H_1}\cos^2 \alpha + \delta Z_{H_2}\sin^2 \alpha \right] -
\delta M^2_{H^0H^0}, \\
\hat\Sigma_{h^0h^0}(q^2) & = & \Sigma_{h^0h^0}(q^2) + 
q^2\left[ \delta Z_{H_1}\sin^2 \alpha + \delta Z_{H_2}\cos^2 \alpha \right] -
\delta M^2_{h^0h^0}, \\
\hat\Sigma_{H^0h^0}(q^2) & = & \Sigma_{H^0h^0}(q^2) + 
q^2\sin\alpha\cos\alpha\left[ \delta Z_{H_2} - \delta Z_{H_1} \right] -
\delta M^2_{H^0h^0}.
\end{eqnarray}

For $H^0A^0$ final states, one has:
\begin{equation}
\Gamma^Z_{H^0A^0}(H.s.e) = -\frac{e\sin(\beta-\alpha)}{s_Wc_W} 
\left[ 
\mbox{cot}(\beta-\alpha)
\frac{\hat\Sigma_{H^0h^0}(M^2_{H^0})}{M^2_{H^0}-M^2_{h^0}}
-\frac{1}{2} \left( \frac{d\hat\Sigma_{H^0H^0}}{dq^2} \right)_{q^2=M^2_{H^0}}
~\right].
\end{equation}

For $h^0A^0$ final states, one has:
\begin{equation}
\Gamma^Z_{h^0A^0}(H.s.e) = +\frac{e\cos(\beta-\alpha)}{s_Wc_W} 
\left[ 
\tan(\beta-\alpha) 
\frac{\hat\Sigma_{H^0h^0}(M^2_{h^0})}{M^2_{h^0}-M^2_{H^0}}
-\frac{1}{2} \left( \frac{d\hat\Sigma_{h^0h^0}}{dq^2} \right)_{q^2=M^2_{h^0}}
~\right].
\end{equation}

\subsubsection{Neutral Higgs counter terms}
\label{nhct}

For $H^0A^0$ final states, the neutral Higgs counter terms give the 
following contribution:
\begin{eqnarray}
\Gamma^Z_{H^0A^0}(H.c.t) & = & \frac{e\cos(\beta-\alpha)}{2s_Wc_W} 
\left[ \cos\beta\sin\beta + \cos\alpha\sin\alpha \right]
(\delta Z_{H_2} - \delta Z_{H_1}) \nonumber \\
                         & - & \frac{e\sin(\beta-\alpha)}{2s_Wc_W}
\left[ (\cos^2\alpha + \sin^2\beta)\delta Z_{H_1} +
(\sin^2\alpha + \cos^2\beta)\delta Z_{H_2} \right].~~~
\end{eqnarray}

For $h^0A^0$ final states, the neutral Higgs counter terms give the 
following contribution:
\begin{eqnarray}
\Gamma^Z_{h^0A^0}(H.c.t) & = & \frac{e\sin(\beta-\alpha)}{2s_Wc_W} 
\left[ \cos\beta\sin\beta - \cos\alpha\sin\alpha \right]
(\delta Z_{H_2} - \delta Z_{H_1}) \nonumber \\
                         & + & \frac{e\cos(\beta-\alpha)}{2s_Wc_W}
\left[ (\sin^2\alpha + \sin^2\beta)\delta Z_{H_1} +
(\cos^2\alpha + \cos^2\beta)\delta Z_{H_2} \right].~~~
\end{eqnarray}

\section{Contribution of box diagrams}

\subsection{Diagram structures for boxes}

Several useful expressions are needed when estimating the contributions of 
the box diagrams. The particles inside the box are ordered clockwise and 
have internal masses $m_1$, $m_2$, $m_3$, $m_4$ starting with $m_1$ 
between the $e^-$ and $e^+$ junctions. In the following, we will use 
the Mandelstam variables $s$, $t$ and $u$, which can be expressed as 
a function of $q^2$ and of the masses $M_1$ and $M_2$ of the two 
outgoing Higgs bosons (i.e. $H^+H^-$ or $H^0A^0$ or $h^0A^0$):
\begin{eqnarray}
s & = & q^2, \\
t & = & \frac{1}{2} (M^2_{2}+M^2_{1}-s) \nonumber \\
  & + & \frac{s\cos\theta}{2} \sqrt{
\left(1+\frac{M_2+M_1}{\sqrt{s}}\right)
\left(1-\frac{M_2+M_1}{\sqrt{s}}\right)
\left(1+\frac{M_2-M_1}{\sqrt{s}}\right)
\left(1-\frac{M_2-M_1}{\sqrt{s}}\right)
}\,,~~~~~~\\
u & = & \frac{1}{2} (M^2_{2}+M^2_{1}-s) \nonumber \\
  & - & \frac{s\cos\theta}{2} \sqrt{
\left(1+\frac{M_2+M_1}{\sqrt{s}}\right)
\left(1-\frac{M_2+M_1}{\sqrt{s}}\right)
\left(1+\frac{M_2-M_1}{\sqrt{s}}\right)
\left(1-\frac{M_2-M_1}{\sqrt{s}}\right)
}\,.~~~~~~
\end{eqnarray}

Note that $u$ and $t$ have an angular dependence and, as a result, so does 
the contribution of the diagram boxes to the one loop cross sections.\\

Let $\ell$ and $\ell'$ be the momenta of the incoming electron and positron,
respectively. Let $P_{f1}$ and $P_{f2}$ be the momenta of the two outgoing 
Higgs bosons. When not otherwise specified, $\ell'$, $P_{f1}$, $P_{f2}$ 
and $\ell$ are oriented clockwise around the box. Then, one has:
\begin{eqnarray}
P^2_{f1} & = & M^2_{1}, \\
P^2_{f2} & = & M^2_{2}, \\
P_{f1}P_{f2} & = & \frac{s-(M^2_{1}+M^2_{2})}{2},\\
\ell'P_{f1} & = & \frac{t-M^2_{1}}{2}, \\
\ell'P_{f2} & = & \frac{u-M^2_{2}}{2}.
\end{eqnarray}

\underline{a) Box7 structures:}
\begin{eqnarray}
\D_7 & = & 6(D_{002}-D_{003})+P^2_{f1}(D_{222}-D_{223})
+P^2_{f2}(D_{332}-D_{333}) \nonumber \\ 
& - &
2P_{f1}P_{f2}(D_{323}-D_{322}) 
-2\ell'P_{f2}(D_{133}-D_{132})-2\ell'P_{f1}(D_{123}-D_{122}) \nonumber \\
& + &
4\left[P^2_{f1}D_{22}+P^2_{f2}D_{23}+2\ell'P_{f1}D_{24}
+2\ell'P_{f2}D_{25} +2P_{f1}P_{f2}D_{26}+4D_{27}\right] \nonumber\\
& - &
4 \left[ P^2_{f2}D_{13}-P^2_{f1}D_{12}-2\ell'P_{f1}D_{11} \right].
\end{eqnarray}

\underline{b) Box8 structures:}
\begin{eqnarray}
\D_8&=&6(D_{002}-D_{003})+P^2_{f1}(D_{222}-D_{223})
+P^2_{f2}(D_{332}-D_{333}) \nonumber \\ 
& - & 2P_{f1}P_{f2}(D_{323}-D_{322})
-2\ell'P_{f2}(D_{133}-D_{132})-2\ell'P_{f1}(D_{123}-D_{122}) \nonumber \\
& + & (2\ell'P_{f1}+P^2_{f1})D_{22}
-(2\ell'P_{f2}+2P_{f1}P_{f2}+P^2_{f2})D_{23}
-(2\ell'P_{f2}+2\ell'P_{f1})D_{25}\nonumber\\
& + &
(2\ell'P_{f2}-2\ell'P_{f1}-2P^2_{f1})D_{26}-2D_{27},
\end{eqnarray}
\begin{eqnarray}
\D'_8 & = & D_{12}-D_{13},
\end{eqnarray}
\begin{eqnarray}
D''_8 & = & D_{0}+D_{12}-D_{13}.
\end{eqnarray}

\underline{c) Box9 structures:}
\begin{eqnarray}
\D_9 & = & D_{12}-D_{13}.
\end{eqnarray}

\underline{d) Box10 structures:}
\begin{eqnarray}
\D_{10} & = & D_{12}+D_{0},
\end{eqnarray}
\begin{eqnarray}
D'_{10} & = & D_{12}.
\end{eqnarray}

Finally, the crossed functions $\bar\D_j$ are obtained from the $\D_j$
functions by making the following changes: $P_{f1}~\to~P_{f2}$, 
i.e. $t~\to~u$ and $M^2_{1}~\to~M^2_{2}$.

\subsection{Charged Higgs sector}

For $e^+e^- \rightarrow H^+H^-$, the box diagrams give the 
following one loop contribution:
\begin{eqnarray}
\label{box-charged}
a^{box}_{L,R}(H^+H^-) & = \displaystyle{\frac{q^2}{2e^2}} \times & 
\Big\{
A^{box7}_{L,R}(H^+H^-) + A^{box8}_{L,R}(H^+H^-) \nonumber \\
&& 
+ A^{box9}_{L,R}(H^+H^-) + A^{box10}_{L,R}(H^+H^-)
\Big\}. 
\end{eqnarray}
 
\subsubsection{Box7 diagrams}

The amplitude $A^{box7}_{L,R}(H^+H^-)$ is obtained by summing various
contributions with gauge and Higgs bosons inside the box:
\begin{eqnarray}
A^{box7}_{L,R}(H^+H^-) 
& = & 
\frac{\alpha^2_{em}}{8s^4_W} \sin^2(\beta-\alpha) P_L 
\D_7(\nu W H^0 W) \nonumber \\
& + & 
\frac{\alpha^2_{em}}{8s^4_W} \cos^2(\beta-\alpha) P_L 
\D_7(\nu W h^0 W) \nonumber \\
& + & 
\frac{\alpha^2_{em}}{8s^4_W} P_L 
\D_7(\nu W A^0 W) \nonumber \\
& - & 
\left(\frac{1-2s^2_W}{2s_Wc_W}\right) \times
\left[
\frac{g_L P_L + g_R P_R}{2s_Wc_W}
\right]
\times
\alpha^2_{em} 
\Big\{\D_7(e \gamma H Z) - \bar{\D}_7(e \gamma H Z) \Big\} \nonumber \\
& - & 
\left(\frac{1-2s^2_W}{2s_Wc_W}\right) \times
\left[
\frac{g_L P_L + g_R P_R}{2s_Wc_W}
\right]
\times
\alpha^2_{em} 
\Big\{\D_7(e Z H \gamma) - \bar{\D}_7(e Z H \gamma) \Big\} \nonumber \\
& + & 
\left(\frac{1-2s^2_W}{2s_Wc_W}\right)^2 \times
\left[
\frac{g^2_L P_L + g^2_R P_R}{4s^2_Wc^2_W}
\right]
\times
\alpha^2_{em} 
\Big\{\D_7(e Z H Z) - \bar{\D}_7(e Z H Z) \Big\} \nonumber \\
& + & 
\left[ P_L + P_R \right] \times 
\alpha^2_{em} 
\Big\{ \D_7(e \gamma H \gamma) - \bar{\D}_7(e \gamma H \gamma) \Big\}.
\end{eqnarray}

\subsubsection{Box8 diagrams}

The amplitude $A^{box8}_{L,R}(H^+H^-)$ is obtained by summing two 
types of box diagrams, which have gauginos inside the box, 
$\tilde{\nu}\tchi\tchi^0\tchi$ and 
$\tilde{e}\tchi^0\tchi\tchi^0$:
\begin{eqnarray}
A^{box8}_{L,R}(H^+H^-) & = & P_{L,R} 
\left[ A^{box8}_{H^+H^-}(\tilde{\nu}\tchi\tchi^0\tchi) + 
A^{box8}_{H^+H^-}(\tilde{e}\tchi^0\tchi\tchi^0) \right].
\end{eqnarray}

For the $\tilde{\nu}\tchi\tchi^0\tchi$ boxes, since there is no
right-handed sneutrino in the MSSM, only a left-handed term is considered:
\begin{eqnarray}
A^{box8}_{H^+H^-}(\tilde{\nu}_L\tchi\tchi^0\tchi)
& = -\displaystyle{\frac{e^2}{16\pi^2s^2_W}} 
\sum_{ijk} Z^{+*}_{1i}Z^+_{1k} \times &
\Big\{
M_{\tchi^{}_i}M_{\tchi^{}_k}c^{L*}_{H ij}c^L_{H kj}
\D''_8(\tilde{\nu}_L\tchi^{}_i\tchi^0_j\tchi^{}_k) \nonumber \\
&& 
+ M_{\tchi^{}_i}M_{\tchi^0_j}c^{L*}_{H ij}c^R_{H kj}
\D'_8(\tilde{\nu}_L\tchi^{}_i\tchi^0_j\tchi^{}_k) \nonumber \\
&& 
+ M_{\tchi^0_j}M_{\tchi^{}_k}c^{R*}_{H ij}c^L_{H kj}
\D'_8(\tilde{\nu}_L\tchi^{}_i\tchi^0_j\tchi^{}_k) \nonumber \\
&& 
+ c^{R*}_{H ij}c^R_{H kj}
\D_8(\tilde{\nu}_L\tchi^{}_i\tchi^0_j\tchi^{}_k) \Big\}.
\end{eqnarray}

For the $\tilde{e}\tchi^0\tchi\tchi^0$ boxes, one has both left-handed and
right-handed terms. The left-handed term is given by:
\begin{eqnarray}
A^{box8}_{H^+H^-}(\tilde{e}_L\tchi^0\tchi\tchi^0)
& = \displaystyle{\frac{e^2}{32\pi^2s^2_Wc^2_W}} & 
\sum_{ijk} (Z^{N*}_{1i}s_W+Z^{N*}_{2i}c_W)(Z^{N}_{1k}s_W+Z^{N}_{2k}c_W) \times 
\nonumber \\
&&
\Big\{
M_{\tchi^0_i}M_{\tchi^0_k}c^{R*}_{H ji}c^R_{H jk}
\bar{\D}''_8(\tilde{e}_L\tchi^0_i\tchi^{}_j\tchi^0_k)
\nonumber \\
&&
+ M_{\tchi^0_i}M_{\tchi^{}_j}c^{R*}_{H ji}c^L_{H jk}
\bar{\D}'_8(\tilde{e}_L\tchi^0_i\tchi^{}_j\tchi^0_k) 
\nonumber \\
&&
+ M_{\tchi^{}_j}M_{\tchi^0_k}c^{L*}_{H ji}c^R_{H jk}
\bar{\D}'_8(\tilde{e}_L\tchi^0_i\tchi^{}_j\tchi^0_k)
\nonumber \\
&&
+ c^{L*}_{H ji}c^L_{H jk}
\bar{\D}_8(\tilde{e}_L\tchi^0_i\tchi^{}_j\tchi^0_k) 
\Big\} \nonumber \\
& -\displaystyle{\frac{e^2}{32\pi^2s^2_Wc^2_W}} & 
\sum_{ijk} (Z^{N*}_{1i}s_W+Z^{N*}_{2i}c_W)(Z^{N}_{1k}s_W+Z^{N}_{2k}c_W) \times 
\nonumber \\
&&
\Big\{
M_{\tchi^0_i}M_{\tchi^0_k}c^{L*}_{H ji}c^L_{H jk}
\D''_8(\tilde{e}_L\tchi^0_i\tchi^{}_j\tchi^0_k) 
\nonumber \\
&&
+ M_{\tchi^0_i}M_{\tchi^{}_j}c^{L*}_{H ji}c^R_{H jk}
\D'_8(\tilde{e}_L\tchi^0_i\tchi^{}_j\tchi^0_k)
\nonumber \\
&&
+ M_{\tchi^{}_j}M_{\tchi^0_k}c^{R*}_{H ji}c^L_{H jk}
\D'_8(\tilde{e}_L\tchi^0_i\tchi^{}_j\tchi^0_k)
\nonumber \\
&&
+ c^{R*}_{H ji}c^R_{H jk}
\D_8(\tilde{e}_L\tchi^0_i\tchi^{}_j\tchi^0_k) 
\Big\} 
\end{eqnarray} 
while the right-handed term is:
\begin{eqnarray}
A^{box8}_{H^+H^-}(\tilde{e}_R\tchi^0\tchi\tchi^0)
& = \displaystyle{\frac{e^2}{8\pi^2c^2_W}} \sum_{ijk} 
Z^{N}_{1i}Z^{N*}_{1k} \times &
\Big\{
M_{\tchi^0_i}M_{\tchi^0_k}c^{L*}_{H ji}c^L_{H jk}
\bar{\D}''_8(\tilde{e}_R\tchi^0_i\tchi^{}_j\tchi^0_k)
\nonumber \\
&&
M_{\tchi^0_i}M_{\tchi^{}_j}c^{L*}_{H ji}c^R_{H jk}
\bar{\D}'_8(\tilde{e}_R\tchi^0_i\tchi^{}_j\tchi^0_k) 
\nonumber \\
&&
+ M_{\tchi^{}_j}M_{\tchi^0_k}c^{R*}_{H ji}c^L_{H jk}
\bar{\D}'_8(\tilde{e}_R\tchi^0_i\tchi^{}_j\tchi^0_k)
\nonumber \\
&&
+ c^{R*}_{H ji}c^R_{H jk}
\bar{\D}_8(\tilde{e}_R\tchi^0_i\tchi^{}_j\tchi^0_k) 
\Big\}
\nonumber \\
& -\displaystyle{\frac{e^2}{8\pi^2c^2_W}} \sum_{ijk} 
Z^{N}_{1i}Z^{N*}_{1k} \times &
\Big\{
M_{\tchi^0_i}M_{\tchi^0_k}c^{R*}_{H ji}c^R_{H jk}
\D''_8(\tilde{e}_R\tchi^0_i\tchi^{}_j\tchi^0_k) 
\nonumber \\
&&
+ M_{\tchi^0_i}M_{\tchi^{}_j}c^{R*}_{H ji}c^L_{H jk}
\D'_8(\tilde{e}_R\tchi^0_i\tchi^{}_j\tchi^0_k) 
\nonumber \\
&&
+ M_{\tchi^{}_j}M_{\tchi^0_k}c^{L*}_{H ji}c^R_{H jk}
\D'_8(\tilde{e}_R\tchi^0_i\tchi^{}_j\tchi^0_k)
\nonumber \\
&&
+ c^{L*}_{H ji}c^L_{H jk}
\D_8(\tilde{e}_R\tchi^0_i\tchi^{}_j\tchi^0_k) 
\Big\}.
\end{eqnarray} 

\subsubsection{Box9 diagrams}

There is no right-handed contribution from box9 diagrams. As for the 
left-handed amplitude $A^{box9}_{L}(H^+H^-)$, it is obtained as follows:
\begin{eqnarray}
A^{box9}_{L}(H^+H^-) & = & \frac{e^2}{32\pi^2s^2_Wc^2_W} \sum_i
|Z^{N}_{1i}s_W+Z^{N}_{2i}c_W|^2 g^2_{H \tilde{e}_L \tilde{\nu}_L} 
\D_9(\tchi^0_i \tilde{e}_L \tilde{\nu}_L \tilde{e}_L) \nonumber \\
& - & \frac{e^2}{16\pi^2s^2_W} \sum_i
|Z^{+}_{1i}|^2 g^2_{H \tilde{e}_L \tilde{\nu}_L} 
\bar{\D}_9(\tchi^{}_i \tilde{\nu}_L \tilde{e}_L \tilde{\nu}_L).
\end{eqnarray}

\subsubsection{Twisted box10 diagrams}

Twisted box10 diagrams contribute only with a left-handed term:
\begin{eqnarray}
A^{box10}_{L}(H^+H^-) & 
= \displaystyle{\frac{e^2}{16\sqrt{2}\pi^2s^2_Wc_W}} &  
\hspace*{-3mm}
\displaystyle{\sum_{ij}}
Z^{+*}_{1i}(Z^{N}_{1j}s_W+Z^{N}_{2j}c_W) \times  
g_{H \tilde{e}_L \tilde{\nu}_L} \times \nonumber \\
&&
\hspace*{-3mm}
\left[
M_{\tchi^{}_i}\,c^{L*}_{H ij} 
\D_{10}(\tilde{\nu}_L \tchi^{}_i \tchi^0_j \tilde{e}_L)+
M_{\tchi^0_j}\,c^{R*}_{H ij} 
\D'_{10}(\tilde{\nu}_L \tchi^{}_i \tchi^0_j \tilde{e}_L)
\right]
\nonumber \\
& - \displaystyle{\frac{e^2}{16\sqrt{2}\pi^2s^2_Wc_W}} &  
\hspace*{-3mm}
\displaystyle{\sum_{ij}}
Z^{+}_{1i}(Z^{N*}_{1j}s_W+Z^{N*}_{2j}c_W) \times
g_{H \tilde{e}_L \tilde{\nu}_L} \times \nonumber \\
&&
\hspace*{-3mm}
\left[
M_{\tchi^0_j}\,c^{R}_{H ij} 
\bar{\D}_{10}(\tilde{e}_L \tchi^0_j \tchi^{}_i \tilde{\nu}_L) +
M_{\tchi^{}_i}\,c^{L}_{H ij} 
\bar{\D}'_{10}(\tilde{e}_L \tchi^0_j \tchi^{}_i \tilde{\nu}_L)
\right].~~~~~~~~~
\end{eqnarray}

\subsection{Neutral Higgs sector}

For $e^+e^- \rightarrow H^0A^0/h^0A^0$, there is no box9 diagram, so one has:
\begin{eqnarray}
\label{box-neutral}
a^{box}_{L,R}(H^0A^0/h^0A^0) & = \displaystyle{\frac{iq^2}{2e^2}} & 
\Big\{A^{box7}_{L,R}(H^0A^0/h^0A^0) + A^{box8}_{L,R}(H^0A^0/h^0A^0) 
\nonumber \\ 
&& + A^{box10}_{L,R}(H^0A^0/h^0A^0)
\Big\}. 
\end{eqnarray}

\subsubsection{Box7 diagrams}

In the neutral Higgs sector, there is no right-handed contribution from box7 
diagrams. The left-handed amplitude $A^{box7}_{L}(H^0A^0/h^0A^0)$ is 
obtained as follows:
\begin{eqnarray}
A^{box7}_{L}(H^0A^0/h^0A^0) 
& = & 
\frac{e^4}{128\pi^2s^4_W} \times \left[ Z_{ab} \right] \times 
\Big\{
\D_7(\nu W H W)+\bar{\D}_7(\nu W H W)
\Big\}.~~
\end{eqnarray}

\subsubsection{Box8 diagrams}

Both $\tilde{\nu}\tchi\tchi\tchi$ and $\tilde{e}\tchi^0\tchi^0\tchi^0$ 
box diagrams contribute to the amplitude $A^{box8}_{L,R}(H^0A^0/h^0A^0)$:
\begin{eqnarray}
A^{box8}_{L,R}(H^0A^0/h^0A^0) & = & P_{L,R} 
\left[ A^{box8}_{H^0A^0/h^0A^0}(\tilde{\nu}\tchi\tchi\tchi) + 
A^{box8}_{H^0A^0/h^0A^0}(\tilde{e}\tchi^0\tchi^0\tchi^0) \right].
\end{eqnarray}

The $\tilde{\nu}\tchi\tchi\tchi$ boxes contribute only with  
left-handed terms:
\begin{eqnarray}
A^{box8}_{H^0A^0}(\tilde{\nu}_L\tchi\tchi\tchi)
& = \displaystyle{\frac{e^2}{16\pi^2s^2_W}} 
\sum_{ijk} Z^{+*}_{1i}Z^+_{1k} \times &
\Big\{
M_{\tchi^{}_i}M_{\tchi^{}_k}c^{R}_{A^0 ji}c^L_{H^0 kj}
\D''_8(\tilde{\nu}_L\tchi^{}_i\tchi^{}_j\tchi^{}_k) 
\nonumber \\
&&
+ M_{\tchi^{}_i}M_{\tchi^{}_j}c^{R}_{A^0 ji}c^R_{H^0 kj}
\D'_8(\tilde{\nu}_L\tchi^{}_i\tchi^{}_j\tchi^{}_k)
\nonumber \\
&&
+M_{\tchi^{}_j}M_{\tchi^{}_k}c^{L}_{A^0 ji}c^L_{H^0 kj}
\D'_8(\tilde{\nu}_L\tchi^{}_i\tchi^{}_j\tchi^{}_k)
\nonumber \\
&&
+ c^{L}_{A^0 ji}c^R_{H^0 kj}
\D_8(\tilde{\nu}_L\tchi^{}_i\tchi^{}_j\tchi^{}_k) 
\Big\}
\nonumber \\
& - \displaystyle{\frac{e^2}{16\pi^2s^2_W}} 
\sum_{ijk} Z^{+*}_{1i}Z^+_{1k} \times &
\Big\{
M_{\tchi^{}_i}M_{\tchi^{}_k}c^{R}_{H^0 ji}c^L_{A^0 kj}
\bar{\D}''_8(\tilde{\nu}_L\tchi^{}_i\tchi^{}_j\tchi^{}_k) 
\nonumber \\
&&
+ M_{\tchi^{}_i}M_{\tchi^{}_j}c^{R}_{H^0 ji}c^R_{A^0 kj}
\bar{\D}'_8(\tilde{\nu}_L\tchi^{}_i\tchi^{}_j\tchi^{}_k) 
\nonumber \\
&&
+ M_{\tchi^{}_j}M_{\tchi^{}_k}c^{L}_{H^0 ji}c^L_{A^0 kj}
\bar{\D}'_8(\tilde{\nu}_L\tchi^{}_i\tchi^{}_j\tchi^{}_k) 
\nonumber \\
&&
+ c^{L}_{H^0 ji}c^R_{A^0 kj}
\bar{\D}_8(\tilde{\nu}_L\tchi^{}_i\tchi^{}_j\tchi^{}_k) 
\Big\},
\end{eqnarray}
\begin{eqnarray}
A^{box8}_{h^0A^0}(\tilde{\nu}_L\tchi\tchi\tchi)
& = \displaystyle{\frac{e^2}{16\pi^2s^2_W}} 
\sum_{ijk} Z^{+*}_{1i}Z^+_{1k} \times &
\Big\{
M_{\tchi^{}_i}M_{\tchi^{}_k}c^{R}_{A^0 ji}c^L_{h^0 kj}
\D''_8(\tilde{\nu}_L\tchi^{}_i\tchi^{}_j\tchi^{}_k) 
\nonumber \\
&&
+ M_{\tchi^{}_i}M_{\tchi^{}_j}c^{R}_{A^0 ji}c^R_{h^0 kj}
\D'_8(\tilde{\nu}_L\tchi^{}_i\tchi^{}_j\tchi^{}_k)
\nonumber \\
&&
+M_{\tchi^{}_j}M_{\tchi^{}_k}c^{L}_{A^0 ji}c^L_{h^0 kj}
\D'_8(\tilde{\nu}_L\tchi^{}_i\tchi^{}_j\tchi^{}_k)
\nonumber \\
&&
+ c^{L}_{A^0 ji}c^R_{h^0 kj}
\D_8(\tilde{\nu}_L\tchi^{}_i\tchi^{}_j\tchi^{}_k) 
\Big\}
\nonumber \\
& - \displaystyle{\frac{e^2}{16\pi^2s^2_W}} 
\sum_{ijk} Z^{+*}_{1i}Z^+_{1k} \times &
\Big\{
M_{\tchi^{}_i}M_{\tchi^{}_k}c^{R}_{h^0 ji}c^L_{A^0 kj}
\bar{\D}''_8(\tilde{\nu}_L\tchi^{}_i\tchi^{}_j\tchi^{}_k) 
\nonumber \\
&&
+ M_{\tchi^{}_i}M_{\tchi^{}_j}c^{R}_{h^0 ji}c^R_{A^0 kj}
\bar{\D}'_8(\tilde{\nu}_L\tchi^{}_i\tchi^{}_j\tchi^{}_k) 
\nonumber \\
&&
+ M_{\tchi^{}_j}M_{\tchi^{}_k}c^{L}_{h^0 ji}c^L_{A^0 kj}
\bar{\D}'_8(\tilde{\nu}_L\tchi^{}_i\tchi^{}_j\tchi^{}_k)
\nonumber \\
&&
+ c^{L}_{h^0 ji}c^R_{A^0 kj}
\bar{\D}_8(\tilde{\nu}_L\tchi^{}_i\tchi^{}_j\tchi^{}_k) 
\Big\}.
\end{eqnarray}

The $\tilde{e}\tchi^0\tchi^0\tchi^0$ boxes contribute with both 
left-handed and right-handed terms:
\begin{eqnarray}
A^{box8}_{H^0A^0}(\tilde{e}\tchi^0\tchi^0\tchi^0)
& \hspace*{-15mm} 
= \displaystyle{\frac{e^2}{32\pi^2s^2_Wc^2_W}} P_L &
\hspace*{-18mm} 
\sum_{ijk} (Z^{N*}_{1i}s_W+Z^{N*}_{2i}c_W)(Z^{N}_{1k}s_W+Z^{N}_{2k}c_W) \times 
\nonumber \\
&&
\hspace*{-18mm} 
\Big\{
M_{\tchi^0_i}M_{\tchi^0_k}n^{R}_{A^0 ij}n^L_{H^0 jk}
\D''_8(\tilde{e}_L\tchi^0_i\tchi^0_j\tchi^0_k) 
\nonumber \\
&&
\hspace*{-18mm} 
+ M_{\tchi^0_i}M_{\tchi^0_j}n^{R}_{A^0 ij}n^R_{H^0 jk}
\D'_8(\tilde{e}_L\tchi^0_i\tchi^0_j\tchi^0_k)
\nonumber \\
&&
\hspace*{-18mm} 
+ M_{\tchi^0_j}M_{\tchi^0_k}n^{L}_{A^0 ij}n^L_{H^0 jk}
\D'_8(\tilde{e}_L\tchi^0_i\tchi^0_j\tchi^0_k)
\nonumber \\
&&
\hspace*{-18mm} 
+ n^{L}_{A^0 ij}n^R_{H^0 jk}
\D_8(\tilde{e}_L\tchi^0_i\tchi^0_j\tchi^0_k) 
\Big\} \nonumber \\
& \hspace*{-15mm}
- \displaystyle{\frac{e^2}{32\pi^2s^2_Wc^2_W}} P_L & 
\hspace*{-18mm} 
\sum_{ijk} (Z^{N*}_{1i}s_W+Z^{N*}_{2i}c_W)(Z^{N}_{1k}s_W+Z^{N}_{2k}c_W) \times 
\nonumber \\
&&
\hspace*{-18mm} 
\Big\{
M_{\tchi^0_i}M_{\tchi^0_k}n^{R}_{H^0 ij}n^L_{A^0 jk}
\bar{\D}''_8(\tilde{e}_L\tchi^0_i\tchi^0_j\tchi^0_k) 
\nonumber \\
&&
\hspace*{-18mm} 
+ M_{\tchi^0_i}M_{\tchi^0_j}n^{R}_{H^0 ij}n^R_{A^0 jk}
\bar{\D}'_8(\tilde{e}_L\tchi^0_i\tchi^0_j\tchi^0_k) 
\nonumber \\
&&
\hspace*{-18mm} 
+ M_{\tchi^0_j}M_{\tchi^0_k}n^{L}_{H^0 ij}n^L_{A^0 jk}
\bar{\D}'_8(\tilde{e}_L\tchi^0_i\tchi^0_j\tchi^0_k)
\nonumber \\
&&
\hspace*{-18mm} 
+ n^{L}_{H^0 ij}n^R_{A^0 jk}
\bar{\D}_8(\tilde{e}_L\tchi^0_i\tchi^0_j\tchi^0_k) 
\Big\}
\nonumber \\
& - \displaystyle{\frac{e^2}{8\pi^2c^2_W}} P_R 
\sum_{ijk} Z^{N}_{1i}Z^{N*}_{1k} \times &
\hspace*{-2mm} 
\Big\{ 
M_{\tchi^0_i}M_{\tchi^0_k}n^{L}_{A^0 ij}n^R_{H^0 jk}
\D''_8(\tilde{e}_R\tchi^0_i\tchi^0_j\tchi^0_k) 
\nonumber \\
&&
\hspace*{-2mm} 
+ M_{\tchi^0_i}M_{\tchi^0_j}n^{L}_{A^0 ij}n^L_{H^0 jk}
\D'_8(\tilde{e}_R\tchi^0_i\tchi^0_j\tchi^0_k) 
\nonumber \\
&&
\hspace*{-2mm} 
+ M_{\tchi^0_j}M_{\tchi^0_k}n^{R}_{A^0 ij}n^R_{H^0 jk}
\D'_8(\tilde{e}_R\tchi^0_i\tchi^0_j\tchi^0_k) 
\nonumber \\
&&
\hspace*{-2mm} 
+ n^{R}_{A^0 ij}n^L_{H^0 jk}
\D_8(\tilde{e}_R\tchi^0_i\tchi^0_j\tchi^0_k) 
\Big\} \nonumber \\
& + \displaystyle{\frac{e^2}{8\pi^2c^2_W}} P_R 
\sum_{ijk} Z^{N}_{1i}Z^{N*}_{1k} \times &
\hspace*{-2mm} 
\Big\{
M_{\tchi^0_i}M_{\tchi^0_k}n^{L}_{H^0 ij}n^R_{A^0 jk}
\bar{\D}''_8(\tilde{e}_R\tchi^0_i\tchi^0_j\tchi^0_k) 
\nonumber \\
&&
\hspace*{-2mm} 
+ M_{\tchi^0_i}M_{\tchi^0_j}n^{L}_{H^0 ij}n^L_{A^0 jk}
\bar{\D}'_8(\tilde{e}_R\tchi^0_i\tchi^0_j\tchi^0_k)
\nonumber \\
&&
\hspace*{-2mm} 
+ M_{\tchi^0_j}M_{\tchi^0_k}n^{R}_{H^0 ij}n^R_{A^0 jk}
\bar{\D}'_8(\tilde{e}_R\tchi^0_i\tchi^0_j\tchi^0_k)
\nonumber \\
&&
\hspace*{-2mm} 
+ n^{R}_{H^0 ij}n^L_{A^0 jk}
\bar{\D}_8(\tilde{e}_R\tchi^0_i\tchi^0_j\tchi^0_k) 
\Big\},
\end{eqnarray} 
\begin{eqnarray}
A^{box8}_{h^0A^0}(\tilde{e}\tchi^0\tchi^0\tchi^0)
& \hspace*{-15mm} 
= \displaystyle{\frac{e^2}{32\pi^2s^2_Wc^2_W}} P_L &
\hspace*{-18mm} 
\sum_{ijk} (Z^{N*}_{1i}s_W+Z^{N*}_{2i}c_W)(Z^{N}_{1k}s_W+Z^{N}_{2k}c_W) \times 
\nonumber \\
&&
\hspace*{-18mm} 
\Big\{
M_{\tchi^0_i}M_{\tchi^0_k}n^{R}_{A^0 ij}n^L_{h^0 jk}
\D''_8(\tilde{e}_L\tchi^0_i\tchi^0_j\tchi^0_k) 
\nonumber \\
&&
\hspace*{-18mm} 
+ M_{\tchi^0_i}M_{\tchi^0_j}n^{R}_{A^0 ij}n^R_{h^0 jk}
\D'_8(\tilde{e}_L\tchi^0_i\tchi^0_j\tchi^0_k)
\nonumber \\
&&
\hspace*{-18mm} 
+ M_{\tchi^0_j}M_{\tchi^0_k}n^{L}_{A^0 ij}n^L_{h^0 jk}
\D'_8(\tilde{e}_L\tchi^0_i\tchi^0_j\tchi^0_k)
\nonumber \\
&&
\hspace*{-18mm} 
+ n^{L}_{A^0 ij}n^R_{h^0 jk}
\D_8(\tilde{e}_L\tchi^0_i\tchi^0_j\tchi^0_k) 
\Big\} \nonumber \\
& \hspace*{-15mm} 
- \displaystyle{\frac{e^2}{32\pi^2s^2_Wc^2_W}} P_L & 
\hspace*{-18mm} 
\sum_{ijk} (Z^{N*}_{1i}s_W+Z^{N*}_{2i}c_W)(Z^{N}_{1k}s_W+Z^{N}_{2k}c_W) \times 
\nonumber \\
&&
\hspace*{-18mm} 
\Big\{
M_{\tchi^0_i}M_{\tchi^0_k}n^{R}_{h^0 ij}n^L_{A^0 jk}
\bar{\D}''_8(\tilde{e}_L\tchi^0_i\tchi^0_j\tchi^0_k) 
\nonumber \\
&&
\hspace*{-18mm} 
+ M_{\tchi^0_i}M_{\tchi^0_j}n^{R}_{h^0 ij}n^R_{A^0 jk}
\bar{\D}'_8(\tilde{e}_L\tchi^0_i\tchi^0_j\tchi^0_k) 
\nonumber \\
&&
\hspace*{-18mm} 
+ M_{\tchi^0_j}M_{\tchi^0_k}n^{L}_{h^0 ij}n^L_{A^0 jk}
\bar{\D}'_8(\tilde{e}_L\tchi^0_i\tchi^0_j\tchi^0_k)
\nonumber \\
&&
\hspace*{-18mm} 
+ n^{L}_{h^0 ij}n^R_{A^0 jk}
\bar{\D}_8(\tilde{e}_L\tchi^0_i\tchi^0_j\tchi^0_k) 
\Big\}
\nonumber \\
& - \displaystyle{\frac{e^2}{8\pi^2c^2_W}} P_R 
\sum_{ijk} Z^{N}_{1i}Z^{N*}_{1k} \times &
\hspace*{-2mm} 
\Big\{ 
M_{\tchi^0_i}M_{\tchi^0_k}n^{L}_{A^0 ij}n^R_{h^0 jk}
\D''_8(\tilde{e}_R\tchi^0_i\tchi^0_j\tchi^0_k) 
\nonumber \\
&&
\hspace*{-2mm} 
+ M_{\tchi^0_i}M_{\tchi^0_j}n^{L}_{A^0 ij}n^L_{h^0 jk}
\D'_8(\tilde{e}_R\tchi^0_i\tchi^0_j\tchi^0_k) 
\nonumber \\
&&
\hspace*{-2mm} 
+ M_{\tchi^0_j}M_{\tchi^0_k}n^{R}_{A^0 ij}n^R_{h^0 jk}
\D'_8(\tilde{e}_R\tchi^0_i\tchi^0_j\tchi^0_k) 
\nonumber \\
&&
\hspace*{-2mm} 
+ n^{R}_{A^0 ij}n^L_{h^0 jk}
\D_8(\tilde{e}_R\tchi^0_i\tchi^0_j\tchi^0_k) 
\Big\} \nonumber \\
& + \displaystyle{\frac{e^2}{8\pi^2c^2_W}} P_R 
\sum_{ijk} Z^{N}_{1i}Z^{N*}_{1k} \times &
\hspace*{-2mm} 
\Big\{
M_{\tchi^0_i}M_{\tchi^0_k}n^{L}_{h^0 ij}n^R_{A^0 jk}
\bar{\D}''_8(\tilde{e}_R\tchi^0_i\tchi^0_j\tchi^0_k) 
\nonumber \\
&&
\hspace*{-2mm} 
+ M_{\tchi^0_i}M_{\tchi^0_j}n^{L}_{h^0 ij}n^L_{A^0 jk}
\bar{\D}'_8(\tilde{e}_R\tchi^0_i\tchi^0_j\tchi^0_k)
\nonumber \\
&&
\hspace*{-2mm} 
+ M_{\tchi^0_j}M_{\tchi^0_k}n^{R}_{h^0 ij}n^R_{A^0 jk}
\bar{\D}'_8(\tilde{e}_R\tchi^0_i\tchi^0_j\tchi^0_k)
\nonumber \\
&&
\hspace*{-2mm} 
+ n^{R}_{h^0 ij}n^L_{A^0 jk}
\bar{\D}_8(\tilde{e}_R\tchi^0_i\tchi^0_j\tchi^0_k) 
\Big\}.
\end{eqnarray} 
 
\subsubsection{Twisted box10 diagrams}

Two types of twisted box10 diagrams must be considered: 
$\tilde{\nu} \tchi \tchi \tilde{\nu}$ and 
$\tilde{e} \tchi^0 \tchi^0 \tilde{e}$.\\ 

The $\tilde{\nu} \tchi \tchi \tilde{\nu}$ 
boxes contribute with left-handed terms only:
\begin{eqnarray}
A^{box10}_{H^0A^0}(\tilde{\nu}_L \tchi \tchi \tilde{\nu}_L)
& = & \frac{e^2}{16\pi^2s^2_W}
\displaystyle{\sum_{ij}}
Z^{+*}_{1i}Z^{+}_{1j} \times 
g_{H^0 \tilde{\nu}_L \tilde{\nu}_L} \times \nonumber \\
&&
\left[
M_{\tchi^{}_i}\,c^{R}_{A^0 ji} 
\D_{10}(\tilde{\nu}_L \tchi^{}_i \tchi^{}_j \tilde{\nu}_L)+
M_{\tchi^{}_j}\,c^{L}_{A^0 ji} 
\D'_{10}(\tilde{\nu}_L \tchi^{}_i \tchi^{}_j \tilde{\nu}_L)
\right],~~~~
\end{eqnarray}
\begin{eqnarray}
A^{box10}_{h^0A^0}(\tilde{\nu}_L \tchi^{} \tchi^{} \tilde{\nu}_L)
& = & \frac{e^2}{16\pi^2s^2_W}
\displaystyle{\sum_{ij}}
Z^{+*}_{1i}Z^{+}_{1j} \times
g_{h^0 \tilde{\nu}_L \tilde{\nu}_L} \times \nonumber \\
&&
\left[
M_{\tchi^{}_i}\,c^{R}_{A^0 ji} 
\D_{10}(\tilde{\nu}_L \tchi^{}_i \tchi^{}_j \tilde{\nu}_L)+
M_{\tchi^{}_j}\,c^{L}_{A^0 ji} 
\D'_{10}(\tilde{\nu}_L \tchi^{}_i \tchi^{}_j \tilde{\nu}_L)
\right].~~~~
\end{eqnarray}

The $\tilde{e} \tchi^0 \tchi^0 \tilde{e}$ boxes contribute with both 
left-handed and right-handed terms. Writing all these terms into one 
single expression leads to:
\begin{eqnarray}
A^{box10}_{H^0A^0}(\tilde{e} \tchi^0 \tchi^0 \tilde{e})
& = \displaystyle{\frac{e^2}{32\pi^2s^2_Wc^2_W}} P_L &
\hspace*{-2mm}
\displaystyle{\sum_{ij}}
(Z^{N*}_{1i}s_W+Z^{N*}_{2i}c_W)(Z^{N}_{1j}s_W+Z^{N}_{2j}c_W) \times 
g_{H^0 \tilde{e}_L \tilde{e}_L} \times \nonumber \\
&&
\hspace*{-2mm}
\left[
M_{\tchi^0_i}\,n^{R}_{A^0 ji} 
\D_{10}(\tilde{e}_L \tchi^0_i \tchi^0_j \tilde{e}_L)+
M_{\tchi^0_j}\,n^{L}_{A^0 ji} 
\D'_{10}(\tilde{e}_L \tchi^0_i \tchi^0_j \tilde{e}_L)
\right] \nonumber \\
& \hspace*{-8mm}
+ \displaystyle{\frac{e^2}{8\pi^2c^2_W}} P_R &
\hspace*{-10mm}
\displaystyle{\sum_{ij}}
Z^{N*}_{1i} Z^{N}_{1j}  \times 
g_{H^0 \tilde{e}_R \tilde{e}_R} \times \nonumber \\
&&
\hspace*{-10mm}
\left[
M_{\tchi^0_i}\,n^{L}_{A^0 ji} 
\D_{10}(\tilde{e}_R \tchi^0_i \tchi^0_j \tilde{e}_R)+
M_{\tchi^0_j}\,n^{R}_{A^0 ji} 
\D'_{10}(\tilde{e}_R \tchi^0_i \tchi^0_j \tilde{e}_R)
\right],~~~~~~
\end{eqnarray}
\begin{eqnarray}
A^{box10}_{h^0A^0}(\tilde{e} \tchi^0 \tchi^0 \tilde{e})
& = \displaystyle{\frac{e^2}{32\pi^2s^2_Wc^2_W}} P_L &
\hspace*{-2mm}
\displaystyle{\sum_{ij}}
(Z^{N*}_{1i}s_W+Z^{N*}_{2i}c_W)(Z^{N}_{1j}s_W+Z^{N}_{2j}c_W) \times 
g_{h^0 \tilde{e}_L \tilde{e}_L} \times \nonumber \\
&&
\hspace*{-2mm}
\left[
M_{\tchi^0_i}\,n^{R}_{A^0 ji} 
\D_{10}(\tilde{e}_L \tchi^0_i \tchi^0_j \tilde{e}_L)+
M_{\tchi^0_j}\,n^{L}_{A^0 ji} 
\D'_{10}(\tilde{e}_L \tchi^0_i \tchi^0_j \tilde{e}_L)
\right] \nonumber \\
& \hspace*{-8mm}
+ \displaystyle{\frac{e^2}{8\pi^2c^2_W}} P_R &
\hspace*{-10mm}
\displaystyle{\sum_{ij}}
Z^{N*}_{1i} Z^{N}_{1j}  \times 
g_{h^0 \tilde{e}_R \tilde{e}_R} \times \nonumber \\
&&
\hspace*{-10mm}
\left[
M_{\tchi^0_i}\,n^{L}_{A^0 ji} 
\D_{10}(\tilde{e}_R \tchi^0_i \tchi^0_j \tilde{e}_R)+
M_{\tchi^0_j}\,n^{R}_{A^0 ji} 
\D'_{10}(\tilde{e}_R \tchi^0_i \tchi^0_j \tilde{e}_R)
\right].~~~~~~
\end{eqnarray}

\section{Conclusion and outlooks}
\label{concl}

In this paper, we discussed all electroweak one loop contributions to the pair 
production cross section for charged and neutral Higgs bosons in $e^+e^-$
collisions, in the theoretical framework of the MSSM. The one loop amplitudes 
of initial vertices and $e^\pm$ self-energy, of $\gamma$ and $Z$ boson 
self-energies, of the corresponding counter terms, of final vertices 
and Higgs self-energies and of box diagrams are respectively given by 
equations (\ref{in-charged}), (\ref{se-charged}), (\ref{ct-charged}), 
(\ref{fin-charged}) and (\ref{box-charged}) for the charged Higgs sector, 
and by equations (\ref{in-neutral}), (\ref{se-neutral}), (\ref{ct-neutral}), 
(\ref{fin-neutral}) and (\ref{box-neutral}) for the neutral Higgs sector. 
The left-handed and right-handed amplitudes of all these electroweak one loop 
contributions are:
\begin{itemize}
\item in the charged Higgs sector:
\begin{eqnarray}
a^{1\,loop}_{L,R}(H^+H^-) & = & a^{in}_{L,R}(H^+H^-) 
\nonumber \\
                                  & + & a^{RG}_{L,R}(H^+H^-) 
                                    + a^{ct}_{L,R}(H^+H^-)
\nonumber \\
                                  & + & a^{fin}_{L,R}(H^+H^-)
\nonumber \\
                                  & + & a^{box}_{L,R}(H^+H^-),
\end{eqnarray} 
\item in the neutral Higgs sector:
\begin{eqnarray}
a^{1\,loop}_{L,R}(H^0A^0/h^0A^0) & = & a^{in}_{L,R}(H^0A^0/h^0A^0) 
\nonumber \\
                                  & + & a^{RG}_{L,R}(H^0A^0/h^0A^0) 
                                    + a^{ct}_{L,R}(H^0A^0/h^0A^0)
\nonumber \\
                                  & + & a^{fin}_{L,R}(H^0A^0/h^0A^0)
\nonumber \\
                                  & + & a^{box}_{L,R}(H^0A^0/h^0A^0).
\end{eqnarray} 
\end{itemize}

The differential production cross section for charged or neutral Higgs bosons 
at the one loop level can then be calculated as follows:
\begin{equation}
\frac{d\sigma}{d\cos\theta} = 
\frac{\pi\alpha_{em}^2\beta^3_H}{8q^2} (1-\cos^2\theta) \times 
\left[
|a^{Born}_{L}|^2 + 2 |a^{Born}_{L} a^{1\,loop}_{L}| + 
|a^{Born}_{R}|^2 + 2 |a^{Born}_{R} a^{1\,loop}_{R}| 
\right].~~
\end{equation}

In the previous equation, $a^{Born}_{L,R}$ is the Born amplitude of 
equation (\ref{born-charged}) or (\ref{born-neutral}). As for $\beta_H$, 
it stands for the velocity of the Higgs bosons, see equation~(\ref{betah}). 
After integration over $\cos \theta$ (which appears in the contributions 
of the box diagrams), one obtains the total pair production cross section 
at the one loop level. 
Note that, in the case of the tree level cross sections for $H^+H^-$ 
and $H^0A^0+h^0A^0$, there is a direct dependence on $M_A$ only and 
not on the other MSSM parameters. However, after having taken into 
account all electroweak one loop contributions, this is not true 
anymore. Indeed, $a^{1\,loop}_{L,R}$ depends on other MSSM parameters 
than just $M_A$.\\

A C++ numerical code has been developed in order to calculate accurately 
all one loop electroweak contributions and, in turn, to compare the pair 
production cross sections for MSSM charged and neutral Higgs bosons at 
tree level and at the one loop level. The relevant Feynman diagrams 
are computed by calling the suitable functions in the LoopTools~2.1 
library~\cite{LoopTools}. The input of the code, in standard notation, is 
the following: $\tan\beta$, $\mu$, $M_A$ (the mass of the $A^0$ Higgs boson), 
the gaugino parameters $M_1$ and $M_2$, the scalar mass scale $M_S$, the 
sfermion mixing matrix parameters $A_u$ and $A_d$. A possible reference 
for these parameters is~\cite{hep9807427}. This input requires a preliminary 
pre-processing using the FeynHiggs~2.1~\cite{FeynHiggs} code. A subset of 
these parameters is then fed into FeynHiggs in order to compute the masses 
of the Higgs bosons $h^0$, $H^0$, $H^\pm$, as well as the mixing angle in the 
neutral sector $\alpha$. These additional parameters do indeed appear in the 
analytical expressions described in the text. The output of the code is the 
cross section for the various processes under consideration. We have 
successfully checked that the MSSM Higgs bosons pair production cross 
section computed by our code at the one loop level remains stable against 
UV divergences, both in the charged and neutral Higgs sectors. Also, we 
have checked that the variation of the computed one loop cross section 
with $q^2$ agrees with our expectations.\\

Note that, in this code, we have included all virtual 
contributions involving particles having electroweak interactions
in the MSSM, but we did not treat pure QED effects, such as Initial 
State Radiation (ISR) and Final State Radiation (FSR). The reason is 
that these effects may depend on the characteristics of
the detectors (for instance, they need specific kinematical cuts) 
and some specific codes exist in order to treat them. Nevertheless, 
in order to be able to test electroweak symmetry properties at high 
energy, in particular those of supersymmetric nature, which is indeed 
the purpose of this work, we have included the virtual photon effects, 
but with a photon mass set to $M_Z$ in order to keep these effects 
finite. In order to have the complete (observable) contribution including 
QED effects, one should compute the following combination: 
{\it 
Our contribution + ISR + FSR + virtual soft photon with zero mass $-$ 
virtual soft photon with $M_Z$.
}
This combination should be calculated at the level of the codes that include 
the ISR and FSR effects.

\newpage

\section*{Appendix A: Vertices and couplings}

\underline{Gauge - Fermion - Fermion}\\ 

Let $Q_f$ and $T^3_f$ be respectively the charge and the third isospin 
component of the fermion $f$, then the couplings of gauge bosons to 
left-handed and right-handed fermions are:
\ba
P_L\,g_{\gamma f f} = Q_f 
& \mbox{and} &
P_R\,g_{\gamma f f} = Q_f,
\nonumber \\
P_L\,g_{Z f f}= \frac{T^3_f-Q_fs^2_W}{s_Wc_W} 
& \mbox{and} &
P_R\,g_{Z f f}= -Q_f\frac{s_W}{c_W}, 
\nonumber \\
P_L\,g_{W f f'}={1\over s_W\sqrt{2}} 
& \mbox{and} & 
P_R\,g_{W f f'}=0. \nonumber
\ea

Note that, in this paper, we have also used a simplified notation for the 
couplings of $\gamma$ or $Z$ to fermion pairs:
\ba
g_{V L f} & \equiv & P_L\,g_{V f f}, \nonumber \\
g_{V R f} & \equiv & P_R\,g_{V f f}, \nonumber
\ea
where $V$ stands for either $\gamma$ or $Z$.\\ 

\underline{Gauge - Gaugino - Gaugino} 
\ba
\O^{\gamma L}_{ij} = -e\delta_{ij}
& \mbox{and} &
\O^{\gamma R}_{ij} = -e\delta_{ij},
\nonumber \\
\O^{ZL}_{ij} = 
-\frac{e\left[Z^{+*}_{1i}Z^{+}_{1j}+\delta_{ij}(c^2_W-s^2_W)\right]}{2s_Wc_W}
& \mbox{and} &
\O^{ZR}_{ij} = 
-\frac{e\left[Z^{-}_{1i}Z^{-*}_{1j}+\delta_{ij}(c^2_W-s^2_W)\right]}{2s_Wc_W},
\nonumber \\
\O^{0L}_{ij}=\frac{e(Z^{N*}_{4i}Z^{N}_{4j}-Z^{N*}_{3i}Z^{N}_{3j})}{2s_Wc_W}
& \mbox{and} &
\O^{0R}_{ij}=-\frac{e(Z^{N}_{4i}Z^{N*}_{4j}-Z^{N}_{3i}Z^{N*}_{3j})}{2s_Wc_W},
\nonumber \\
\O^{WL}_{ij}=\frac{e}{s_W}
\left( Z^{N}_{2j}Z^{+*}_{1i}-{1\over\sqrt{2}}Z^{N}_{4k}Z^{+*}_{2i} \right)
& \mbox{and} & 
\O^{WR}_{ij}=\frac{e}{s_W}
\left( Z^{N*}_{2j}Z^{-}_{1i}+{1\over\sqrt{2}}Z^{N*}_{3j}Z^{-}_{2i} \right).
\nonumber
\ea

Here, the $Z_{ij}$ terms correspond to the various elements of the unitary 
mixing matrices of the charginos and neutralinos. They are derived from 
the diagonalization of the MSSM gaugino mass matrix, see for instance 
reference~\cite{zij} for details.\\

\underline{Gauge - Sfermion - Sfermion}\\ 

The couplings of gauge bosons to unmixed left-handed and 
right-handed sfermions are:
\ba
g^0_{\gamma \tilde{f}_L \tilde{f}_L} = -eQ_f 
& \mbox{and} &
g^0_{\gamma \tilde{f}_R \tilde{f}_R} = -eQ_f,
\nonumber \\
g^0_{Z \tilde{f}_L \tilde{f}_L} = -\frac{e(T^3_f-Q_fs^2_W)}{s_Wc_W} 
& \mbox{and} &
g^0_{Z \tilde{f}_R \tilde{f}_R} = eQ_f\frac{s_W}{c_W}, 
\nonumber \\
g^0_{W \tilde{f}_L \tilde{f}'_L}=-{e\over s_W\sqrt{2}}
& \mbox{and} & 
g^0_{W \tilde{f}_R \tilde{f}'_R}=0.
\nonumber
\ea

Let $\theta_{\tilde{f}}$ be the mixing angle of the sfermion 
$\tilde{f}$ (generally a third generation squark). Let us also 
define $c_{\tilde{f}} \equiv \cos\theta_{\tilde{f}}$ and 
$s_{\tilde{f}} \equiv \sin\theta_{\tilde{f}}$. The coupling 
between a gauge boson and two sfermions with mixing is 
then given by:
\ba
g_{\gamma\tilde{f}^{}_{1}\tilde{f}^{}_{1}} = 
g_{\gamma\tilde{f}^{}_{2}\tilde{f}^{}_{2}} = 
g^0_{\gamma\tilde{f}^{}_{L}\tilde{f}^{}_{L}} = 
g^0_{\gamma\tilde{f}^{}_{R}\tilde{f}^{}_{R}} = -eQ_{f}, \nonumber
\ea
\ba
g_{Z \tilde{f}^{}_{1}\tilde{f}^{}_{1}} & = &
c^2_{\tilde{f}} g^0_{Z \tilde{f}^{}_{L}\tilde{f}^{}_{L}}+
s^2_{\tilde{f}} g^0_{Z \tilde{f}^{}_{R}\tilde{f}^{}_{R}}, \nonumber\\
g_{Z \tilde{f}^{}_{2}\tilde{f}^{}_{2}} & = &
s^2_{\tilde{f}} g^0_{Z \tilde{f}^{}_{L}\tilde{f}^{}_{L}}+
c^2_{\tilde{f}} g^0_{Z \tilde{f}^{}_{R}\tilde{f}^{}_{R}}, \nonumber\\
g_{Z \tilde{f}^{}_{1}\tilde{f}^{}_{2}} & = &
g_{Z \tilde{f}^{}_{2}\tilde{f}^{}_{1}} =
c_{\tilde{f}}s_{\tilde{f}} (g^0_{Z \tilde{f}^{}_{R}\tilde{f}^{}_{R}} -
g^0_{Z \tilde{f}^{}_{L}\tilde{f}^{}_{L}}), \nonumber
\ea
\ba
g_{W \tilde{f}_{1}\tilde{f}'_{1}} & = &
c_{\tilde{f}}c_{\tilde{f}'} g^0_{W \tilde{f}_L \tilde{f}'_L},
\nonumber \\
g_{W \tilde{f}_{2}\tilde{f}'_{2}} & = &
s_{\tilde{f}}s_{\tilde{f}'} g^0_{W \tilde{f}_L \tilde{f}'_L},
\nonumber \\ 
g_{W \tilde{f}_{1}\tilde{f}'_{2}} & = &
-c_{\tilde{f}}s_{\tilde{f}'} g^0_{W \tilde{f}_L \tilde{f}'_L},
\nonumber \\ 
g_{W \tilde{f}_{2}\tilde{f}'_{1}} & = &
-c_{\tilde{f}'}s_{\tilde{f}} g^0_{W \tilde{f}_L \tilde{f}'_L}.
\nonumber
\ea

\underline{Gauge - Gauge - Higgs}
\ba
g_{ZZH^0} = \frac{eM_W}{s_Wc^2_W}\cos(\beta-\alpha) 
& \mbox{and} &
g_{ZZh^0} = \frac{eM_W}{s_Wc^2_W}\sin(\beta-\alpha), \nonumber \\
g_{WWH^0} = \frac{eM_W}{s_W}     \cos(\beta-\alpha) 
& \mbox{and} &
g_{WWh^0} = \frac{eM_W}{s_W}     \sin(\beta-\alpha), \nonumber \\
g_{\gamma W G} = eM_W 
& \mbox{and} &
g_{Z W G} = -eM_W \frac{s_W}{c_W}. \nonumber
\ea

\underline{Gauge - Higgs - Higgs} 
\ba  
g_{Z H^0 A^0} = -{e\over2s_Wc_W}\sin(\beta-\alpha)
& \mbox{and} &
g_{Z h^0 A^0} = +{e\over2s_Wc_W}\cos(\beta-\alpha),
\nonumber \\
g_{Z H^0 G^0} = +{e\over2s_Wc_W}\cos(\beta-\alpha)
& \mbox{and} &
g_{Z h^0 G^0} = +{e\over2s_Wc_W}\sin(\beta-\alpha),
\nonumber \\
g_{W^\pm H^\pm H^0}  = +\,Q_W \times {e\over2s_W}\sin(\beta-\alpha)
& \mbox{and} &
g_{W^\pm H^\pm h^0}  = -\,Q_W \times {e\over2s_W}\cos(\beta-\alpha),
\nonumber \\
g_{W^\pm G^\pm H^0}  = -\,Q_W \times {e\over2s_W}\cos(\beta-\alpha)
& \mbox{and} &
g_{W^\pm G^\pm h^0}  = -\,Q_W \times {e\over2s_W}\sin(\beta-\alpha),
\nonumber \\
g_{W H A^0}  = +{e\over2s_W}
& \mbox{and} &
g_{W H G^0}  = 0,
\nonumber \\
g_{W G A^0}  = 0
& \mbox{and} &
g_{W G G^0}  = +{e\over2s_W},
\nonumber \\
g_{\gamma H H}   = -e
& \mbox{and} &
g_{Z H H}   = -e\frac{1-2s^2_W}{2s_Wc_W},
\nonumber \\
g_{\gamma G G}   = -e
& \mbox{and} &
g_{Z G G}   = -e\frac{1-2s^2_W}{2s_Wc_W}.
\nonumber 
\ea

\underline{Higgs - Fermion - Fermion}\\

The coupling constant between a Higgs boson and a fermion 
pair is proportional to the mass of the fermion(s). Thus, 
only the third generation quarks are usually considered.\\

In the charged Higgs sector:
\begin{itemize}
\item for $b\to t~H^-$: 
$c^L_{b \to tH^-}=\displaystyle{\frac{e}{\sqrt{2}s_WM_W}\, 
M_t\,\mbox{cot}\beta}$ 
and $c^R_{b \to tH^-}=\displaystyle{\frac{e}{\sqrt{2}s_WM_W}\, 
M_b \tan\beta}$,
\item for $t\to b~H^+$: 
$c^L_{t \to bH^+}=\displaystyle{\frac{e}{\sqrt{2}s_WM_W}\, 
M_b \tan\beta}$ 
and $c^R_{t \to bH^+}=\displaystyle{\frac{e}{\sqrt{2}s_WM_W}\, 
M_t\,\mbox{cot}\beta}$.
\end{itemize}

In the neutral Higgs sector, the left-handed coupling constants are:
\begin{itemize}
\item $c^L_{H^0t}=-\displaystyle{\frac{eM_t}{2s_WM_W} \times 
\frac{\sin\alpha}{\sin\beta}}$
and 
$c^L_{H^0b}=-\displaystyle{\frac{eM_b}{2s_WM_W} \times 
\frac{\cos\alpha}{\cos\beta}}$,
\item $c^L_{h^0t}=-\displaystyle{\frac{eM_t}{2s_WM_W} \times 
\frac{\cos\alpha}{\sin\beta}}$
and 
$c^L_{H^0b}=+\displaystyle{\frac{eM_b}{2s_WM_W} \times 
\frac{\sin\alpha}{\cos\beta}}$,
\item $c^L_{A^0t}=\displaystyle{\frac{eM_t}{2s_WM_W} \times 
\mbox{cot}\beta}$
and 
$c^L_{A^0b}=\displaystyle{\frac{eM_b}{2s_WM_W} \times 
\tan\beta}$.
\end{itemize}

As for the right-handed couplings constants, one simply has:
\begin{itemize}
\item $c^R_{H^0f}=c^L_{H^0f}$,
\item $c^R_{h^0f}=c^L_{h^0f}$,
\item $c^R_{A^0f}=-c^L_{A^0f}$.
\end{itemize}
~~~

\underline{Higgs - Gaugino - Gaugino}\\

In the charged Higgs sector, there are two types of vertex to consider: 
$\tchi^0_j \to \tchi^+_i ~H^-$ and $\tchi^+_j \to \tchi^0_i ~H^+$. 
Let us choose the case where $i$ and $j$ label respectively a chargino 
and a neutralino, then the corresponding left-handed and right-handed
coupling constants are:
\begin{itemize} 
\item $c^L_{H{ij}}=
\displaystyle{\frac{e \sin\beta}{s_Wc_W}}
\left[
{1\over\sqrt{2}}
Z^-_{2i}(Z^N_{1j}s_W+Z^N_{2j}c_W)-Z^-_{1i}Z^N_{3j}c_W
\right]$,
\item $c^R_{H{ij}}=-
\displaystyle{\frac{e \cos\beta}{s_Wc_W}}
\left[
{1\over\sqrt{2}}
Z^{+*}_{2i}(Z^{N*}_{1j}s_W+Z^{N*}_{2j}c_W)+Z^{+*}_{1i}Z^{N*}_{4j}c_W
\right]$.
\end{itemize}

For the other vertex (where $i$ and $j$ label respectively a 
neutralino and a chargino), one should instead use the left-handed 
and right-handed coupling constants $c^{R*}_{H{ji}}$ 
and $c^{L*}_{H{ji}}$, respectively.\\

In the neutral Higgs sector, one must consider the coupling between a 
neutral Higgs boson and either two charginos or two neutralinos.\\

For the coupling between a neutral Higgs boson and two charginos:
\ba
c^L_{H^0ij}=-\displaystyle{
{e\over\sqrt{2}s_W}
\left[
\cos\alpha Z^{-}_{2i}Z^{+}_{1j}+\sin\alpha Z^{-}_{1i}Z^{+}_{2j}
\right]} 
& \mbox{and} & 
c^{R}_{H^0ij}=c^{L*}_{H^0ji}, \nonumber \\
c^L_{h^0ij}=-\displaystyle{
{e\over\sqrt{2}s_W}
\left[
-\sin\alpha Z^{-}_{2i}Z^{+}_{1j}+\cos\alpha Z^{-}_{1i}Z^{+}_{2j}
\right]}
& \mbox{and} & 
c^{R}_{h^0ij}=c^{L*}_{h^0ji}, \nonumber \\
c^L_{A^0ij}=-\displaystyle{
{e\over\sqrt{2}s_W}
\left[
\sin\beta Z^{-}_{2i}Z^{+}_{1j}+\cos\beta Z^{-}_{1i}Z^{+}_{2j}
\right]}
& \mbox{and} & 
c^{R}_{A^0ij}=-c^{L*}_{A^0ji}. \nonumber
\ea

Let us now consider the coupling between a neutral Higgs boson and 
two neutralinos. For the left-handed components, one has:
\ba
n^L_{H^0ij} & = \displaystyle{{e\over2s_Wc_W}} \times &
\Big\{
(\cos\alpha Z^{N}_{3j}-\sin\alpha Z^{N}_{4j})
(Z^{N}_{1i}s_W-Z^{N}_{2i}c_W) \nonumber \\
&&
+(\cos\alpha Z^{N}_{3i}-\sin\alpha Z^{N}_{4i})
(Z^{N}_{1j}s_W-Z^{N}_{2j}c_W)
\Big\}, \nonumber \\
n^L_{h^0ij} & = -\displaystyle{{e\over2s_Wc_W}} \times &
\Big\{
(\sin\alpha Z^{N}_{3j}+\cos\alpha Z^{N}_{4j})
(Z^{N}_{1i}s_W-Z^{N}_{2i}c_W) \nonumber \\
&&
+(\sin\alpha Z^{N}_{3i}+\cos\alpha Z^{N}_{4i})
(Z^{N}_{1j}s_W-Z^{N}_{2j}c_W)
\Big\}, \nonumber \\ 
n^L_{A^0ij} & = \displaystyle{{e\over2s_Wc_W}} \times &
\Big\{
(\sin\beta Z^{N}_{3j}-\cos\beta Z^{N}_{4j})
(Z^{N}_{1i}s_W-Z^{N}_{2i}c_W) \nonumber \\
&&
+(\sin\beta Z^{N}_{3i}-\cos\beta Z^{N}_{4i})
(Z^{N}_{1j}s_W-Z^{N}_{2j}c_W)
\Big\}. \nonumber 
\ea

As for the right-handed components, one simply has:
\ba
n^{R}_{H^0ij} & = & n^{L*}_{H^0ji}, \nonumber \\
n^{R}_{h^0ij} & = & n^{L*}_{h^0ji}, \nonumber \\
n^{R}_{A^0ij} & = & -n^{L*}_{A^0ji}. \nonumber
\ea

\underline{Higgs - Sfermion - Sfermion}\\ 

Let us first consider the light unmixed sfermions. Their coupling constant to 
the charged and neutral Higgs bosons is not proportional to the mass of the 
corresponding fermion(s) and it can thus not be neglected.\\

In the charged Higgs sector, if $\tilde{f}$ and $\tilde{f}'$ represent 
respectively up-squarks and down-squarks of the first and second generations, 
or sneutrinos and charged sleptons, one has:
\ba
g_{H \tilde{f}^{}_L \tilde{f}'_L} = 
-\displaystyle{{eM_W \over s_W\sqrt{2}}\sin2\beta}
& \mbox{and} &
g_{H \tilde{f}^{}_R \tilde{f}'_R} = 0. \nonumber
\ea

In the neutral Higgs sector, one has:
\ba
g_{H^0 \tilde{f}^{}_L \tilde{f}^{}_L} = -{eM_W\over s_Wc^2_W}
(T^3_f-Q_fs^2_W)\cos(\alpha+\beta)
& \mbox{and} &
g_{H^0 \tilde{f}^{}_R \tilde{f}^{}_R} = -{eM_W\over s_Wc^2_W}
Q_fs^2_W \cos(\alpha+\beta) ,\nonumber \\
g_{h^0 \tilde{f}^{}_L \tilde{f}^{}_L} = {eM_W\over s_Wc^2_W}
(T^3_f-Q_fs^2_W)\sin(\alpha+\beta)
& \mbox{and} &
g_{h^0 \tilde{f}^{}_R \tilde{f}^{}_R} = {eM_W\over s_Wc^2_W}
Q_fs^2_W \sin(\alpha+\beta). \nonumber 
\ea

Note that the couplings between $A^0$ and a pair of light unmixed 
sfermions are proportional to the fermion mass and are thus negligible.\\

Let us now consider the heavy sfermions, i.e. the third generation squarks, 
and, in a first step, let us assume that there is no mixing. The coupling 
constants between the MSSM Higgs boson and a pair of unmixed heavy sfermions 
are given below.\\ 

For the charged Higgs bosons $H$, one has:
\ba
g^0_{H \tilde{t}^{}_L \tilde{b}^{}_L} & = & 
-{eM_W\over s_W\sqrt{2}}
\left[
\sin2\beta-{M^2_b \tan\beta+ M^2_t\,\mbox{cot}\beta \over M^2_W}
\right], \nonumber \\
g^0_{H \tilde{t}^{}_R \tilde{b}^{}_R} & = & 
{eM_tM_b\over s_WM_W\sqrt{2}}
\left[ \tan\beta+\mbox{cot}\beta \right], \nonumber \\
g^0_{H \tilde{t}^{}_L \tilde{b}^{}_R} & = & 
-{eM_b\over s_WM_W\sqrt{2}}
\left[ \mu - A_b \tan\beta \right]. \nonumber \\
g^0_{H \tilde{t}^{}_R \tilde{b}^{}_L} & = & 
-{eM_t\over s_WM_W\sqrt{2}}
\left[ \mu - A_t \mbox{cot}\beta \right]. \nonumber
\ea

For the neutral Higgs boson $A^0$, there are only off-diagonal terms:
\ba
g^0_{A^0 \tilde{t}^{}_L \tilde{t}^{}_L} = 
g^0_{A^0 \tilde{t}^{}_R \tilde{t}^{}_R} = 0
~\mbox{and}~
g^0_{A^0 \tilde{b}^{}_L \tilde{b}^{}_L} = 
g^0_{A^0 \tilde{b}^{}_R \tilde{b}^{}_R} = 0, 
\nonumber \\
g^0_{A^0 \tilde{t}^{}_L \tilde{t}^{}_R} = 
g^0_{A^0 \tilde{t}^{}_R \tilde{t}^{}_L} = 
-{eM_t\over2s_WM_W}
\left[-\mu -A_t\,\mbox{cot}\beta \right], 
\nonumber \\
g^0_{A^0 \tilde{b}^{}_L \tilde{b}^{}_R} = 
g^0_{A^0 \tilde{b}^{}_R \tilde{b}^{}_L} = 
-{eM_b\over2s_WM_W}
\left[-\mu -A_b \tan\beta \right]. \nonumber
\ea

For the neutral Higgs boson $H^0$, one has:
\ba
g^0_{H^0 \tilde{t}^{}_{L} \tilde{t}^{}_{L}} & = & -{eM_W\over s_Wc^2_W}
\left(
{1\over2}-{2\over3}s^2_W
\right) 
\cos(\alpha+\beta)
-{eM^2_t \over s_WM_W}{\sin\alpha \over \sin\beta}, \nonumber \\
g^0_{H^0 \tilde{t}^{}_{R} \tilde{t}^{}_{R}} & = & -{eM_W\over s_Wc^2_W}
\left(
{2\over3}s^2_W 
\right) 
\cos(\alpha+\beta)
-{eM^2_t \over s_WM_W}{\sin\alpha \over \sin\beta}, \nonumber \\
g^0_{H^0 \tilde{t}^{}_{L} \tilde{t}^{}_{R}} & = & 
g^0_{H^0 \tilde{t}^{}_{R} \tilde{t}^{}_{L}} = -{eM_t\over2s_WM_W}
\left( 
{-\mu \cos\alpha + A_t \sin\alpha \over \sin\beta}
\right). \nonumber
\ea
\ba
g^0_{H^0 \tilde{b}^{}_{L} \tilde{b}^{}_{L}} & = & {eM_W\over s_Wc^2_W}
\left(
{1\over2}-{1\over3}s^2_W
\right) 
\cos(\alpha+\beta)
-{eM^2_b \over s_WM_W}{\cos\alpha \over \cos\beta}, \nonumber \\
g^0_{H^0 \tilde{b}^{}_{R} \tilde{b}^{}_{R}} & = & {eM_W\over s_Wc^2_W}
\left(
{1\over3}s^2_W 
\right)  
\cos(\alpha+\beta)
-{eM^2_b \over s_WM_W}{\cos\alpha \over \cos\beta}, \nonumber \\
g^0_{H^0 \tilde{b}^{}_{L} \tilde{b}^{}_{R}} & = & 
g^0_{H^0 \tilde{b}^{}_{R} \tilde{b}^{}_{L}} = -{eM_b\over2s_WM_W}
\left( 
{-\mu \sin\alpha + A_b \cos\alpha \over \cos\beta}
\right). \nonumber
\ea
~~~

For the neutral Higgs boson $h^0$, one has:
\ba
g^0_{h^0 \tilde{t}^{}_{L} \tilde{t}^{}_{L}} & = & {eM_W\over s_Wc^2_W}
\left(
{1\over2}-{2\over3}s^2_W
\right) 
\sin(\alpha+\beta)
-{eM^2_t \over s_WM_W}{\cos\alpha \over \sin\beta}, \nonumber \\
g^0_{h^0 \tilde{t}^{}_{R} \tilde{t}^{}_{R}} & = & {eM_W\over s_Wc^2_W}
\left(
{2\over3}s^2_W 
\right) 
\sin(\alpha+\beta)
-{eM^2_t \over s_WM_W}{\cos\alpha \over \sin\beta}, \nonumber \\
g^0_{h^0 \tilde{t}^{}_{L} \tilde{t}^{}_{R}} & = & 
g^0_{h^0 \tilde{t}^{}_{R} \tilde{t}^{}_{L}} = {eM_t\over2s_WM_W}
\left( 
{-\mu \sin\alpha - A_t \cos\alpha \over \sin\beta}
\right). \nonumber 
\ea
\ba
g^0_{h^0 \tilde{b}^{}_{L} \tilde{b}^{}_{L}} & = & -{eM_W\over s_Wc^2_W}
\left(
{1\over2}-{1\over3}s^2_W
\right)
\sin(\alpha+\beta)
+{eM^2_b \over s_WM_W}{\sin\alpha \over \cos\beta}, \nonumber \\
g^0_{h^0 \tilde{b}^{}_{R} \tilde{b}^{}_{R}} & = & -{eM_W\over s_Wc^2_W}
\left(
{1\over3}s^2_W 
\right) 
\sin(\alpha+\beta)
+{eM^2_b \over s_WM_W}{\sin\alpha \over \cos\beta}, \nonumber \\
g^0_{h^0 \tilde{b}^{}_{L} \tilde{b}^{}_{R}} & = & 
g^0_{h^0 \tilde{b}^{}_{R} \tilde{b}^{}_{L}} = -{eM_b\over2s_WM_W}
\left( 
{-\mu \cos\alpha - A_b \sin\alpha \over \cos\beta}
\right). \nonumber
\ea

Let us know take the sfermion mixing into account and let $R^{\tilde{t}}$ 
and $R^{\tilde{b}}$ be the rotation matrices for $\tilde{t}$ and $\tilde{b}$ 
squarks, respectively. If $\tilde{f}^0_{1,2} \equiv \tilde{f}_{L,R}$, then:
\ba
\tilde{f}_{i} & = & 
R^{\tilde{f}}_{ij}\tilde{f}^0_{j}~\mbox{with}~R^{\tilde{f}}_{ij} = 
\left(
\begin{array}{cc}
c_{\tilde{f}} & s_{\tilde{f}} \\
-s_{\tilde{f}} & c_{\tilde{f}} 
\end{array}
\right). \nonumber
\ea

In the charged Higgs sector, one has:
\ba
g_{H \tilde{t}^{}_{i} \tilde{b}^{}_{j}} & = & 
\sum_{i'j'}R^{\tilde{t}}_{ii'}R^{\tilde{b}}_{jj'}
g^0_{H \tilde{t}^{}_{i'} \tilde{b}^{}_{j'}}. 
\nonumber
\ea

Similarly, in the neutral Higgs sector, for 
$\tilde{f}$ stands for either $\tilde{t}$ 
or $\tilde{b}$, then one obtains the 
following coupling constants:
\ba
g_{A^0 \tilde{f}^{}_{i} \tilde{f}^{}_{j}} & = &  
\sum_{i'j'}R^{\tilde{f}}_{ii'}R^{\tilde{f}}_{jj'}
g^0_{A^0 \tilde{f}^{}_{i'} \tilde{f}^{}_{j'}}
\nonumber
\ea

\ba
g_{H^0 \tilde{f}^{}_{i} \tilde{f}^{}_{j}} & = &  
\sum_{i'j'}R^{\tilde{f}}_{ii'}R^{\tilde{f}}_{jj'}
g^0_{H^0 \tilde{f}^{}_{i'} \tilde{f}^{}_{j'}} 
\nonumber
\ea

\ba
g_{h^0 \tilde{f}^{}_{i} \tilde{f}^{}_{j}} & = &  
\sum_{i'j'}R^{\tilde{f}}_{ii'}R^{\tilde{f}}_{jj'}
g^0_{h^0 \tilde{f}^{}_{i'} \tilde{f}^{}_{j'}}. 
\nonumber
\ea

Note that, in the case of the neutral boson $A^0$, the 
coupling to a pair of sfermions is the same with or 
without mixing.\\

~~

\underline{Higgs - Higgs - Higgs}
\ba
&&g_{H^0HH} =  
{eM_W\over s_W}
\left[
{\cos2\beta\cos(\beta+\alpha)\over2c^2_W}-\cos(\beta-\alpha)
\right], \nonumber \\
&&g_{h^0HH} =  
-{eM_W\over s_W}
\left[
{\cos2\beta\sin(\beta+\alpha)\over2c^2_W}+\sin(\beta-\alpha)
\right], \nonumber \\
&&g_{H^0GG} =  
-{eM_W\over 2s_Wc^2_W}
\left[
\cos2\beta\cos(\beta+\alpha)
\right], \nonumber \\
&&g_{h^0GG} =  
{eM_W\over 2s_Wc^2_W}
\left[
\cos2\beta\sin(\beta+\alpha)
\right], \nonumber \\
&&g_{H^0GH} =  
-{eM_W\over2s_W}
\left[
\sin(\beta-\alpha)-{\sin2\beta\cos(\alpha+\beta)\over c^2_W}
\right], \nonumber \\
&&g_{h^0GH} =  
{eM_W\over2s_W}
\left[
\cos(\beta-\alpha)-{\sin2\beta\sin(\alpha+\beta)\over c^2_W}
\right], \nonumber \\
&&g_{A^0G^{\pm}H^{\pm}} = Q_G \times {eM_W\over2s_W}, \nonumber \\
&&g_{H^0H^0H^0} = -{3eM_W\over2s_Wc^2_W}
\cos2\alpha \cos(\beta+\alpha), \nonumber \\
&&g_{h^0h^0h^0} = -{3eM_W\over2s_Wc^2_W}
\cos2\alpha \sin(\beta+\alpha), \nonumber \\
&&g_{h^0H^0H^0} = {eM_W\over2s_Wc^2_W}
\left[
2\sin2\alpha\cos(\beta+\alpha)+\cos2\alpha\sin(\beta+\alpha)
\right], \nonumber \\
&&g_{H^0h^0h^0} = -{eM_W\over2s_Wc^2_W}
\left[
2\sin2\alpha\sin(\beta+\alpha)-\cos2\alpha\cos(\beta+\alpha)
\right], \nonumber \\
&&g_{H^0G^0G^0} = -{eM_W\over2s_Wc^2_W}
\cos2\beta \cos(\beta+\alpha), \nonumber \\
&&g_{h^0G^0G^0} = {eM_W\over2s_Wc^2_W}
\cos2\beta \sin(\beta+\alpha), \nonumber \\
&&g_{H^0A^0A^0} = {eM_W\over2s_Wc^2_W}
\cos2\beta \cos(\beta+\alpha), \nonumber\\ 
&&g_{h^0A^0A^0} = -{eM_W\over2s_Wc^2_W}
\cos2\beta \sin(\beta+\alpha), \nonumber \\ 
&&g_{H^0A^0G^0} = {eM_W\over2s_Wc^2_W}
\sin2\beta \cos(\beta+\alpha), \nonumber \\
&&g_{h^0A^0G^0} = -{eM_W\over2s_Wc^2_W}
\sin2\beta \sin(\beta+\alpha). \nonumber 
\ea

\underline{Higgs - Higgs - Sfermion - Sfermion}\\

For light unmixed sfermions, the coupling constants are not 
proportional to the mass of the fermion(s) and can not be 
neglected. The coupling constants for the heavy sfermions 
are then obtained by adding a term proportional to the mass 
of the corresponding fermion(s).\\

If $\tilde{f}$ is a slepton 
($\tilde{\ell}$ or $\tilde{\nu}$) or a squark from the first or 
second generation ($\tilde{q}$), one has:
\ba
g_{H^0H^0\tilde f_L\tilde f_L} & = & {e^2\over2s^2_W} 
\left[
-{\cos2\alpha\over c^2_W} \left( T^3_f - Q_f s^2_W \right)
\right], \nonumber \\
g_{H^0H^0\tilde f_R\tilde f_R} & = & {e^2\over2s^2_W} 
\left[
-{\cos2\alpha\over c^2_W} \left( Q_f s^2_W \right)
\right], \nonumber \\
g_{h^0h^0\tilde f_L\tilde f_L} & = & {e^2\over2s^2_W} 
\left[
{\cos2\alpha\over c^2_W} \left( T^3_f - Q_f s^2_W \right)
\right], \nonumber \\
g_{h^0h^0\tilde f_R\tilde f_R} & = & {e^2\over2s^2_W} 
\left[
{\cos2\alpha\over c^2_W} \left( Q_f s^2_W \right)
\right], \nonumber \\
g_{H^0h^0\tilde f_L\tilde f_L} & = & {e^2\over2s^2_W} 
\left[ 
{\sin2\alpha\over c^2_W} \left( T^3_f - Q_f s^2_W \right)
\right], \nonumber \\
g_{H^0h^0\tilde f_R\tilde f_R} & = & {e^2\over2s^2_W} 
\left[ 
{\sin2\alpha\over c^2_W} \left( Q_f s^2_W \right)
\right], \nonumber \\
g_{A^0A^0\tilde f_L\tilde f_L} & = & {e^2\over2s^2_W} 
\left[
{\cos2\beta\over c^2_W} \left( T^3_f - Q_f s^2_W \right)
\right], \nonumber \\
g_{A^0A^0\tilde f_R\tilde f_R} & = & {e^2\over2s^2_W} 
\left[
{\cos2\beta\over c^2_W} \left( Q_f s^2_W \right)
\right].\nonumber
\ea

For the third generation squarks (stop and sbottom), one has:
\ba
g_{H^0H^0\tilde t_{L,R}\tilde t_{L,R}} & = & 
g_{H^0H^0\tilde u_{L,R}\tilde u_{L,R}} -
{e^2\over2s^2_W} \left ( {M_t\sin\alpha \over M_W\sin\beta} \right)^2,
\nonumber \\
g_{h^0h^0\tilde t_{L,R}\tilde t_{L,R}} & = &
g_{h^0h^0\tilde u_{L,R}\tilde u_{L,R}} -
{e^2\over2s^2_W} \left ( {M_t\cos\alpha \over M_W\sin\beta} \right)^2,
\nonumber \\
g_{H^0h^0\tilde t_{L,R}\tilde t_{L,R}} & = & 
g_{H^0h^0\tilde u_{L,R}\tilde u_{L,R}} -
{e^2\sin2\alpha\over4s^2_W} \left ( {M_t \over M_W\sin\beta} \right)^2,
\nonumber \\
g_{A^0A^0\tilde t_{L,R}\tilde t_{L,R}} & = & 
g_{A^0A^0\tilde u_{L,R}\tilde u_{L,R}} -
{e^2\over2s^2_W} \left ( {M_t\cos\beta \over M_W\sin\beta} \right)^2
\nonumber
\ea
and
\ba
g_{H^0H^0\tilde b_{L,R}\tilde b_{L,R}} & = & 
g_{H^0H^0\tilde d_{L,R}\tilde d_{L,R}} -
{e^2\over2s^2_W} \left ( {M_b\cos\alpha \over M_W\cos\beta} \right)^2,
\nonumber \\
g_{h^0h^0\tilde b_{L,R}\tilde b_{L,R}} & = &
g_{h^0h^0\tilde d_{L,R}\tilde d_{L,R}} -
{e^2\over2s^2_W} \left ( {M_b\sin\alpha \over M_W\cos\beta} \right)^2,
\nonumber \\
g_{H^0h^0\tilde b_{L,R}\tilde b_{L,R}} & = & 
g_{H^0h^0\tilde d_{L,R}\tilde d_{L,R}} -
{e^2\sin2\alpha\over4s^2_W} \left ( {M_b \over M_W\cos\beta} \right)^2,
\nonumber \\
g_{A^0A^0\tilde b_{L,R}\tilde b_{L,R}} & = & 
g_{A^0A^0\tilde d_{L,R}\tilde d_{L,R}} -
{e^2\over2s^2_W} \left ( {M_b\sin\beta \over M_W\cos\beta} \right)^2.
\nonumber
\ea

\newpage

\section*{Appendix B: Passarino-Veltman functions}

The calculation of one loop Feynman diagrams can be performed by 
combining propagators using the following formula:
$$
\frac 1 {A_1^{\alpha_1}\cdots A_n^{\alpha_n}} = 
\frac{\Gamma(\alpha_1+\cdots+\alpha_n)}{\Gamma(\alpha_1)\cdots\Gamma(\alpha_n)}
\int_0^1 dx_1\cdots dx_n\delta(x_1+\cdots
+x_n-1)\frac{x_1^{\alpha_1-1}\cdots x_n^{\alpha_n-1}}{(A_1 x_1+\cdots +A_n
x_n)^{\alpha_1+\cdots +\alpha_n}}. \nonumber
$$

However, it is often convenient to reduce each one loop diagram to the sum 
of standard contributions, the so-called Passarino-Veltman (PV) functions.

\subsection*{B.1~Standard definitions}

Let us define the 1, 2, 3 and 4 point functions according to: 
\ba
A_0(a) &=& \int \frac{d^Dk}{i\pi^2}\frac{1}{N_a}, \nonumber\\
\{B_0,B^\mu, B^{\mu\nu}\}(ab) &=& \int \frac{d^Dk}{i\pi^2}
\frac{\{1,k^\mu,k^\mu k^\nu\}}{N_a N_b}, \nonumber\\
\{C_0, C^\mu, C^{\mu\nu}\}(abc) &=& \int \frac{d^Dk}{i\pi^2}
\frac{\{1, k^\mu, k^\mu k^\nu\}}{N_a N_b N_c}, \nonumber\\
\{D_0, D^\mu, D^{\mu\nu}, D^{\mu\nu\rho}\}(abcd) &=& \int \frac{d^Dk}{i\pi^2}
\frac{\{1, k^\mu, k^\mu k^\nu, k^\mu k^\nu k^\rho\}}
{N_a N_b N_c N_d}, \nonumber
\ea
where the denominators are:
\ba
N_1 &=& k^2-m_1^2+i\epsilon, \nonumber\\
N_2 &=& (k+p_1)^2-m_2^2+i\epsilon, \nonumber\\
N_3 &=& (k+p_1+p_2)^2-m_3^2+i\epsilon, \nonumber\\
N_4 &=& (k+p_1+p_2+p_3)^2-m_4^2+i\epsilon. \nonumber
\ea

Here, all integrals are kept $D$-dimensional. However, the rest of the 
calculations will be performed in four dimensions.\\

In one loop diagrams, the following conventions are used:
\begin{itemize}
\item the external momenta $p_{1 \dots N}$ are oriented clockwise,
\item the internal masses $m_{1 \dots N}$ are oriented clockwise as well, 
with $m_1$ between $p_N$ and $p_1$.
\end{itemize}

Let $K$ be a multi-index such that:
\ba
B_K(12) &=& B_K(p_1^2, m_1^2, m_2^2), \nonumber \\
C_K(123) &=& C_K(p_1^2, p_2^2, (p_1+p_2)^2, m_1^2, m_2^2, m_3^2),
\nonumber \\
D_K(1234) &=& D_K(p_1^2, p_2^2, p_3^2, (p_1+p_2+p_3)^2,
(p_1+p_2)^2, (p_2+p_3)^2, m_1^2, m_2^2, m_3^2, m_4^2). \nonumber
\ea

The reduction of tensorial functions into scalar functions can then be done 
according to the following standard notations:
\ba
B^\mu(12) &=& p_1^\mu B_1(12), \nonumber\\
B^{\mu\nu}(12) &=& p_1^\mu p_1^\nu B_{21}(12)+g^{\mu\nu} B_{22}(12), 
\nonumber
\ea
\ba
C^\mu(123) &=& p_1^\mu C_{11}(123)+p_2^\mu C_{12}(123), \nonumber\\
C^{\mu\nu}(123) &=& p_1^\mu p_1^\nu C_{21}(123)+ p_2^\mu p_2^\nu
C_{22}(123) + p_1^{\{\mu} p_2^{\nu\}} C_{23}(123)+g^{\mu\nu} C_{24}(123),
\nonumber\\
C^{\mu\nu\rho}(123) &=& 
(g^{\mu\nu}p_1^\rho + g^{\mu\rho}p_1^\nu+g^{\nu\rho}p_1^\mu) C_{001}(123) 
\nonumber \\
&+&
(g^{\mu\nu}p_2^\rho + g^{\mu\rho}p_2^\nu+g^{\nu\rho} p_2^\mu) C_{002}(123) 
\nonumber\\
&+&
p_1^\mu p_1^\nu p_1^\rho C_{111}(123) + 
p_2^\mu p_2^\nu p_2^\rho C_{222}(123) \nonumber\\
&+& 
(p_1^\mu p_1^\nu p_2^\rho + p_1^\mu p_2^\nu p_1^\rho + 
p_2^\mu p_1^\nu p_1^\rho  ) C_{112}(123) \nonumber\\  
&+& (p_2^\mu p_2^\nu p_1^\rho + p_2^\mu p_1^\nu p_2^\rho 
+ p_1^\mu p_2^\nu p_2^\rho  ) C_{122}(123), \nonumber
\ea
\ba
D^\mu(1234) &=& p_1^\mu D_{11}(1234)+p_2^\mu D_{12}(1234)+
p_3^\mu D_{13}(1234),
\nonumber\\
D^{\mu\nu}(1234) &=& p_1^\mu p_1^\nu D_{21}(1234)+ p_2^\mu p_2^\nu
D_{22}(1234) + p_3^\mu p_3^\nu D_{23}(1234) \nonumber\\
&+& p_1^{\{\mu} p_2^{\nu\}} D_{24}(1234)+
p_1^{\{\mu} p_3^{\nu\}} D_{25}(1234)+ 
p_2^{\{\mu} p_3^{\nu\}} D_{26}(1234)+
g^{\mu\nu} D_{27}(1234), \nonumber\\
D^{\mu\nu\rho}(1234) &=& 
(g^{\mu\nu}p_1^\rho+g^{\nu\rho}p_1^\mu +
g^{\mu\rho}p_1^\nu)D_{001}(1234) \nonumber \\
&+&
(g^{\mu\nu}p_2^\rho+g^{\nu\rho}p_2^\mu+g^{\mu\rho}p_2^\nu)D_{002}(1234)
\nonumber \\
&+& 
(g^{\mu\nu}p_3^\rho+g^{\nu\rho}p_3^\mu+g^{\mu\rho}p_3^\nu)D_{003}(1234)
\nonumber \\
&+& 
\sum_{1\le ijk\le 3} p_i^\mu p_j^\nu p_k^\rho D_{ijk}(1234).
 \nonumber 
\ea

In the reduction of $D^{\mu\nu\rho}$, the sum is over all triplets
$(i, j, k)$ with repetitions, i.e. $3^3=27$ terms. By construction, 
the coefficients $D_{ijk}$ are invariant under index permutations.\\

More details about this approach and about the reduction of the PV tensorial 
integrals into scalar ones can be found in~\cite{pv}.

\subsection*{B.2~{\tt LoopTools} definitions}

Sometimes, as for instance in the {\tt LoopTools} library~\cite{LoopTools}, 
it is convenient to use another notation and to introduce momenta $k_i$ 
given by:
\ba
k_1 &=& p_1, \nonumber\\
k_2 &=& p_1+p_2, \nonumber\\
k_3 &=& p_1+p_2+p_3... \nonumber \\
k_N &=& \sum_{i=1}^{N} p_i \nonumber
\ea

In that case, the tensorial coefficients are:
\ba
B^\mu &=& k_1^\mu B_1^L, \nonumber\\
B^{\mu\nu} &=& k_1^\mu k_1^\nu B_{11}^L+g^{\mu\nu} B_{00}^L. \nonumber
\ea
\ba
C^\mu &=& k_1^\mu C_1^L+k_2^\mu C_2^L, \nonumber\\
C^{\mu\nu} &=& \sum_{ij=1,2} k_i^\mu k_j^\nu C_{ij}^L+
g^{\mu\nu} C_{00}^L, \nonumber\\
C^{\mu\nu\rho} &=& \sum_{ijl=1,2} k_i^\mu k_j^\nu k_l^\rho C_{ijl}^L + 
\sum_{i=1,2} (g^{\mu\nu}k_i^\rho + g^{\mu\rho}k_i^\nu + g^{\nu\rho}k_i^\mu) 
C_{00i}^L, \nonumber
\ea
\ba
D^\mu &=& k_1^\mu D_1^L + k_2^\mu D_2^L + k_3^\mu D_3^L, \nonumber\\
D^{\mu\nu} &=& \sum_{1\le ijk\le 3} k_i^\mu k_j^\nu D_{ij}^L + 
g^{\mu\nu} D_{00}^L, \nonumber 
\ea
where $C^L_{ij}$, $C^L_{ijl}$ and $D^L_{ij}$ are completely symmetric.\\

By expanding these equations and by then comparing all their terms to those 
arising from the reduction of standard tensorial functions, one can find 
the relations that exist between the standard PV functions and those which 
are computed in the {\tt LoopTools} library.\\

For the 2 point functions, one obtains:
\ba
B_1 &=& B_1^L \nonumber\\
&\bullet&\nonumber\\
B_{21} &=& B_{11}^L \nonumber\\
B_{22} &=& B_{00}^L \nonumber
\ea

For the 3 point functions, one obtains:
\ba
C_{11} &=& C_1^L + C_2^L \nonumber\\
C_{12} &=& C_2^L \nonumber\\
&\bullet&\nonumber\\
C_{21} &=& C_{11}^L+2C_{12}^L+C_{22}^L \nonumber\\
C_{22} &=& C_{22}^L \nonumber\\
C_{23} &=& C_{12}^L + C_{22}^L \nonumber\\
C_{24} &=& C_{00}^L \nonumber\\
&\bullet&\nonumber\\
C_{001} &=& C_{001}^L+C_{002}^L\nonumber\\
C_{002} &=& C_{002}^L\nonumber\\
&\bullet&\nonumber\\
C_{111} &=& C_{111}^L+3C_{112}^L+3C_{122}^L+C_{222}^L\nonumber\\
C_{222} &=& C_{222}^L\nonumber\\
C_{112} &=& C_{112}^L+2C_{122}^L+C_{222}^L\nonumber\\
C_{122} &=& C_{122}^L+C_{222}^L \nonumber
\ea
~~\\

For the 4 point functions, one obtains:
\ba
D_{11} &=& D_1^L+D_2^L+D_3^L \nonumber\\
D_{12} &=& D_2^L+D_3^L \nonumber\\
D_{13} &=& D_3^L \nonumber \\
&\bullet&\nonumber\\
D_{21} &=& D_{11}^L+D_{22}^L+D_{33}^L+2(D_{12}^L+D_{13}^L+D_{23}^L) \nonumber\\
D_{22} &=& D_{22}^L+2D_{23}^L+D_{33}^L\nonumber\\
D_{23} &=& D_{33}^L\nonumber\\
D_{24} &=& D_{12}^L+D_{13}^L+D_{22}^L+2D_{23}^L+D_{33}^L\nonumber\\
D_{25} &=& D_{13}^L+D_{23}^L+D_{33}^L\nonumber\\
D_{26} &=& D_{23}^L+D_{33}^L\nonumber\\
D_{27} &=& D_{00}^L\nonumber\\
&\bullet&\nonumber\\
D_{001} &=& D^L_{001}+D^L_{002}+D^L_{003}\nonumber\\
D_{002} &=& D^L_{002}+D^L_{003}\nonumber\\
D_{003} &=& D^L_{003}\nonumber\\
&\bullet&\nonumber\\
D_{111} &=& D^L_{111} +3 D^L_{112}+3 D^L_{113}+ 3 D^L_{122} + 
6 D^L_{123}+3 D^L_{133}+ D^L_{222} + 3 D^L_{223} + 3 D^L_{233} + D^L_{333}
\nonumber\\
D_{112} &=& D^L_{112} +D^L_{113}+2 D^L_{122}+ 4 D^L_{123}+ 
2 D^L_{133}+ D^L_{222} + 3 D^L_{223} + 
3 D^L_{233} + D^L_{333}\nonumber\\
D_{113} &=& D^L_{113} + 2 D^L_{123} + 2 D^L_{133} + D^L_{223} + 
2 D^L_{233} + D^L_{333}\nonumber\\
D_{122} &=& D^L_{122} + 2 D^L_{123} + D^L_{133} + D^L_{222} + 
3 D^L_{223} + 3 D^L_{233} + D^L_{333}\nonumber\\
D_{133} &=& D^L_{133} + D^L_{233} + D^L_{333}\nonumber\\
D_{123} &=& D^L_{123} + D^L_{133} + D^L_{223} + 2 D^L_{233} + 
D^L_{333}\nonumber\\
&\bullet&\nonumber\\
D_{222} &=& D^L_{222} + 3 D^L_{223} + 3 D^L_{233} + D^L_{333}\nonumber\\
D_{223} &=& D^L_{223} + 2 D^L_{233} + D^L_{333}\nonumber\\
D_{233} &=& D^L_{233} + D^L_{333}\nonumber\\
&\bullet&\nonumber\\
D_{333} &=& D^L_{333} \nonumber
\ea

\end{document}